\documentclass[twocolumn,amsmath,amssymb,aps,prb,reprint,longbibliography]{revtex4-1}
\usepackage{graphicx}
\begin{document}

\title{Dynamics and Nonmonotonic Drag for Individually Driven Skyrmions  }
 
\author{C. Reichhardt and C. J. O. Reichhardt}
\affiliation{Theoretical Division and Center for Nonlinear Studies,
Los Alamos National Laboratory, Los Alamos, New Mexico 87545, USA}

\date{\today}

\begin{abstract}
  We examine the motion of an individual skyrmion driven through an assembly of other skyrmions by a constant or increasing force in the absence of quenched disorder. The skyrmion behavior is determined by the ratio of the damping and Magnus terms, as expressed in terms of the intrinsic skyrmion Hall angle. For a fixed driving force in the damping dominated regime, the effective viscosity decreases monotonically with increasing skyrmion density, similar to what is observed in overdamped systems where it becomes difficult for the driven particle to traverse the surrounding medium at high densities. In contrast, in the Magnus dominated regime the velocity dependence on the density is nonmonotonic, and there is a regime in which the skyrmion moves faster with increasing density, as well as a pronounced speed-up effect in which a skyrmion traveling through a dense medium moves more rapidly than it would at low densities or in the single particle limit. At higher densities, the effective damping increases and the velocity decreases. The velocity-force curves in the Magnus-dominated regime show marked differences from those in the damping-dominated regimes. Under an increasing drive we find that there is a threshold force for skyrmion motion which increases with density. Additionally, the skyrmion Hall angle is drive dependent, starting near zero at the threshold for motion and increasing with increasing drive before reaching a saturation value, similar to the behavior found for skyrmions driven over quenched disorder. We map dynamic phase diagrams showing the threshold for motion, nonlinear flow, speed-up, and saturation regimes. We also find that in some cases, increasing the density can reduce the skyrmion Hall angle while producing a velocity boost, which could be valuable for applications.
\end{abstract}

\maketitle

\vskip 2pc

\section{Introduction}

Skyrmions are particle-like magnetic textures
\cite{Muhlbauer09,Yu10,Nagaosa13,EverschorSitte18,Reichhardt21} 
that have been identified in a growing number of materials, 
including numerous systems
in which they are stable at room temperature
\cite{Woo16,Soumyanarayanan17,Montoya18}. 
Skyrmions can be set into motion with applied currents
\cite{Nagaosa13,EverschorSitte18,Woo16,Montoya18,Jonietz10,Yu12,Iwasaki13},
magnetic field gradients \cite{Zhang18},
and thermal gradients \cite{Mochizuki14}.
Due to their size scale, room temperature stability,  
and mobility, skyrmions are promising candidates for various applications 
including memory
\cite{Fert13,Tomasello14,Fert17}
and novel computing architectures
\cite{Zazvorka19,Zhang20a,Pinna20}.
In terms of basic science, 
skyrmions represent a new class of systems that exhibit 
collective dynamics under an applied drive
\cite{Fisher98,Reichhardt17},
similar to vortices in superconductors
\cite{Bhattacharya93,Blatter94},
colloids \cite{Tierno12a}, Wigner crystals \cite{Reichhardt01}
frictional systems \cite{Vanossi13},
and other soft matter systems \cite{OHern03}.
A unique feature which distinguishes skyrmions from 
these other systems is the strong gyroscopic or Magnus force
component of their dynamics,
which generates velocities perpendicular to the 
net force experienced by the skyrmion
\cite{Nagaosa13,EverschorSitte18,Jonietz10,Iwasaki13,EverschorSitte14}.
One consequence of this
is that a skyrmion moves at an
angle, known as the skyrmion Hall angle,
with respect to an applied driving force
\cite{Nagaosa13,EverschorSitte14,Reichhardt15a,Reichhardt15,Jiang17,Litzius17}.
The Magnus term also produces many other effects,
including accelerations or speed-ups of skyrmions
interacting with barriers or walls
\cite{Iwasaki14,Zhang15a,Reichhardt16a,CastellQueralt19,Xing20},
ratchet effects \cite{Reichhardt15aa,Ma17,Chen20,Gobel21},
and spiraling motion in a confined potential 
\cite{Reichhardt21,Buttner15,Fernandes20} or after a quench \cite{Brown18}.
Such effects can be strongly 
modified by pinning or skyrmion-skyrmion interactions \cite{Reichhardt21}. 

Beyond moving skyrmions
with currents, other methods that can be used to manipulate individual skyrmions
include scanning tunneling microscope tips
\cite{Hanneken16}, local magnetic fields \cite{Wang17,Casiraghi19}, 
and optical traps \cite{Wang20b}. 
Extensive experimental and theoretical studies have been performed
in systems such as colloids
\cite{Hastings03,Habdas04,Squires05,Gazuz09,Winter12},
granular matter \cite{Drocco05,Candelier10,Reichhardt10,Kolb13,Reichhardt19a},
active matter \cite{Reichhardt15aaa}, 
superconducting vortices
\cite{Straver08,Reichhardt09a,Auslaender09,Veshchunov16,Kremen16,Crassous11},
various soft matter systems
\cite{Reichhardt08,Wulfert17,Zia18,Wang19}, and metallic glasses \cite{Yu20}
examining the dynamics of individually driven
particles moving through an assembly of other
particles
under either a constant force or an increasing force. 
Measurements of the velocity-force relations or
the drag on the driven particle provide information 
about how the effective viscosity of the
system changes across density-induced transitions such as solid to liquid
\cite{Reichhardt04a,Dullens11}, 
liquid to glass \cite{Habdas04,Gazuz09}, or unjammed to jammed
\cite{Drocco05,Candelier10,Reichhardt19a}.
For example, the viscosity can increase strongly at the onset of
a glass or solid phase produced by an increase in the density.
When the probe particle is driven with
an increasing force,
there can be a force threshold below which the particle is
unable to move
\cite{Hastings03,Habdas04,Senbil19,Gruber20},
as well as distinct features in the velocity-force curves
\cite{Benichou13a,Illien18}
which change as the density increases.

In soft matter systems, driving a single probe
particle through a background of other particles  
is known as active rheology
\cite{Gazuz09,Reichhardt19a,Reichhardt15aaa,Zia18,Gruber20}.
Similar probes have been applied in
type-II superconductors using individually dragged vortices
\cite{Straver08,Reichhardt09a,Auslaender09,Veshchunov16,Kremen16,Crassous11}. 
Existing work on
active rheology has involved overdamped systems, where the viscosity
generally increases with increasing density.
It is not clear how the addition of a Magnus term
would impact active rheology.
For example, it could change the threshold force for motion, 
increase or decrease the drag,
modify the Hall angle, or
alter the way in which the probe particle moves with respect to the
background particles.
Beyond skyrmions,
there are many other chiral systems in which Magnus or gyroscopic effects
can be important
for active rheology, including chiral soft matter
fluids and solids, active chiral systems
\cite{Kokot17,Banerjee17,Soni19,Scholz21,Reichhardt20b}, 
charged particles in a magnetic field
\cite{Schecter99,Schirmacher15}, 
magnetic colloids in oscillating fields \cite{Tierno07},
spinning particles in fluids \cite{Grzybowski01,Denisov18},
fluid vortices \cite{Ryzhov13,Aref07},
and fracton systems \cite{Doshi21}.   

In this work we study active rheology for skyrmion systems
in the absence of quenched disorder using a
particle-based model. We apply an external driving force to one
skyrmion and measure the velocity components and drag
both parallel and perpendicular to the driving direction.
For a constant drive,
the velocity decreases
monotonically with increasing skyrmion density
in the damping-dominated regime.
This is
similar to what has been found for the active rheology of overdamped systems,
where 
it becomes more difficult for the particle to pass through a denser medium
due to the increased interactions with other particles.
When the Magnus force is dominant, the
driven skyrmion velocity has a strongly
non-monotonic density dependence and
{\it increases} with increased density.
In this regime, the velocity can be boosted or accelerated beyond
the expected velocity for an isolated
driven skyrmion in the absence of surrounding skyrmions.
As the density increases,
the skyrmion Hall angle decreases and
the velocity reaches a peak value
before decreasing at higher densities.  
The velocity boost effect arises
due to the creation of a locally asymmetric density profile around
the driven skyrmion oriented perpendicular to the driving direction.
This 
creates an additional repulsive force on the driven skyrmion
which does not cancel out due to the lack of symmetry and
is converted by the Magnus term
into a velocity component aligned with the driving direction.
The behavior
is similar to the acceleration effect found for skyrmions interacting 
with walls and barriers
\cite{Iwasaki14,Zhang15a,Reichhardt16a,CastellQueralt19,Xing20}.  

For the case of a driving force that is increased from zero,
the driven skyrmion exhibits
a critical threshold for the onset of motion.
The skyrmion Hall angle
is zero at the threshold and increases
with increasing velocity
until it saturates to a constant value at high drives,
similar to the behavior found
for skyrmions driven over quenched disorder
\cite{Reichhardt21,Reichhardt15,Jiang17,Litzius17,Reichhardt16,Legrand17,Diaz17,Juge19,Zeissler20,Litzius20}. 
In the Magnus dominated regime, the boost effect is nonmonotonic, with
reduced boost at low and high drives and
maximum boost at intermediate drives. 
This drive dependence
originates from the reaction of the surrounding skyrmions to the motion
of the driven skyrmion.
Local density fluctuations of the background skyrmions have time to relax
when the driven skyrmion velocity is low,
but do not have time to form when the velocity is high.
In each case the boost effect is reduced.
In contrast, for intermediate driven skyrmion velocities,
density fluctuations in the surrounding skyrmions
form and relax on the same time scale as the motion of the
driven skyrmion.

In the strongly damped regime, the velocity-force curves are similar to
those observed for the active rheology of overdamped systems,
while in the Magnus-dominated regime,
negative differential conductivity can appear when
part of the velocity in the driving direction gets transferred
into the direction perpendicular to the drive.
We also find that under constant driving, there are regimes in which
an increase in density produces a velocity boost with a reduction of the
skyrmion Hall angle.  This indicates that
harnessing skyrmion-skyrmion interactions
may be a viable method for reducing the skyrmion Hall angle without
reducing the skyrmion speed, which could be important for applications.
We map dynamic phase diagrams as a function of skyrmion density,
driving force, and the ratio of the damping term to the Magnus force.
We discuss
possible experimental realizations and
the application of our results
to the broader class of systems with gyroscopic forces.

\section{Simulation}

We consider an assembly of $N$ skyrmions in a 
two-dimensional system with periodic boundary conditions
in the $x$ and $y$-directions.
A single skyrmion is coupled to an external drive and is
driven through the other skyrmions
in the absence of quenched disorder.
The skyrmion dynamics are obtained using a particle-based model
from the modified Thiele equation
\cite{Reichhardt21,Reichhardt15a,Reichhardt15,Brown18,Lin13}. 
The equation of motion for skyrmion $i$ is given by 
\begin{equation} 
\alpha_d {\bf v}_{i} + \alpha_m {\hat z} \times {\bf v}_{i} =
{\bf F}^{ss}_{i} + {\bf F}^{D}_{i}  
\end{equation}
where
${\bf v}_{i} = {d {\bf r}_{i}}/{dt}$
is the skyrmion velocity,
$\alpha_{d}$ is the damping coefficient,
and $\alpha_{m}$ is the coefficient for the Magnus force.
The damping term
aligns the velocities in the direction of the net external forces,
while the Magnus term creates a perpendicular velocity component.
The skyrmion-skyrmion interaction force is
${\bf F}^{ss}_{i} = \sum^{N}_{j=1}K_{1}(r_{ij}){\hat {\bf r}_{ij}}$,
where $r_{ij} = |{\bf r}_{i} - {\bf r}_{j}|$
is the distance between skyrmions $i$ and $j$
and $K_1$ is the modified Bessel function. 
The driving force ${\bf F}^{D}_i=F^D{\bf \hat{x}}$ is applied
only to the driven skyrmion
and is zero for all other skyrmions.

In the absence of other skyrmions,
the driven skyrmion
would move in the direction of
the intrinsic Hall angle,
$\theta^{\rm int}_{sk} = -\arctan(\alpha_{m}/\alpha_{d})$.
In the overdamped limit where $\alpha_{m} = 0$,
$\theta^{\rm int}_{sk} = 0^\circ$.
When skyrmion-skyrmion interactions occur,
the skyrmion Hall angle is reduced below
its intrinsic value and is defined as
$\theta_{sk} = \arctan(\langle V_{\perp}\rangle/\langle V_{||}\rangle)$,
where
$\langle V_{||}\rangle$ and
$\langle V_{\perp}\rangle$
are the average velocities of
the driven skyrmion
parallel and perpendicular to the direction of the drive, respectively.
In this work, each individual simulation is performed at a
constant value of $F_D$ and lasts
$3\times 10^5$ to $1 \times 10^6$
simulation time steps to avoid
transient effects.
For drives very near the critical threshold for motion,
much larger time intervals are needed to reach a steady state
\cite{Reichhardt09}. 
The density of the system is given by $\rho = N/L^2$, where
the system size $L = 36$.    

\section{Results}

\begin{figure}
\includegraphics[width=3.5in]{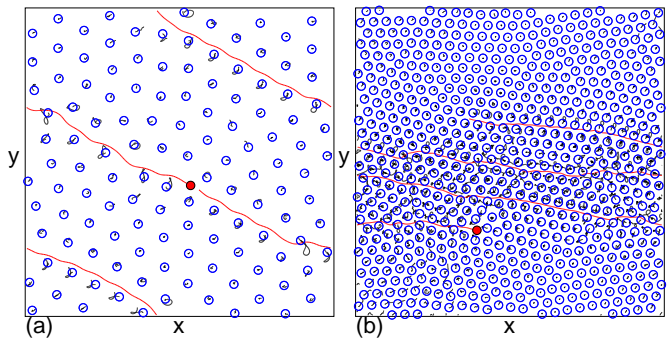}
\caption{Driven skyrmion and trajectory (red) along with the surrounding
  skyrmions (blue) and their trajectories (black) in a sample with
  $\alpha_m/\alpha_d=0.98$ and an
  intrinsic skyrmion Hall angle of $\theta^{\rm int}_{sk}=-44.4^\circ$ at
  a driving force of $F_D=0.5$.
  (a)
  At $\rho = 0.1$,
  the driven skyrmion moves at an angle of
  $\theta_{sk}\approx -29^\circ$. 
  (b)
  At $\rho = 0.5$,
  the magnitude of the skyrmion Hall angle $\theta_{sk}$ is reduced due to
  the increasing number of collisions with the other skyrmions.}
\label{fig:1}
\end{figure}

In Fig.~\ref{fig:1}(a) we show an example of our system
for a collection of skyrmions at a density of $\rho = 0.1$.
The red particle is being 
driven at a constant force $F_{D} = 0.5$,
while the blue particles are the non-driven or bath skyrmions.
For this system, $\theta^{\rm int}_{sk} = -44.4^\circ$,
so the ratio between the damping and Magnus forces
is close to one.
The driven skyrmion
moves on average at an angle of $\theta_{sk}=-29^\circ$.
This is smaller in magnitude than the intrinsic skyrmion Hall angle,
indicating that collisions with
the bath skyrmions are reducing $\theta_{sk}$.
The magnitude of  $\theta_{sk}$ 
increases for higher values of $F_{D}$
but decreases with increasing $\rho$.
Figure~\ref{fig:1}(b) shows the same
system at a higher density of $\rho = 0.5$, where
the driven skyrmion moves at
an even smaller $\theta_{sk}$.
The skyrmion Hall angle and net velocity depend
strongly on the skyrmion density, the driving force, and the
ratio of the Magnus force to the damping term.

If the driven skyrmion did not collide with any other skyrmions, it would
move at the 
intrinsic skyrmion Hall angle of
$\theta_{sk}^{\rm int} = -\arctan(\alpha_{m}/\alpha_{d})$
with an average absolute velocity of
$|V| = F_{D}/\sqrt{\alpha_{d}^2 + \alpha^2_{m}}$. 
We first consider
systems in which we constrain
$\alpha^2_{d} + \alpha^{2}_{m} = 1.0$.
Under this condition, we can define a velocity $V_0=|V|_{N=1}=F_D$ to be the
velocity in the single particle limit.
Once skyrmion-skyrmion interactions are introduced,
in overdamped systems under fixed driving we expect to find
$|V| < V_{0}$.
The velocity 
decreases with increasing $\rho$ and there
is a critical density $\rho_{c}$ at which $|V| = 0$. 

\begin{figure}
\includegraphics[width=3.5in]{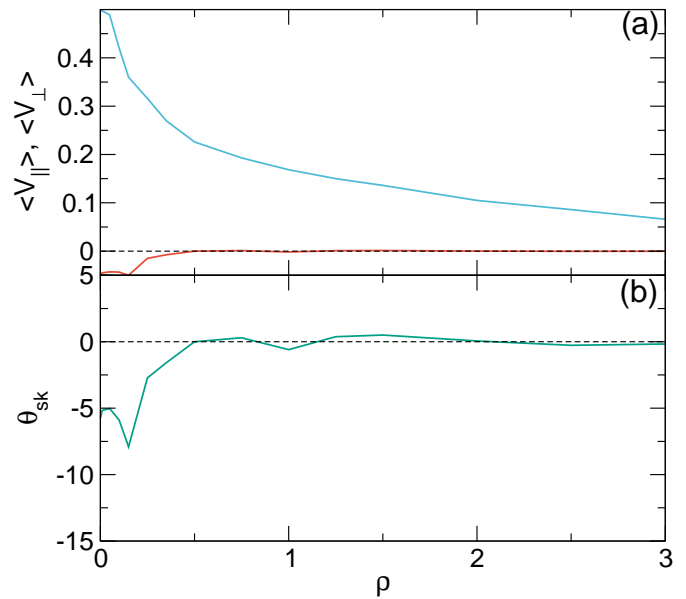}
\caption{(a) $\langle V_{||}\rangle$ (blue) and $\langle V_{\perp}\rangle$
(red) versus $\rho$ for a system with  
$\alpha_{m}/\alpha_{d} = 0.1$,  $\alpha^2_{d} + \alpha^{2}_{m} = 1.0$,  
$\theta^{\rm int}_{sk} = -5.739^\circ$, and $F_{D} = 0.5$.
In this damping dominated regime,
$\langle V_{||}\rangle$ decreases monotonically with
increasing $\rho$.
For all values of $\rho$, $\langle V_{||}\rangle \leq V_0 = 0.5$.
(b) The corresponding $\theta_{sk} = \arctan(\langle V_{\perp}\rangle/
\langle V_{||}\rangle)$.} 
\label{fig:2}
\end{figure}

\begin{figure}
\includegraphics[width=3.5in]{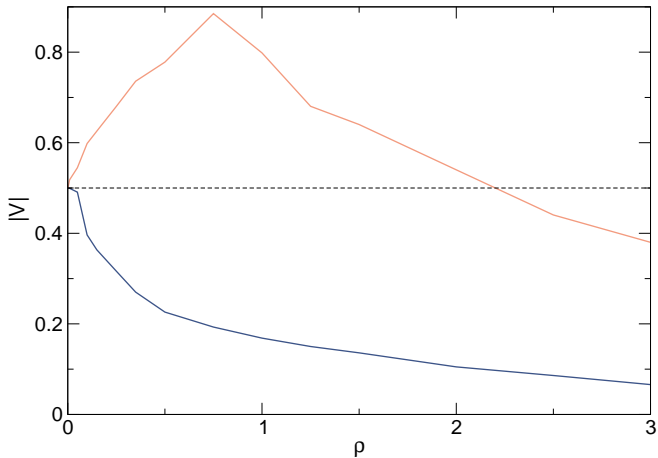}
\caption{
$|V|  = \sqrt{\langle V_{||}\rangle^2 + \langle V_{\perp}\rangle^2}$
versus $\rho$ for the system in Fig.~\ref{fig:2} (blue)
in the damping dominated regime
with $\alpha_m/\alpha_d=0.1$, $\alpha_m^2+\alpha_d^2=1.0$,
$\theta_{sk}^{\rm int}=-5.739^\circ$, and $F_D=0.5$, 
and for the system in Fig.~\ref{fig:3} (orange)
in the Magnus dominated regime 
with $\alpha_m/\alpha_d=9.95$ and $\theta_{sk}^{\rm int}=-84.26^\circ$.
The dashed line indicates the expected response for
an isolated particle with $|V|=V_0=F_{D}= 0.5$. 
In the Magnus dominated regime, there is a range of density over which
$|V|$
increases with increasing $\rho$ and is higher than $V_{0}$,
indicating the existence of a velocity boost.}     
\label{fig:4}
\end{figure}

In Fig.~\ref{fig:2}(a) we plot $\langle V_{||}\rangle$ and
$\langle V_{\perp}\rangle$ versus $\rho$ for
a system
in the damping dominated regime 
with $\alpha_{m}/\alpha_{d} = 0.1$, 
$\theta^{\rm int}_{sk} = -5.739^\circ$, and $F_{D} = 0.5$.
As $\rho$ increases, $\langle V_{\perp}\rangle$ and $\theta_{sk}$
both decrease in magnitude and approach zero
for $\rho > 0.5$. 
In Fig.~\ref{fig:4}
we plot $|V| = \sqrt{\langle V_{||}\rangle^2 + \langle V_{\perp}\rangle^2}$
for the same system,
where the dashed line indicates that $|V|\leq V_0$ for all values of $\rho$.
Here $|V|$ monotonically deceases with increasing $\rho$,
similar to the behavior found in fully overdamped systems.
When $\alpha_{m}= 0$,
$\langle V_{\perp}\rangle = 0$,
$\theta_{sk} = 0$,
and $\langle V_{||}\rangle$ has
the same shape
shown
in Fig.~\ref{fig:2} but its magnitude is slightly reduced.

\begin{figure}
\includegraphics[width=3.5in]{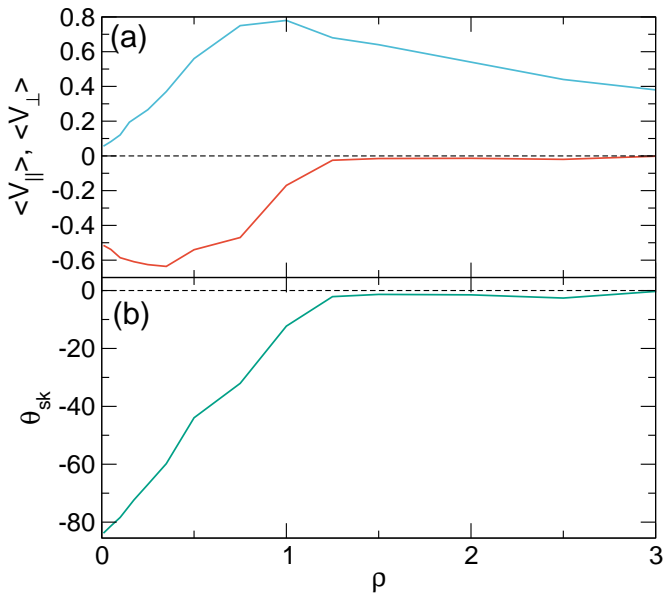}
\caption{
(a) $\langle V_{||}\rangle$ (blue) and $\langle V_{\perp}\rangle$ (red)
versus $\rho$ for the same system in Fig.~\ref{fig:2}
but in a Magnus dominated regime, where   
$\alpha_{m}/\alpha_{d} = 9.95$,  $\alpha^2_{d} + \alpha^{2}_{m} = 1.0$,  
$\theta^{\rm int}_{sk} = -84.26^\circ$, and $F_{D} = 0.5$. 
There is a range of $\rho$ over which
$\langle V_{||}\rangle>V_0=0.5$,
indicating a velocity boost in the driving direction.
(b) The corresponding $\theta_{sk} = \arctan(\langle V_{\perp}\rangle/\langle V_{||}\rangle)$, showing a linear decrease in magnitude with increasing
$\rho$ followed by a saturation regime for $\rho > 1.25$. }
\label{fig:3}
\end{figure}

In Fig.~\ref{fig:3} we plot $\langle V_{||}\rangle$, $\langle V_{\perp}\rangle$,
and $\theta_{sk}$ versus $\rho$ for
the same system from Fig.~\ref{fig:2} in the Magnus dominated regime with
$\alpha_{m}/\alpha_{d} = 9.95$ and $\theta^{\rm int}_{sk} = -84.26^\circ$.
Here the velocities are highly nonmonotonic.
$\langle V_{||}\rangle$ initially increases with
increasing $\rho$, reaches a maximum of $\langle V_{||}\rangle=0.78$
near $\rho  = 1.25$, 
and then decreases again,
while $\langle V_{\perp}\rangle$ has a small initial increase in magnitude
followed by a decrease in magnitude to a value of zero
near $\rho=1.6$.
The Hall angle has the value
$\theta_{sk}=-84.26^\circ$ at low $\rho$,
decreases linearly in magnitude
with increasing $\rho$, and
reaches zero for $\rho > 1.25$.
The maximum value of $\langle V_{||}\rangle = 0.78$ is higher than
$V_{0} = 0.5$, indicating
that the velocity in the driving direction
is being boosted as the skyrmion density increases.
There is also a small boost in $|\langle V_{\perp}\rangle|$
near $\rho = 0.3$, where $|\langle V_{\perp}\rangle| \gtrsim V_0$.
In Fig.~\ref{fig:4}, the plot of $|V|$ versus $\rho$
for the system in Fig.~\ref{fig:3} 
indicates that $|V|$ reaches a maximum value close to $0.9$.
The dashed line represents $V_{0} = 0.5$, so that
a velocity boost is occurring
whenever
$|V|>V_0$.
We observe velocity boosting up to $\rho = 2.0$,
while for $\rho > 2.0$, the net velocity decreases,
indicating an increase in the effective damping. 

\begin{figure}
\includegraphics[width=3.5in]{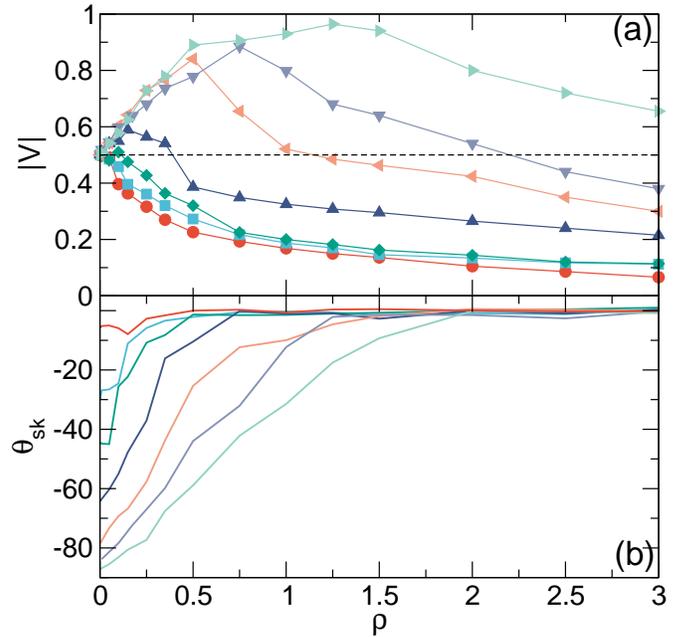}
\caption{ (a) $|V|$
versus $\rho$ for the systems in Figs.~\ref{fig:2} and \ref{fig:3} with
$\alpha_d^2+\alpha_m^2=1.0$ and $F_D=0.5$ at
$\alpha_{m}/\alpha_{d} = 0.1$ (red circles), 0.57 (light blue squares),
0.98 (dark green diamonds), 2.065 (dark blue up triangles),
4.924 (purple down triangles), 9.95 (orange left triangles),
and $19.97$ (light green right triangles), from bottom to top.
The dashed line corresponds to $V_{0} = 0.5$.
For $\alpha_{m}/\alpha_{d} > 1.0$, there is an overshoot regime with $|V|>V_0$.
(b) The corresponding values of
$\theta_{sk}$ versus $\rho$, from top to bottom.
In each case, $\theta_{sk}$ saturates to zero for large enough values of
$\rho$.} 
\label{fig:5}
\end{figure}

In Fig.~\ref{fig:5}(a) we plot $|V|$ versus $\rho$
for the systems in Figs.~\ref{fig:2} and \ref{fig:3} at
$\alpha_{m}/\alpha_{d} = 0.1$, 0.57, 0.98, 2.065, 4.924, 9.95, and $19.97$.
The images in Fig.~\ref{fig:1} are taken from the sample with
$\alpha_{m}/\alpha_{d} = 0.98$. 
For $\alpha_{m}/\alpha_{d} \leq 1.0$,
there is no overshoot in $|V|$
and
we always find 
$|V|<V_{0}$.
When $\alpha_{m}/\alpha_{d} > 1.0$,
an overshoot emerges with a
peak velocity which
increases and shifts to larger $\rho$ with increasing Magnus force.
Figure~\ref{fig:5}(b) shows the corresponding
$\theta_{sk}$ versus $\rho$. In every case,
$\theta_{sk}$ starts at its intrinsic value for small $\rho$,
decreases in magnitude with increasing $\rho$, and reaches
a saturation at high $\rho$.  

\begin{figure}
\includegraphics[width=3.5in]{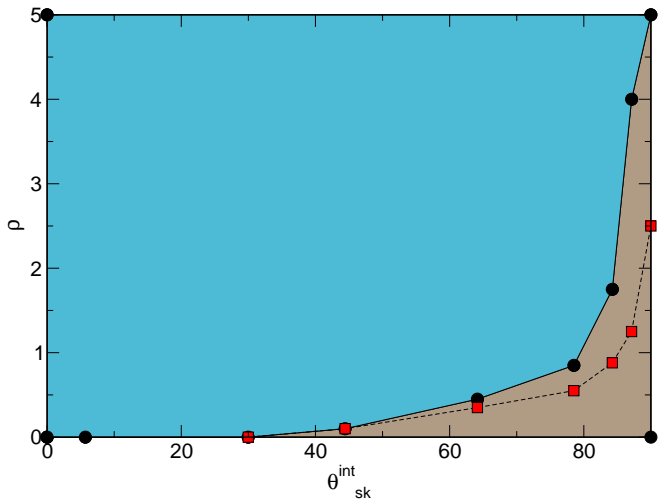}
\caption{Dynamic phase diagram as a function of $\rho$ versus
intrinsic skyrmion Hall angle $\theta^{\rm int}_{sk}$ for the system in
Fig.~\ref{fig:5} with $\alpha_d^2+\alpha_m^2=1.0$ and $F_D=0.5$.
The velocity boost is present in the brown region and absent in the blue
region.
The red squares indicate the points at which
the boost of $|V|$ is maximized.
}
\label{fig:6}
\end{figure}

In Fig.~\ref{fig:6} we map the regions where a velocity boost is present and
absent as a function of
$\rho$ versus 
$\theta^{\rm int}_{sk}$
for the system in Fig.~\ref{fig:5}.
As $\theta^{\rm int}_{sk}$ increases, the upper edge of the velocity boost
window
shifts to higher values of $\rho$. The red squares
indicate the
values of $\rho$ at which the boost of $|V|$ takes its maximum value for
each choice of $\theta^{\rm int}_{sk}$.
The boost disappears for all $\rho$ when $\theta^{\rm int}_{sk} \leq 44^\circ$.

\section{Drive Dependence}

\begin{figure}
\includegraphics[width=3.5in]{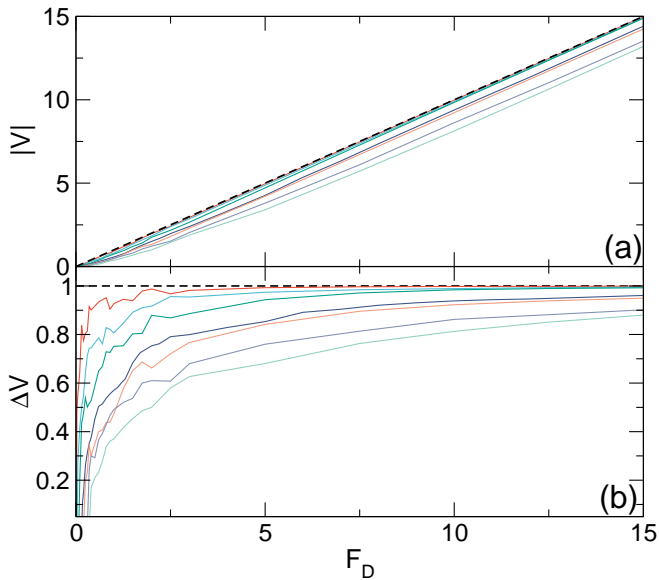}
\caption{ 
(a) $|V|$ versus $F_{D}$ for the system in Fig.~\ref{fig:2}
with $\alpha_m/\alpha_d=0.1$, $\alpha_d^2+\alpha_m^2=1.0$,
and $\theta_{sk}^{\rm int}=-5.739^\circ$
at 
$\rho = 0.05$, 0.1, 0.25, 0.5, 1.0, 1.5, and $2.0$,
from top to bottom.
The dashed line is the expected curve for the single particle limit. Here
$|V|$ is always lower than the single particle limit.
(b) $\Delta V = |V|/V_{0}$ versus $F_D$,
where $V_{0}$ is the expected velocity in
the single particle limit at a given drive. 
Here $\Delta V < 1.0$, indicating increased damping compared to the single
particle limit. At higher drives $\Delta V$
approaches the single particle limit.
}
\label{fig:7} 
\end{figure}

\begin{figure}
\includegraphics[width=3.5in]{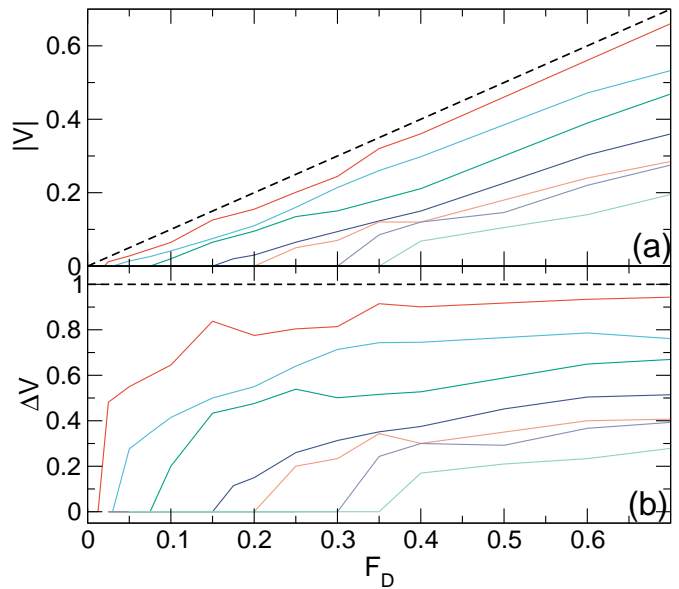}
\caption{ 
A blowup of Fig.~\ref{fig:7} at low values of $F_D$ for the system in
Fig.~\ref{fig:2} with $\alpha_m/\alpha_d=0.1$, $\alpha_d^2+\alpha_m^2=1.0$,
and $\theta^{\rm int}_{sk}=-5.739^\circ$ at $\rho=0.05$, 0.1, 0.25, 0.5, 1.0,
1.5, and 2.0, from top to bottom. The dashed line is the expected curve for
the single particle limit.
(a) $|V|$ versus $F_D$, showing that there is a threshold force $F_c$
with $|V| = 0.0$ for $F_{D} < F_{c}$.
$F_{c}$ shifts to higher values of $F_D$ with increasing $\rho$. 
(b) The corresponding $\Delta V$ versus $F_{D}$.  
}
\label{fig:8} 
\end{figure}

The size of the velocity overshoot
and the value of the skyrmion Hall angle
at a given skyrmion density
is also a function of
the magnitude of
the driving force. 
In Fig.~\ref{fig:7}(a) we plot $|V|$
versus $F_{D}$ for the system in Fig.~\ref{fig:2}
with $\alpha_{m}/\alpha_{d} = 0.1$ in the damping dominated limit
at 
$\rho = 0.05$, 0.1, 0.25, 0.5, 1.0, 1.5, and $2.0$.
The dashed lines indicate 
the expected behavior in the single particle limit. 
For any given value of $F_D$, Fig.~\ref{fig:7}(a) shows that
$|V|$ decreases monotonically with 
increasing $\rho$.  
In Fig.~\ref{fig:7}(b) we plot
$\Delta V = |V|/V_{0}$ versus $F_D$, where $V_{0}$ is the
velocity in the single particle limit at each value of $F_D$.
When $\Delta V = 1.0$, the driven skyrmion is moving with the same velocity 
it would have in the single particle limit.
$\Delta V < 1.0$ indicates increased damping,
while $\Delta V > 1.0$ is a signature of a boosted velocity.  
Here we find
$\Delta V < 1.0$ over the entire range of $\rho$,
indicating that the driven skyrmion is moving slower
than the free particle limit.
The damping effect is strongest at low drives, and
the overall damping increases with increasing $\rho$.
There is also a threshold $F_{c}$ below which
the driven skyrmion does not move,
as shown in Fig.~\ref{fig:8}(a)
where we plot a blow up of the low drive region from
Fig.~\ref{fig:7}(a).
When $F_{D} < F_{c}$,
$|V| = 0.0$.
$F_{c}$
starts from zero in the
single particle limit and
increases with increasing $\rho$.
The plot of $\Delta V$ versus $F_D$ in Fig.~\ref{fig:8}(b) indicates that
$\Delta V$ is zero below $F_c$
and rises to a saturation value which
approaches the single particle limit as $\rho$ decreases.
In general, $\Delta V$ increases towards $\Delta V=1.0$
with increasing $F_{D}$ since the surrounding skyrmions have less time
to respond to a rapidly moving driven skyrmion.
For overdamped systems with quenched disorder, the 
velocity of the probe particle has the power law form
$V = (F_{d} - F_{c})^\alpha$,
so the particle moves the most slowly
just above the threshold depinning force
\cite{Reichhardt21,Fisher98,Reichhardt17}, while
at high drives the velocity generally approaches
the pin-free limit \cite{Fisher98,Reichhardt17}.
In our system,
there is no quenched disorder
but the background bath of skyrmions
effectively serves as a type of
deformable quenched disorder.
At high velocities,
where the background
skyrmions do not have time to respond before the driven skyrmion
has passed beyond them,
the system looks more like a single
skyrmion moving through an array of fixed defects or scattering sites. 

\begin{figure}
\includegraphics[width=3.5in]{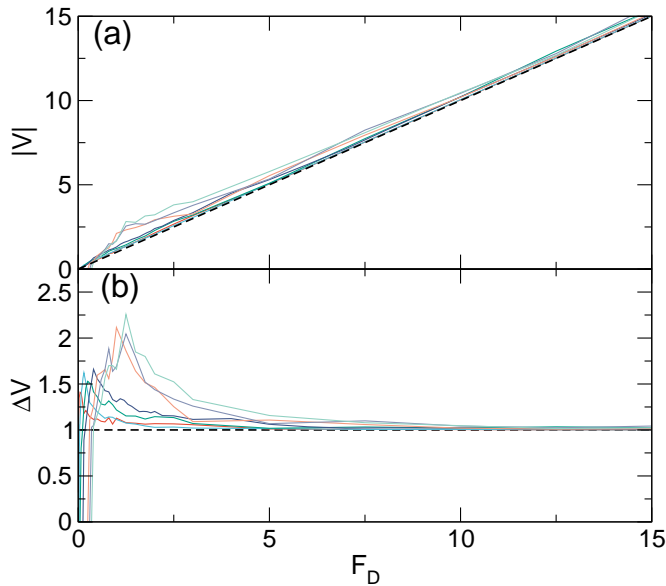}
\caption{ 
(a) $|V|$ versus $F_{D}$ for
a system  
with $\alpha_m/\alpha_d=9.95$ and $\alpha^2_d+\alpha_m^2=1.0$
at
$\rho = 0.05$, 0.1, 0.25, 0.5, 1.0, 1.5, and $2.0$,
from bottom to top.
$|V|$ is generally above
the dashed line,
which is the expected behavior in the single particle limit.
(b) The corresponding $\Delta V = |V|/V_{0}$ versus $F_D$,
where $V_{0}$ is the expected velocity for
the single particle limit at a given drive.
Here $\Delta V > 1.0$ indicates
the occurrence of velocity boosting,
with $\Delta V$ approaching the single particle limit at higher drives.
}
\label{fig:9}
\end{figure}

\begin{figure}
\includegraphics[width=3.5in]{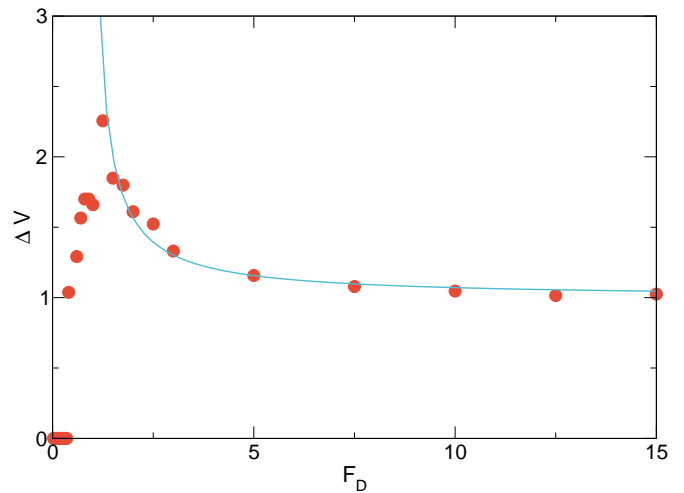}
\caption{
Circles: $\Delta V$ versus $F_D$ from the system with $\rho=2.0$,
$\alpha_m/\alpha_d=9.95$, and $\alpha_d^2+\alpha_m^2=1.0$  
in Fig.~\ref{fig:9}.
The line indicates
a fit to $\Delta V = [A/(F_{D}-F_{\rm max})] + 1.0$ with $A = 0.65$
and $F_{\rm max}=0.85$.
}
\label{fig:10}
\end{figure}

In Fig.~\ref{fig:9}
we plot $|V|$ and $\Delta V$ versus $F_{D}$ for varied $\rho$
in the Magnus dominated regime with
$\alpha_{m}/\alpha_{d} = 9.95$ and $\alpha_d^2+\alpha_m^2=1.0$
for $\rho = 0.05$, 0.1, 0.25, 0.5, 1.0, 1.5, and $2.0$.
The dashed line indicates the expected behavior in the single particle
limit. Here we find that
$|V|$ is generally larger than the
single particle limit, indicating a velocity boost.   
In Fig.~\ref{fig:9}(b), $\Delta V$ increases with
increasing $\rho$ and has a peak at $F_{\rm max}$ which shifts to higher values 
of $F_{D}$ as $\rho$ becomes larger.
For large $F_D$, $\Delta V$
decreases
back to the single particle limit.
When $\rho = 2.0$,
the peak value of
$\Delta V$ 
is almost $2.5$ times larger
than the free particle limit.
We still observe a threshold force $F_c$ for motion which is
almost the same as that found for the overdamped case.
In general we do not observe a significant difference
in the threshold for motion 
as a function of the ratio of the Magnus term to the damping.
In systems with quenched disorder,
it has been shown that the depinning threshold
often decreases with increasing Magnus force
since the motion of the skyrmions
requires a longer time to relax when the Magnus force is larger
\cite{Reichhardt21,Brown18}.
This increases the time required
in the presence of disorder
for the spiraling motion of the skyrmions to settle 
into equilibrium after a finite applied drive is increased
\cite{Reichhardt21,Reichhardt15a,Reichhardt15,Brown19}. 
In our work we apply a constant drive and wait for the system to settle into 
a steady state.
The transient time to reach that state increases near the threshold,
suggesting that the observation of a reduced pinning threshold
in other studies with quenched disorder could be the result of the
increased relaxation time for larger Magnus forces.
In studies where the drive is continuously increased
from zero to a finite value, the sweeping rate above which transient
effects become important
should depend on both the damping and Magnus force. 
Such rate effects will be the subject of a future work.

In Fig.~\ref{fig:9}(b), we find that $\Delta V$ decreases with
increasing drive
for $F_D>F_{\rm max}$ 
and approaches $\Delta V=1.0$ at high drives.
The decrease
follows the form
$\Delta V =  [A/(F_{D} -F_{\rm max})] + 1.0$.
In Fig.~\ref{fig:10} we plot $\Delta V$ versus $F_D$
for the specific case of $\rho = 2.0$
from the system in Fig.~\ref{fig:9}(b) along with a fit using
$A=0.65$ and $F_{\rm max}=0.85$.
A similar fit can be performed for the other $\Delta V$ versus $F_D$ curves.
This type of $1/F_{D}$ dependence on fluctuations or dynamic disorder
is also often observed in driven systems with quenched disorder
\cite{Reichhardt21,Reichhardt15a,Reichhardt15}.  

The speed up effect as a function of $\rho$ and $F_{D}$
arises due to the Magnus force.
In systems where a single driven skyrmion interacts with a wall 
or barrier which is parallel to the $x$ direction, an applied drive
parallel to the
$x$ direction generates a skyrmion Hall component
which deflects the driven particle in the $y$ direction.
This increases the skyrmion-wall interaction
force in the $y$ direction, which results in
a perpendicular skyrmion velocity component
along the $x$ direction.
The wall-induced velocity contribution
adds to the $x$ direction velocity from the driving, generating
a velocity boost effect.
The magnitude of the boost velocity
depends on the nature of the pairwise interaction between the wall and the
skyrmion.
If the interaction is of the power law form
$V(r) \approx 1/r^{\alpha}$,
then as the skyrmion approaches the wall more closely, $r$ decreases
and the resulting velocity in the $x$ direction increases
\cite{Iwasaki14,Zhang15a,Reichhardt16a,CastellQueralt19,Xing20}.  
The boost effect can also occur for skyrmions moving over
a two-dimensional periodic substrate \cite{Reichhardt15a} or over random 
pinning \cite{Gong20},
since the effect arises whenever
the component of the
substrate force which is perpendicular to the direction of drive
is not balanced by a
compensating substrate force in the other direction.
This is due to the skyrmion Hall angle which 
pushes the skyrmion to one side of the pinning sites.

\begin{figure}
\includegraphics[width=3.5in]{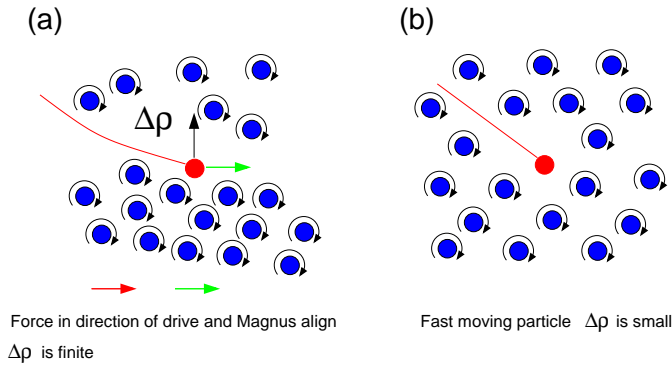}
\caption{
(a) A schematic showing the mechanism creating the velocity boost.
The driven skyrmion (red) is driven in the positive $x$-direction
through a background assembly of skyrmions (blue)
and moves at a negative angle with respect to the $x$ direction due to the
skyrmion Hall effect. The red line is the driven skyrmion trajectory.
The motion creates a local density gradient $\Delta \rho$
which generates a repulsive force on the
driven skyrmion in the positive $y$-direction
(black arrow). The Magnus term converts this force into a
velocity in the positive $x$ direction (green arrow), parallel to the
drive.
(b) The same at higher $F_{D}$ where the surrounding skyrmions
do not have time to respond to the driven
skyrmion. Here $\Delta \rho$ and hence the boost velocity
are reduced. 
}
\label{fig:11}
\end{figure}

For our single driven skyrmion,
there is no pinning or barrier wall;
however, a velocity boost effect can still occur
due to the creation of a local density gradient
in the surrounding medium. 
The mechanism of this effect
is outlined in the schematic of
Fig.~\ref{fig:11}(a), where a driven skyrmion is moving under a
finite driving force $F_{D}$.
For intermediate drives, the
skyrmion tries to translate along $\theta_{sk}^{\rm int}$,
which is at a
negative angle with respect to the $x$ axis.
In the process, it displaces some of the surrounding skyrmions
in the negative $y$ direction, creating
a local density imbalance $\Delta \rho$
such that the density is higher below the driven
skyrmion and lower above it.
Due to the repulsive pairwise interactions with the surrounding skyrmions,
there is an unbalanced force in the positive $y$ direction
on the driven skyrmion,
as indicated by the black arrow in the schematic.
The Magnus term transforms this repulsive force
into a velocity
in the positive $x$-direction, parallel to the drive,
as indicated by the green arrow.
Thus, the boost velocity
$V_{\rm boost} = |V|-V_0 \propto \Delta \rho$. 
Here
the $x$-direction velocity from the damping term
is aligned with but opposite to the velocity induced by the Magnus force.
In contrast, when $F_D$ is larger,
the surrounding skyrmions
do not have time to respond to
the rapidly moving driven
skyrmion,
producing a small $\Delta \rho$
and reducing 
$V_{\rm boost}$, as shown schematically in Fig.~\ref{fig:11}(b).
For drives that are
only slightly above the threshold for motion, the
driven skyrmion
is moving sufficiently slowly
that the
surrounding skyrmions
have enough time to completely relax
any density perturbations,
making $\Delta \rho$ small.
The maximum boost velocity changes as a function of
$F_{D}$ and $\rho$ due to the different relaxation times.
In general, at a fixed drive 
$|V|$
has a damping dependence of 
$|V_{\rm damp}| \propto 1/\rho$
and a boost velocity contribution of
$|V_{\rm boost}| \propto (\alpha_{m}/\alpha_{d})\Delta \rho$,
where $\Delta \rho$ is a function of both the
unperturbed density and the drive. 
For increasing $\rho$, larger values of $\Delta \rho$
can appear but the threshold driving force for the formation of
density inhomogeneities also increases.
For small $\rho$, $\Delta \rho$
is reduced and the boosting effect is diminished.
The boosting effect occurs for any finite value of the Magnus term,
but in general only when $\alpha_{m|}/\alpha_{d} > 1.0$ do we
find a regime in which the
net velocity is higher than the single particle limit.

\begin{figure}
\includegraphics[width=3.5in]{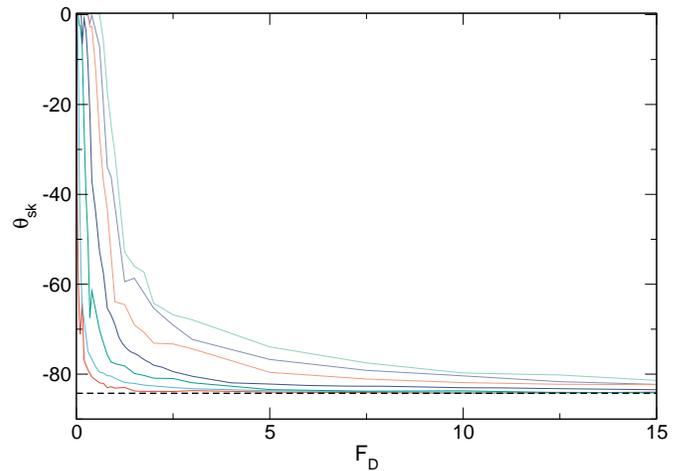}
\caption{
$\theta_{sk}$ versus $F_{D}$ for the system in Fig.~\ref{fig:9}
with $\alpha_m/\alpha_d=9.95$ and $\alpha_d^2+\alpha_m^2=1.0$ at
$\rho = 0.05$, 0.1, 0.25, 0.5, 1.0, 1.5, and $2.0$, from bottom to top.
The dashed line is the single particle limit. 
}
\label{fig:12}
\end{figure}

Another effect of the emergence of a density inhomogeneity
$\Delta \rho$ is a reduction in
the skyrmion Hall angle
at lower drives,
since the skyrmions that accumulate below the
driven particle partially block the motion in the
$-y$ direction.
In Fig.~\ref{fig:12} we plot $\theta_{sk}$ versus $F_{D}$
for the system in Fig.~\ref{fig:9} with
$\alpha_{m}/\alpha_{d} = 9.95$ for $\rho = 0.05$,
0.1, 0.25, 0.5, 1.0, 1.5, and $2.0$. 
The dashed line indicates the single particle limit, where
$\theta_{sk} = -\arctan(\alpha_{m}/\alpha_{d})$ for all values of $F_{D}$. 
We find a finite interval of $F_{D}$
over which the magnitude of $\theta_{sk}$ increases with
increasing $F_{D}$,
followed by a saturation close
to the intrinsic value at higher drives.
As the driven skyrmion approaches the saturation regime,
it is moving fast enough that 
that surrounding skyrmions cannot respond to its presence,
giving a small $\Delta \rho$
and a reduced boost effect,
as shown in
Fig.~\ref{fig:11}(b). The increase of 
$\theta _{sk}$ with increasing $F_{D}$ has also
been observed
for skyrmion assemblies driven over
random disorder, where at higher drives the
skyrmions are only weakly perturbed
by the pinning \cite{Reichhardt15a,Reichhardt15,Jiang17,Litzius17,Reichhardt16,Legrand17,Diaz17,Juge19,Zeissler20,Litzius20}. 

\begin{figure}
\includegraphics[width=3.5in]{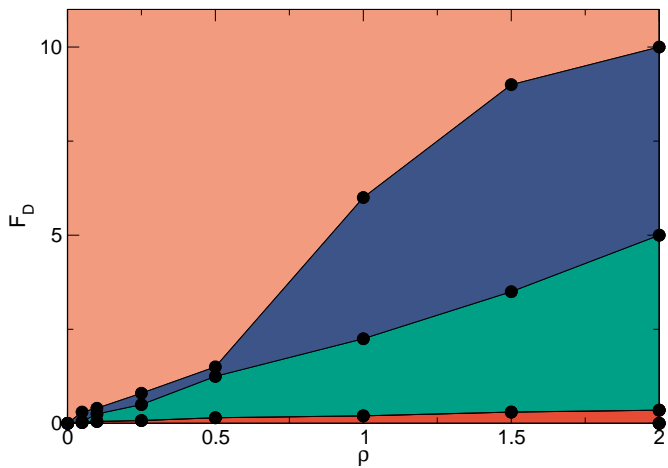}
\caption{
Dynamic phase diagram as a function of $F_{D}$ versus
$\rho$ for a damping dominated system with $\alpha_{m}/\alpha_{d} = 0.1$
and $\alpha_m^2+\alpha_d^2=1.0$.
Red: pinned.
Green: the driven skyrmion moves at
$\theta_{sk} =0.0^\circ$.
Blue: the skyrmion Hall angle is increasing
in magnitude with drive.
Orange: the saturation regime. 
}
\label{fig:13}
\end{figure}

From the features in $\langle V_{\perp}\rangle$, $\langle V_{||}\rangle$,
$|V|$, and $\theta_{sk}$,
we can construct a dynamic phase diagram as a function of $F_{D}$ versus 
$\rho$ for the strongly damped and Magnus dominated limits. 
In Fig.~\ref{fig:13} we show the phase diagram for the damping
dominated system with $\alpha_m/\alpha_d=0.1$
in Figs.~\ref{fig:7} and \ref{fig:8}.
Red indicates the pinned regime where
$F_{D} < F_{c}$.
In the green region,
the driven skyrmion is moving
but its velocity is strictly along
the driving direction so that $\theta_{sk} = 0.0^\circ$.
In the blue region, there is a finite but growing
$\theta_{sk}$, while in the orange region, the skyrmion Hall angle
has saturated.
The thresholds for motion parallel and
perpendicular to the driving direction
both increase with increasing $\rho$.
Previous work 
in systems with quenched disorder also
showed that
there can be separate
thresholds for the onset of motion parallel and
perpendicular to the drive,
with a region
above the first depinning threshold
where the skyrmions can flow
at $\theta_{sk}=0^\circ$
for small intrinsic skyrmion Hall angles
\cite{Jiang17,Reichhardt19,Reichhardt21a}.
As the ratio of the Magnus force to the damping term increases,
the threshold for transverse
motion shifts to smaller $F_D$.
In the saturation regime,
the response
resembles the single particle limit
and $\theta_{sk}$ 
approaches the intrinsic skyrmion Hall value of
$\theta_{sk}^{\rm int} = 5.7^\circ$.     

\begin{figure}
\includegraphics[width=3.5in]{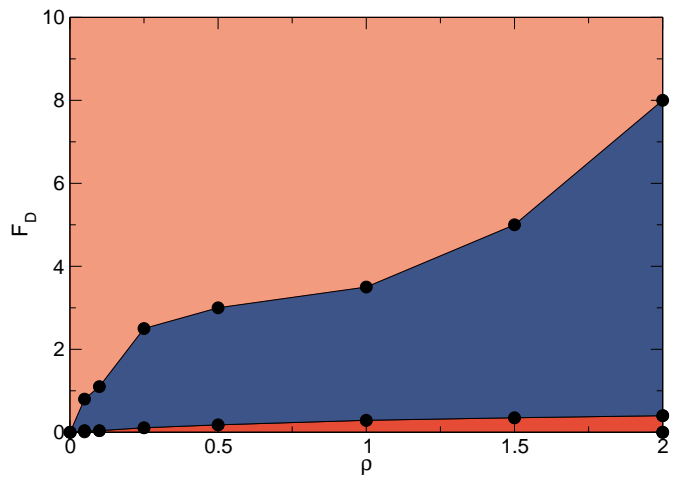}
\caption{
Dynamic phase diagram as a function of $F_{D}$ versus $\rho$
for the Magnus dominated system with $\alpha_{m}/\alpha_{d} = 9.95$
and $\alpha_m^2+\alpha_d^2=1.0$.
Red: pinned.
Blue: the skyrmion Hall angle is increasing
in magnitude with drive and a velocity boost
occurs.
Orange: the saturation regime.
}
\label{fig:14}
\end{figure}

In Fig.~\ref{fig:14} we show the dynamic phase diagram
as a function of $F_D$ versus $\rho$
for the
Magnus dominated system with $\alpha_m/\alpha_d=9.95$.
There is still a pinned regime,
but the region in which
there is finite motion
with $\theta_{sk}=0.0^\circ$
is absent or too small to
detect.
In the blue region,
the velocity is significantly boosted beyond the single particle limit
and
the skyrmion Hall angle increases with increasing drive.
In the orange region,
both $\theta_{sk}$ and the velocity
approach the single particle limit.
Between the
damping dominated and Magnus dominated limits,
the dynamic phase diagram includes a combination
of the features of the phase
diagrams in Fig.~\ref{fig:13} and Fig.~\ref{fig:14}.  

\section{Varied Magnus to Damping Ratio and Velocity-Force Curves}

\begin{figure}
\includegraphics[width=3.5in]{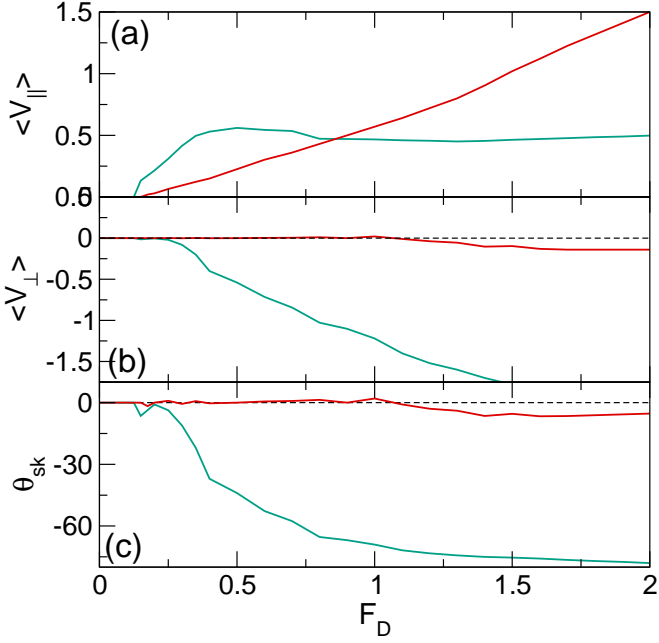}
\caption{
(a) $\langle V_{||}\rangle$ versus $F_{D}$ for
systems with $\rho=0.5$ and $\alpha_m^2+\alpha_d^2=1.0$ at
$\alpha_{m}/\alpha_{d} = 9.95$ (green) and $0.1$ (red). 
(b) The corresponding $\langle V_{\perp}\rangle$ versus $F_{D}$. 
(c) The corresponding $\theta_{sk}$ versus $F_{D}$. 
}
\label{fig:15}
\end{figure}

We next fix the density $\rho$ and vary the ratio of
the Magnus and damping terms while maintaining
the normalization relation $\alpha^2_{d} + \alpha^2_{m} = 1.0$.
In Fig.~\ref{fig:15}(a) we plot
$\langle V_{||}\rangle$ versus $F_{D}$ for
systems with $\rho=0.5$ at
$\alpha_{m}/\alpha_{d} = 9.95$ and $0.1$.
Figures~\ref{fig:15}(b) and (c) show the corresponding values of 
$\langle V_{\perp}\rangle$ and $\theta_{sk}$ versus $F_D$,
respectively.
For the Magnus dominated case
of $\alpha_m/\alpha_d=9.95$,
$\langle V_{||}\rangle$ and
$\langle V_{\perp}\rangle$ become finite at almost the same value
of $F_D$, which also corresponds to
the appearance of a nonzero $\theta_{sk}$.
For the damping dominated sample with $\alpha_{m}/\alpha_{d} = 0.1$,
$\langle V_{\perp}\rangle$ and $\theta_{sk}$ remain zero
up to
$F_{D} \approx  1.0$. 

\begin{figure}
\includegraphics[width=3.5in]{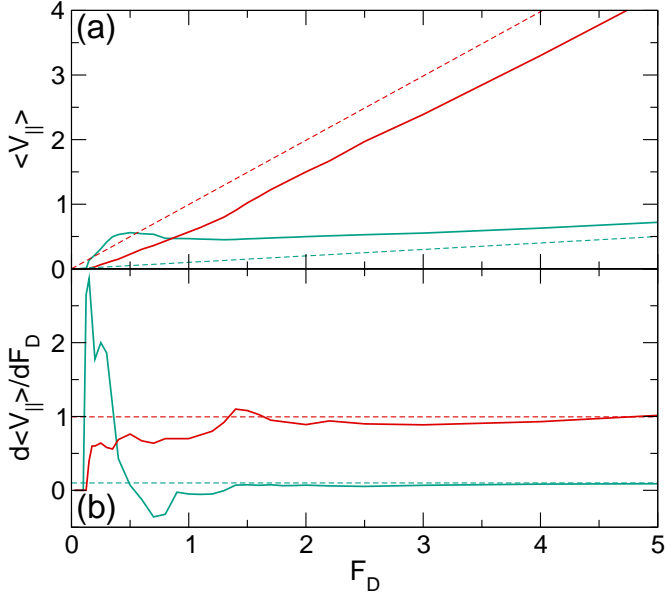}
\caption{
(a) $\langle V_{||}\rangle$ versus $F_{D}$ for
the system in Fig.~\ref{fig:15} with $\rho=0.5$ and $\alpha_m^2+\alpha_d^2=1.0$
at  
$\alpha_{m}/\alpha_{d} = 9.95$ (green) and $0.1$ (red).
The dashed lines are the
expected velocity-force curves
in the single particle limit for these two cases. 
(b) The corresponding $d\langle V_{||}\rangle/dF_{D}$ versus
$F_{D}$, where there is a region of negative differential conductivity with
$d\langle V_{||}\rangle/dF_{D} < 0.0$ for the
$\alpha_m/\alpha_d=9.95$ sample.  
}
\label{fig:16}
\end{figure}

The ratio of the Magnus to the damping term determines
the shape of the velocity-force curves.
For $\alpha_{m}/\alpha_{d} = 0.1$, both
$\langle V_{||}\rangle$ and $\langle V_{\perp}\rangle$
increase monotonically with $F_D$ according to the
linear behavior $V \propto (F_{D} -F_{c})$, where $F_{c}$ is
the threshold for motion in
either the parallel or the perpendicular direction.
When $\alpha_{m}/\alpha_{d} = 9.95$,
$\langle V_{\perp}\rangle$ still increases
linearly with increasing $F_D$ above the depinning threshold;
however,
$\langle V_{||}\rangle$ exhibits a nonmonotonic behavior in which
it initially rises rapidly with increasing $F_D$ 
but then decreases again.
In this regime,
the velocity decreases even though the drive is increasing,
giving $d\langle V_{||}\rangle/dF_{D} < 0$, which is known
as negative differential conductivity \cite{Reichhardt17}.
In Fig.~\ref{fig:16}(a) we plot
$\langle V_{||}\rangle$ versus $F_{D}$ for the
samples from Fig.~\ref{fig:15}.
The green dashed line shows the expected value for
$\langle V_{||}\rangle$ in the single particle limit
at $\alpha_m/\alpha_d=9.95$,
which increases linearly according to
$\langle V_{||}\rangle = 0.995F_{D}$.
For the $\alpha_m/\alpha_d=0.1$ sample,
the red dashed line is the single particle limit
$\langle V_{||}\rangle=0.1F_{D}$.
In the damping dominated system
with $\alpha_m/\alpha_d=0.1$,
$\langle V_{||}\rangle$ is below
the single particle limit,
while in the Magnus dominated system with
$\alpha_m/\alpha_d=9.95$,
$\langle V_{||}\rangle$ is higher than the single particle
limit due to the boosting effect.
In Fig.~\ref{fig:16}(b) we show the corresponding values of
$d\langle V_{||}\rangle/dF_{D}$,
with the
single particle limits marked by dashed lines.
For the 
$\alpha_{m}/\alpha_{d} = 9.95$
system, the initial peak in $d\langle V_{||}\rangle/dF_D$
corresponding to the depinning transition
is followed by
a region of negative differential 
conductivity where
$d\langle V_{||}\rangle/dF_{D} < 0.0$.
At higher drives,
$d\langle V_{||}\rangle/dF_{D}$ saturates to the value expected in the
single particle limit.
For the damping dominated system
with $\alpha_m/\alpha_d=0.1$,
$d\langle V_{||}\rangle/dF_{D}$ is initially
below the single particle value due to the increased damping
from the surrounding skyrmions,
while at higher drives it
approaches the single particle limit.
The negative differential conductivity
in the Magnus-dominated system
is affected by the density,
and for low $\rho$ the shape of the velocity-force curve
approaches that found in the
single particle limit.     

\begin{figure}
\includegraphics[width=3.5in]{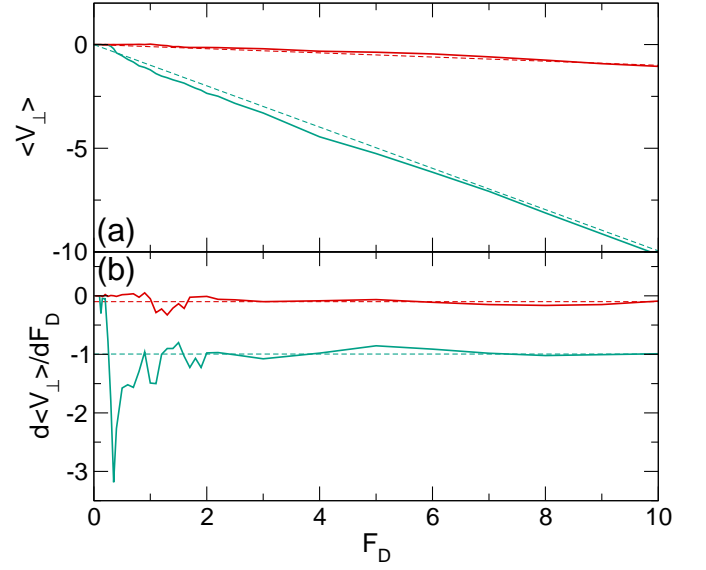}
\caption{
(a) $\langle V_{\perp}\rangle$ versus $F_{D}$ for
the system in Fig.~\ref{fig:15} with $\rho=0.5$ and
$\alpha_m^2+\alpha_d^2=1.0$ at
$\alpha_{m}/\alpha_{d} = 9.95$ (green) and $0.1$ (red).
The dashed lines are the expected velocity-force curves
in the single particle limit for these two cases.  
(b) The corresponding $d\langle V_{\perp}\rangle/dF_{D}$ versus $F_{D}$.
}
\label{fig:17}
\end{figure}

In Fig.~\ref{fig:17}(a) we plot
$\langle V_{\perp}\rangle$ versus $F_D$ for the system in Fig.~\ref{fig:15}
at $\rho=0.5$ with $\alpha_{m}/\alpha_{d} = 9.95$ and $0.1$.
The dashed lines are the expected values for the single particle limit,
which obey $\langle V_{\perp}\rangle = -0.1F_{D}$ for
the $\alpha_{m}/\alpha_{d} = 0.1$ system              
and $\langle V_{\perp}\rangle = -0.995F_{D}$ for
the $\alpha_{m}/\alpha_{d} = 9.95$ system.
In the Magnus dominated regime,
$|\langle V_{\perp}\rangle|$ is slightly larger than
it would be in the single
particle limit,
while in the damping dominated regime
it is slightly lower.
Most of the velocity boost in the Magnus dominated regime is
parallel to the driving direction
since the induced density gradient
is perpendicular to the drive.
In Fig.~\ref{fig:17}(b) we show the corresponding
$d\langle V_{\perp}\rangle/dF_{D}$ versus $F_D$ curves where,
unlike the parallel velocity plotted in
Fig.~\ref{fig:16}, there is no regime of negative differential mobility. 
There is still a peak
in $|d\langle V_{\perp}\rangle/dF_D|$ near the depinning threshold,
while at high drives the
curves approach
the single particle limit. 

\begin{figure}
\includegraphics[width=3.5in]{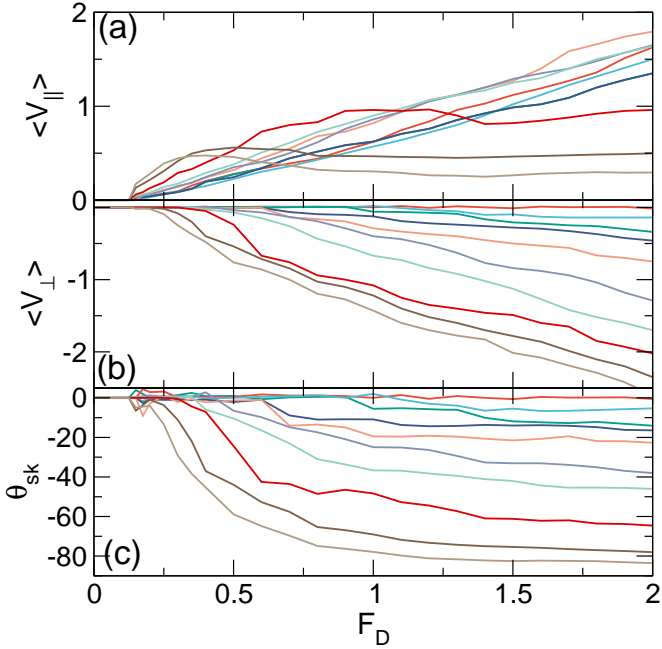}
\caption{ 
(a) $\langle V_{||}\rangle$ versus $F_{D}$ for the system in
Fig.~\ref{fig:15} with
$\rho = 0.5$ and $\alpha_m^2 + \alpha_d^2=1.0$ at
$\alpha_{m} = 0.0$, 0.1, 0.3, 0.5, 0.7, 0.8, 0.9, 0.97, 0.995, and $0.998$,
from bottom to top. 
(b) The corresponding $\langle V_{\perp}\rangle $ versus $F_{D}$.
(c) The corresponding $\theta_{sk}$ versus $F_{D}$. 
}
\label{fig:18}
\end{figure}

In Fig.~\ref{fig:18}(a) we plot $\langle V_{||}\rangle$ versus $F_{D}$
for the system in Fig.~\ref{fig:15} with fixed $\rho = 0.5$
and $\alpha_m^2+\alpha_d^2=1.0$
at $\alpha_{m} = 0$, 0.1, 0.3, 0.5, 0.7, 0.8, 0.9, 0.97, 0.995, and $0.998$.
Figures \ref{fig:18}(b) and (c) show the corresponding
$\langle V_{\perp}\rangle$ and $\theta_{sk}$,
respectively, versus $F_{D}$.
We find
negative differential conductivity for
$\alpha_{m} > 0.9$ or $\alpha_{m}/\alpha_{d} > 5.0$,
indicating that this effect appears only when
the Magnus force
is sufficiently large.
There is no negative differential conductivity in the
$\langle V_{\perp}\rangle$ versus $F_{D}$ curves,
while the plots of $\theta_{sk}$ versus $F_{D}$
indicate that as
$\alpha_{m}$ decreases, the
threshold drive above which the skyrmion
Hall angle becomes finite increases.

\begin{figure}
\includegraphics[width=3.5in]{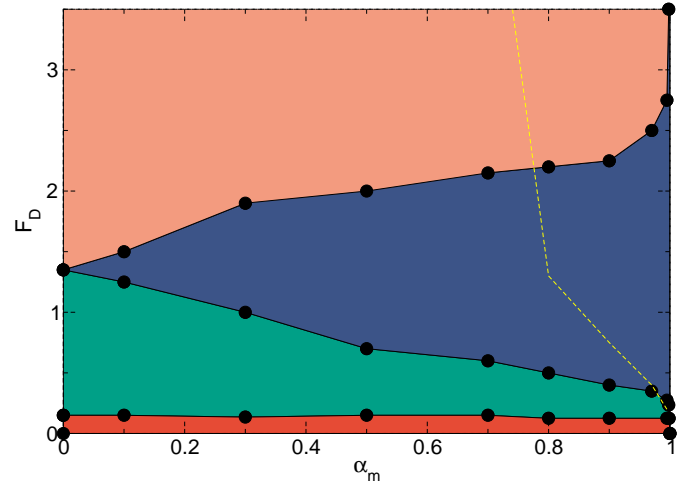}
\caption{Dynamic phase diagram
as a function of $F_D$ versus $\alpha_m$  
for the system in Fig.~\ref{fig:18} with
$\rho=0.5$ and $\alpha_m^2+\alpha_d^2=1.0$
Red: pinned.
Green: the driven skyrmion moves at $\theta_{sk}=0.0^\circ$.
Blue: the skyrmion Hall angle is increasing
in magnitude with drive.
Orange: the saturation regime.
For $F_{D}$ above the dashed line, there
is a velocity boost compared to the single particle limit.  
}
\label{fig:19}
\end{figure}

From the features in Fig.~\ref{fig:18} we can construct a
dynamic phase diagram
as a function of $F_{D}$ versus $\alpha_{m}$,
as shown in Fig.~\ref{fig:19}.
Here we outline the pinned phase,
the flowing regime with $\theta_{sk} = 0.0^\circ$,
the region in which $\theta_{sk}$ increases with
increasing $F_{D}$,
and the saturation regime where there is little change in
$\theta_{sk}$.
For drives above the dashed line,
the velocities are boosted compared to the single particle limit.
The boost is strongly reduced in the saturation regime. 
Figure~\ref{fig:19} indicates that
the depinning threshold is approximately constant
as a function of increasing $\alpha_{m}$,
while the regime
in which
the velocity is locked in the driving
direction
grows in extent with decreasing $\alpha_{m}$.
At $\alpha_{m} = 0$, the onset of saturation
coincides with the point
where the velocity-force curves start to grow linearly with $F_{D}$. 
The transition demarcating the onset of a velocity boost
shifts to lower $\alpha_m$ as $\rho$ increases.

\begin{figure}
\includegraphics[width=3.5in]{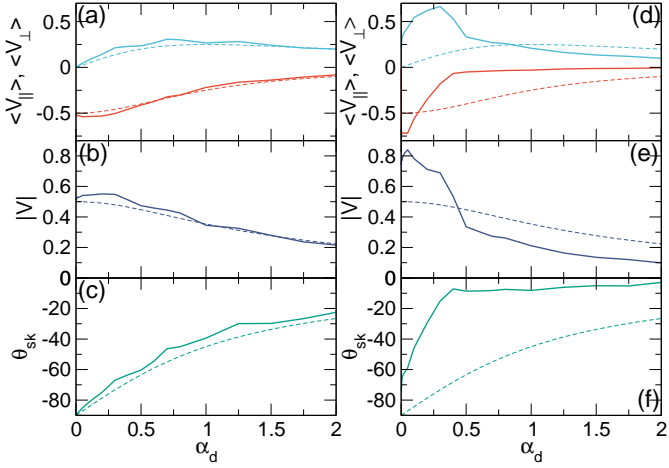}
\caption{ 
(a)  $\langle V_{||}\rangle$ (blue) and $\langle V_{\perp}\rangle$ (red)
versus $\alpha_{d}$ for a system with
$\rho = 0.05$, $F_{D} = 0.5$ and $\alpha_{m} = 1.0$.  
The blue dashed line is the expected single particle
behavior which goes as
$\langle V_{||}\rangle = F_{D}\alpha_{d}/(1 + \alpha^2_{d})$
and the red dashed line is the single particle behavior of
$\langle V_{\perp}\rangle = -F_{D}/(1 + \alpha^2_{d})$.
(b) The corresponding $|V|$ versus $\alpha_{d}$,
where the dashed line is the single
particle limit of $|V| = F_{D}/(1 + \alpha^2_{d})^{1/2}$.
(c) The corresponding skyrmion Hall angle
$\theta_{sk}$ versus $\alpha_{d}$. The dashed line
is the single particle limit of
$\theta_{sk} = -F_{D}\arctan(1/\alpha_{d})$.  
(d) $\langle V_{||}\rangle$ (blue) and $\langle V_{\perp}\rangle$ (red)
versus $\alpha_d$
for the same system but with $\rho = 0.5$.
(e) $|V|$ versus $\alpha_{d}$ for the system in panel (d).
(f) $\theta_{sk}$ versus $\alpha_{d}$ for the system in panel (d).
The denser system exhibits a speed up effect.}
\label{fig:20}
\end{figure}

\begin{figure}
\includegraphics[width=3.5in]{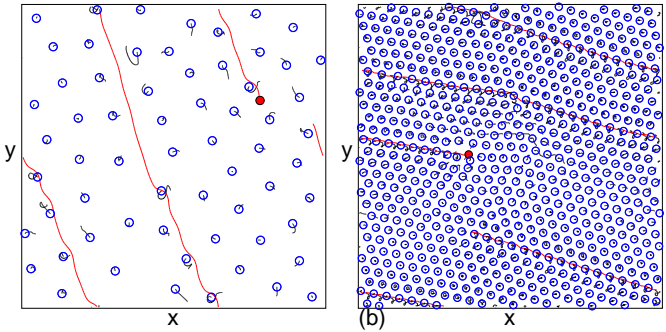}
\caption{
Driven skyrmion and trajectory (red) along with the surrounding skyrmions
(blue) and their trajectories (black)
for the system in Fig.~\ref{fig:20}(a) with $\rho = 0.05$, 
$\alpha_{d} = 0.3$, and $\alpha_{m} = 1.0$,
where the
skyrmion Hall angle is large.
(b) The same but at $\rho = 0.5$,
where the skyrmion Hall angle is smaller and the velocity of the
driven skyrmion
is higher.}
\label{fig:21}
\end{figure}

We next
relax the constraint of $\alpha^2_d + \alpha^2_{m} = 1.0$ and
instead hold either the Magnus or damping term constant while varying
the other quantity.
For a fixed drive, this means that in the single particle limit,
$|V|$ obeys
$|V| \propto F_{D}/\sqrt{\alpha^2_{m} + \alpha^2_{d}}$. 
In Fig.~\ref{fig:20}(a) we plot
$\langle V_{||}\rangle$ and $\langle V_{\perp}\rangle$
versus $\alpha_{d}$ for a system with 
$\rho = 0.05$, $F_{D} = 0.5$, and $\alpha_{m} = 1.0$. 
Here $\langle V_{||}\rangle$ is zero for
$\alpha_{d} = 0.0$, increases to a
maximum value near $\alpha_{d} = 1.0$, and then
decreases with increasing $\alpha_d$,
while $\langle V_{\perp}\rangle$ gradually approaches zero as $\alpha_d$
increases.
In the single particle limit,
$\langle V_{||}\rangle = F_{D}\alpha_{d}/(\alpha^2_{m} + \alpha^2_{d})$
and
$\langle V_{\perp}\rangle  = -F_{D}\alpha_{m}/(\alpha^2_{m} + \alpha^2_{d})$, 
so for a fixed $\alpha_{m} = 1.0$,
$\langle V_{||}\rangle = F_{D}\alpha_{d}/(1 + \alpha^2_{d})$, plotted as a
dashed line.
In this case,
$\langle V_{||}\rangle = 0.0$
when $\alpha_{d} = 0.0$,
and
the parallel velocity also 
approaches zero at large
$\alpha_d$.
Similarly, in the single particle limit,
$\langle V_{\perp}\rangle = -F_{D}/(1 + \alpha^2_{d})$, 
so that
when $\alpha_{d} = 0.0$,
$\langle V_{\perp}\rangle = -F_{D} = -0.5$.     
For $\rho = 0.05$, the density is low enough
that the behavior is close to the single particle 
limit.
In Fig.~\ref{fig:20}(b)
we plot $|V|$
versus $\alpha_{d}$ for the system in
Fig.~\ref{fig:20}(a), where the dashed line
is the
single 
particle limit of $|V| = F_{D}/(1 + \alpha^2_{d})^{1/2}$.
There is a small boost in $|V|$ for $\alpha_{d} < 1.0$,
while $|V|$ drops below the free particle limit 
for $\alpha_{d} > 1.5$.
Figure~\ref{fig:20}(c) shows the corresponding
$\theta_{sk}$ versus $\alpha_d$ for the $\rho=0.05$ system,
where the magnitude of $\theta_{sk}$ gradually decreases with increasing $\alpha_{d}$. 
The dashed line is
the single particle limit,
$\theta_{sk} = -F_{D}\arctan(1/\alpha_{d})$.
The 
measured skyrmion Hall angle is
smaller in magnitude
than the single particle value due to the collisions
with the background skyrmions. 
In Fig.~\ref{fig:21}(a) we show the positions and trajectories of the
driven skyrmion and the background skyrmions
for the system in Fig.~\ref{fig:20}(a) at
$\rho = 0.05$ and $\alpha_{d} = 0.3$. The driven skyrmion
is moving at a skyrmion Hall angle close to
$\theta_{sk}=-65^\circ$. Although there are some collisions with the
background skyrmions,
there is only a small
boost in the velocity.
 
In Fig.~\ref{fig:20}(d) we show
$\langle V_{||}\rangle$ and $\langle V_{\perp}\rangle$
versus $\alpha_d$ for the same system in
Fig.~\ref{fig:20}(a) but at a higher density of $\rho = 0.5$. The dashed
lines are the expected behavior in the single particle limit.
Here there is a boost in
$\langle V_{||}\rangle$ for $\alpha_{d} < 0.75$, while
$\langle V_{||}\rangle$ drops below
the single particle limit
at
higher $\alpha_d$.
There is also a small boost in
$\langle V_{\perp}\rangle$ for $\alpha_{d} < 0.3$.
We note that unlike the low density case,
$\langle V_{||}\rangle$ is finite
at $\alpha_{d} = 0.0$
due to collisions with the background skyrmions. 
Figure~\ref{fig:20}(e) shows the corresponding
$|V|$ versus $\alpha_{d}$
and the expected single particle limit,
indicating that there is a net velocity boost for $\alpha_{d} < 0.5$
followed by
increased
damping for $\alpha_{d} > 0.5$.
In Fig.~\ref{fig:20}(f) we plot the corresponding
$\theta_{sk}$ versus $\alpha_{d}$
and the single particle limit.
The skyrmion 
Hall angle rapidly decreases
in magnitude with increasing $\alpha_d$,
reaching a saturation for $\alpha_{d} > 0.5$. 
Figure~\ref{fig:21}(b) shows the positions and trajectories of the
driven and bath skyrmions for the system in
Fig.~\ref{fig:20}(d) at $\rho = 0.5$ 
and $\alpha_{d} = 0.3$,
where the skyrmion Hall angle
is much smaller than
that found in the $\rho = 0.05$ system
illustrated in Fig.~\ref{fig:21}(a);
however, the velocity of the driven skyrmion is much larger for
the $\rho=0.5$ system.

\begin{figure}
\includegraphics[width=3.5in]{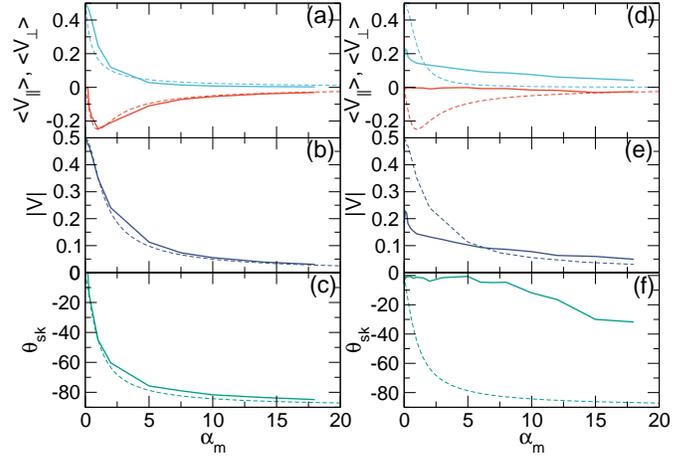}
\caption{
(a) $\langle V_{||}\rangle$ (blue) and $\langle V_{\perp}\rangle$ (red)
versus $\alpha_{m}$ for a system with
$\rho = 0.05$, $F_{D} = 0.5$, and $\alpha_{d} = 1.0$.
The blue dashed line is the expected single particle behavior 
$\langle V_{||}\rangle = F_{D}/(1 + \alpha_{m}^2)$ 
and the red dashed line is the single particle behavior 
$\langle V_{\perp}\rangle = -F_{D}\alpha_{m}/(1 + \alpha^2_{m})$.
(b) $|V|$ versus $\alpha_{d}$ for the system in
panel (a), where the dashed line is the single
particle limit of 
$|V| = F_{D}/(1 + \alpha^2_m)^{1/2}$. 
(c) The corresponding skyrmion Hall angle
$\theta_{sk}$ versus $\alpha_{m}$, where the dashed line
is the single particle limit of 
$\theta_{sk} = -\arctan(\alpha_{m})$. 
(d) $\langle V_{||}\rangle$ (blue) and $\langle V_{\perp}\rangle$
versus $\alpha_d$
for the same system
but with $\rho = 0.5$.
(e) The corresponding $|V|$ versus $\alpha_{m}$ and
(f) the corresponding $\theta_{sk}$ versus $\alpha_m$. } 
\label{fig:22}
\end{figure}

In Fig.~\ref{fig:22}(a) we plot
$\langle V_{||}\rangle$ and
$\langle V_{\perp}\rangle$ versus $\alpha_{m}$
for a system with fixed $\alpha_{d} = 1.0$, 
$\rho = 0.05$, and $F_{D} = 0.5$.
In the single particle limit with $\alpha_{d} = 1.0$, 
$\langle V_{||}\rangle = F_{D}/(1 + \alpha_{m}^2)$
and $\langle V_{\perp}\rangle = -F_{D}\alpha_{m}/(1 + \alpha^2_{m})$.
We find that $\langle V_{||}\rangle$ monotonically decreases and changes from
being slightly higher
than the free particle limit
for $\alpha_{m} < 5.0$ to being slightly lower than the free particle limit
for $\alpha_{m} > 5.0$.
$\langle V_{\perp}\rangle$ starts from zero at $\alpha_{m} = 0.0$,
reaches a maximum value near $\alpha_{m} = 1.0$,
and gradually drops back to zero
with increasing $\alpha_{m}$,
closely following the single particle limit.
In Fig.~\ref{fig:22}(b) we show the corresponding
$|V|$ versus $\alpha_{m}$ along with a dashed line indicating
the single particle limit
of $|V| = F_{D}/(1 + \alpha^2_m)^{1/2}$, which decreases monotonically
with increasing $\alpha_m$.
There is a small boost in the velocity due to the collisions
with the background skyrmions. 
Figure~\ref{fig:21}(c) shows
$\theta_{sk}$ versus $\alpha_{m}$ for the same system
as well as the single particle limit of 
$\theta_{sk} = -\arctan(\alpha_{m})$.
Here the measured skyrmion Hall angle is slightly
smaller in magnitude
than the single particle limit due to the collisions
with bath skyrmions.  

In Fig.~\ref{fig:22}(d) we plot
$\langle V_{||}\rangle$,
$\langle V_{\perp}\rangle$, and the single particle limits for the same
system from Fig.~\ref{fig:22}(a) but at
a higher density of $\rho = 0.5$.
Here $\langle V_{||}\rangle$ is
considerably below
the single particle limit at $\alpha_{m} = 0.0$ due to the 
increased frequency of collisions;
however, for $\alpha_{m} > 2.5$,
it is considerately higher than the single particle limit.
$\langle V_{\perp}\rangle$ is close to zero for
$\alpha_{m} < 7.0$ but begins to increase at larger $\alpha_{m}$.
In Fig.~\ref{fig:22}(e) we show the corresponding
$|V|$ versus $\alpha_{m}$.
$|V|$ falls below the
dashed line representing the single particle limit up to $\alpha_{m} = 7.0$
and exhibits a
boost for higher $\alpha_{m}$.
Figure~\ref{fig:22}(f) illustrates the corresponding
$\theta_{sk}$ versus $\alpha_{m}$ and the single particle limit.
$\theta_{sk}$ is close to zero for $\alpha_{m} < 7.0$,
while for higher $\alpha_m$,
the magnitude of $\theta_{sk}$ increases and
a velocity boost appears.
For higher values of $F_{D}$,
all of the quantities gradually approach the single particle limit. 

\begin{figure}
\includegraphics[width=3.5in]{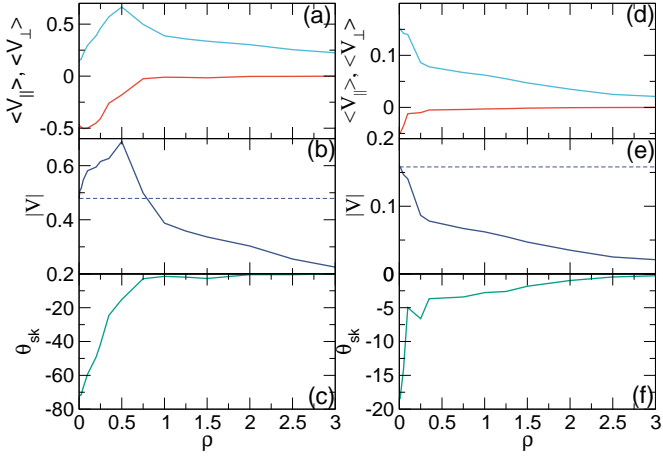}
\caption{
(a)  $\langle V_{||}\rangle$ (blue) and $\langle V_{\perp}\rangle$
(red) versus $\rho$ for a system with
$\alpha_{d} = 0.3$, $\alpha_{m} = 1.0$, and $F_{D} = 0.5$.
(b) The corresponding $|V|$ versus
$\rho$.
(c) The corresponding skyrmion Hall angle $\theta_{sk}$ versus
$\rho$.
(d)
$\langle V_{||}\rangle$ (blue) and
$\langle V_{\perp}\rangle$ (red) versus $\rho$ for the same system with
a higher
$\alpha_{d} = 3.0$.
(e) The corresponding $|V|$ versus
$\rho$.
(f) The corresponding $\theta_{sk}$ versus $\rho$.
} 
\label{fig:23}
\end{figure}

In Fig.~\ref{fig:23}(a) we plot $\langle V_{||}\rangle$ and
$\langle V_{\perp}\rangle$ versus $\rho$ for a
system with fixed $\alpha_{m} = 1.0$,
$\alpha_d = 0.3$, and $F_{D} = 0.5$.  
The single particle limit for these parameters gives
$\langle V_{||}\rangle =0.14$ and  $\langle V_{\perp}\rangle = -0.48$. 
$\langle V_{||}\rangle$ initially increases with increasing $\rho$ to a
peak value of $0.65$, a strong boost that is
four times
larger than the single particle limit.
As $\rho$ increases further, $\langle V_{||}\rangle$
gradually decreases; however,
even at the high density of $\rho = 3.0$,
$\langle V_{||}\rangle$ is still almost twice as large
as the single particle limit, indicating the
continuing effectiveness
of the boost effect. 
$\langle V_{\perp}\rangle$ decreases monotonically in magnitude
with increasing density,
approaching a value close to zero for $\rho >0.7$.   
In Fig.~\ref{fig:23}(b) we show the corresponding
$|V|$ versus $\rho$, where the dashed line is the 
single particle limit of $ |V| = F_{D}/(\alpha^2_{d} + \alpha^2_{m})^{1/2}$
which for these coefficients is $|V|=0.4789$. 
Here there is a boost in the velocity up to $\rho = 0.7$,
while at higher densities the velocities are much more strongly damped.
This indicates that it is possible for the parallel velocity to be boosted
while
the overall velocity is not boosted. 
Figure~\ref{fig:23}(c) illustrates the corresponding $\theta_{sk}$
versus $\rho$.
For low density, the skyrmion Hall angle is near the intrinsic value
of $\theta_{sk}^{\rm int}=-73.3^\circ$, and as $\rho$ increases,
$\theta_{sk}$ approaches zero once $\rho > 0.7$. 

In Fig.~\ref{fig:23}(d) we plot $\langle V_{||}\rangle$ and
$\langle V_{\perp}\rangle$ versus $\rho$ for a
system with a larger $\alpha_d=3.0$ at
$\alpha_{m} = 1.0$ and
$F_{D} = 0.5$.  
For these parameters, the single particle limit gives
$\langle V_{||}\rangle =0.15$ and $\langle V_{\perp}\rangle = -0.05$.
We find that both
$\langle V_{||}\rangle$ and
$\langle V_{\perp}\rangle$ monotonically decrease in magnitude with increasing
$\rho$.
In Fig.~\ref{fig:23}(e) we show the corresponding $|V|$
versus $\rho$, where the dashed line is the  
single particle limit of $|V|=0.158$.
Here there is no boost and $|V|$ drops off rapidly with increasing $\rho$,
showing
a change in slope
near $\rho = 0.3$  to a slower decline.
Figure~\ref{fig:23}(f) illustrates the corresponding $\theta_{sk}$
versus $\rho$, which starts off near 
$\theta_{sk}=-18.4^\circ$ and approaches zero for $\rho > 0.3$.
The change in the skyrmion Hall angle and velocity
across $\rho = 0.3$ occurs
because 
the density is low enough
for $\rho<0.3$ that
the bath skyrmions act like a fluid which is not strongly coupled to the
driven skyrmion, while
when $\rho > 0.3$,
the bath skyrmions act
more like a solid, increasing the drag on the
driven skyrmion.

\section{Discussion}
In our work, the driven skyrmion is free to move in any direction.
In certain chiral soft matter systems, a
similar probe particle could be
implemented using a magnetic or charged particle coupled to a uniform
magnetic or electric field
which does not couple to the remaining particles, allowing the probe
particle
to move at any angle.
In skyrmion systems, it is possible to have samples containing
multiple species of skyrmions, some of which could couple more strongly
than others to
an externally imposed field.
The closest experimental realization of our system
for skyrmions would be to drag a single skyrmion using some form of
localized trap. Such an arrangement
would constrain the driven skyrmion to move only in the direction the trap
is being translated, and would not allow the driven skyrmion to move
at a speed greater than that of the trap.
In this case, changes in the effective
viscosity could be deduced by measuring the force or the fluctuations
exerted by the skyrmion on the trap.
The case of
a trap moving at a constant velocity
will be studied in another work; however,
the results of the present study
can be used as 
a guide to understand
which different velocity regimes could arise.

In our work we have only 
considered a particle-based model, which neglects internal degrees of
freedom and shape changes of the skyrmion.
Such modes
could increase or decrease the damping
experienced by the driven skyrmion or change the
nature of the skyrmion motion.
We have also assumed a simple pairwise repulsion between skyrmions,
but it is possible for skyrmions
to have more complex interactions,
such as competing interactions at different length scales. 
This could produce additional coupling/decoupling or depinning transitions.

In constant velocity experiments,
the force the skyrmion experiences
could have a periodic signature
if the
motion occurs
through a skyrmion solid,
or a broad band noise signature if the skyrmion is moving through a glass or
liquid state.
Numerical work
\cite{Reichhardt16,Diaz17}
and experiments \cite{Sato19,Sato20} on collectively moving skyrmions
have shown the presence of both
broad and narrow band noise,
so it would be interesting to study
the fluctuations exerted on a single skyrmion as it
moves through a bath of other skyrmions.     
We can also compare our results to active rheology in
overdamped chiral granular systems,
where a disk with short range repulsive interactions is pushed through
an assembly of spinning grains \cite{Reichhardt19a}.
The granular system generally does not exhibit any velocity boost
due to its overdamped nature.
The probe particle in the granular system
has no intrinsic Hall angle,
but as function of driving force
shows a finite Hall angle at intermediate drives, with no Hall
angle at low or high drives.
The finite Hall angle arises as a result of
collisions between the probe particle and the spinning
disks, which create a deflection of the probe
particle perpendicular to the driving direction.
This deflection decreases in magnitude as the velocity of the probe
particle increases.

\section{Summary}
We have numerically examined the active rheology of
a single skyrmion driven through a bath of other skyrmions
in the absence of quenched disorder.
Active rheology 
has been used to study the changes in drag
on driven probe particles in various soft matter
and superconducting
vortex systems where the
dynamics is overdamped.
In those systems, the velocity of the probe particle under a
constant driving force rapidly decreases with increasing bath particle
density
due to an increased dragging effect.
For skyrmions, which have a strong Magnus force,
we find that the behavior differs strongly
from what is observed in the damping dominated limit.
The driven skyrmion velocity in the driving direction
is highly non-monotonic as a function of
density, and can increase rather than decreasing 
when the density is increased.
This effect
appears as a boost in the net velocity.
At higher densities,
the velocity decreases with increasing density.
The skyrmion Hall angle also decreases as the bath density increases.
The magnitude
of the velocity boost depends on the system density,
the strength of the Magnus term, 
and the applied drive.
For a fixed density, as we increase the driving force we
find a critical threshold force below which the driven skyrmion does
not move,
a regime in which
the magnitude of the skyrmion Hall angle
increases with drive,
and a regime at higher drive
where the skyrmion Hall angle saturates.
The drive dependence of the skyrmion Hall angle is similar to that
observed for skyrmions driven over quenched disorder.
If the Magnus force is dominant,
a velocity boost appears
which is maximum for an intermediate drive and diminishes at
higher drives.
When the damping force is strong,
the velocities are reduced
but approach the single particle limit at higher drives. 
The velocity-force curves in the damped regime
have
linear or monotonic behavior,
while in the Magnus dominated regime,
the velocity in the driving direction
can decrease with increasing drive,
leading to negative differential conductivity.
The maximum velocity boost 
shifts to higher drives with increased density. 
The velocity boost originates when the driven skyrmion moves at a finite
skyrmion Hall angle and
creates a localized density inhomogeneity in the background skyrmions,
which generate an unbalanced pairwise repulsive force on the
driven skyrmion
perpendicular to the driving direction.
The Magnus force then converts this force into an additional
velocity component in the
direction of drive.
At low bath densities, the localized density fluctuation relaxes quickly
and the velocity boost is small, while
at high drives the
driven skyrmion moves too quickly past the bath skyrmions
for a localized density fluctuation to form,
so the velocity boost is again reduced.
We find regimes in which the skyrmion Hall angle decreases with a
simultaneous increase in the skyrmion velocity, suggesting that
skyrmion-skyrmion interactions can be useful for producing effects that
are of value for use in devices.
We discuss possible experimental realizations of this system
where a single skyrmion could be driven with
some form of tip or optical trap while deflection forces on the
skyrmion are measured.
Beyond skyrmions, our results should be relevant
to any kind of active rheology in systems with gyroscopic forces,
such as active chiral matter,
fluid vortices, electrons in a magnetic field,
fractons, and other gyroscopic systems.

\begin{acknowledgments}
We gratefully acknowledge the support of the U.S. Department of
Energy through the LANL/LDRD program for this work.
This work was supported by the US Department of Energy through
the Los Alamos National Laboratory.  Los Alamos National Laboratory is
operated by Triad National Security, LLC, for the National Nuclear Security
Administration of the U. S. Department of Energy (Contract No. 892333218NCA000001).
\end{acknowledgments}

\bibliography{mybib}

\begin{thebibliography}{103}%
\makeatletter
\providecommand \@ifxundefined [1]{%
 \@ifx{#1\undefined}
}%
\providecommand \@ifnum [1]{%
 \ifnum #1\expandafter \@firstoftwo
 \else \expandafter \@secondoftwo
 \fi
}%
\providecommand \@ifx [1]{%
 \ifx #1\expandafter \@firstoftwo
 \else \expandafter \@secondoftwo
 \fi
}%
\providecommand \natexlab [1]{#1}%
\providecommand \enquote  [1]{``#1''}%
\providecommand \bibnamefont  [1]{#1}%
\providecommand \bibfnamefont [1]{#1}%
\providecommand \citenamefont [1]{#1}%
\providecommand \href@noop [0]{\@secondoftwo}%
\providecommand \href [0]{\begingroup \@sanitize@url \@href}%
\providecommand \@href[1]{\@@startlink{#1}\@@href}%
\providecommand \@@href[1]{\endgroup#1\@@endlink}%
\providecommand \@sanitize@url [0]{\catcode `\\12\catcode `\$12\catcode
  `\&12\catcode `\#12\catcode `\^12\catcode `\_12\catcode `\%12\relax}%
\providecommand \@@startlink[1]{}%
\providecommand \@@endlink[0]{}%
\providecommand \url  [0]{\begingroup\@sanitize@url \@url }%
\providecommand \@url [1]{\endgroup\@href {#1}{\urlprefix }}%
\providecommand \urlprefix  [0]{URL }%
\providecommand \Eprint [0]{\href }%
\providecommand \doibase [0]{http://dx.doi.org/}%
\providecommand \selectlanguage [0]{\@gobble}%
\providecommand \bibinfo  [0]{\@secondoftwo}%
\providecommand \bibfield  [0]{\@secondoftwo}%
\providecommand \translation [1]{[#1]}%
\providecommand \BibitemOpen [0]{}%
\providecommand \bibitemStop [0]{}%
\providecommand \bibitemNoStop [0]{.\EOS\space}%
\providecommand \EOS [0]{\spacefactor3000\relax}%
\providecommand \BibitemShut  [1]{\csname bibitem#1\endcsname}%
\let\auto@bib@innerbib\@empty
\bibitem [{\citenamefont {M{\" u}hlbauer}\ \emph {et~al.}(2009)\citenamefont
  {M{\" u}hlbauer}, \citenamefont {Binz}, \citenamefont {Jonietz},
  \citenamefont {Pfleiderer}, \citenamefont {Rosch}, \citenamefont {Neubauer},
  \citenamefont {Georgii},\ and\ \citenamefont {B{\" o}ni}}]{Muhlbauer09}%
  \BibitemOpen
  \bibfield  {author} {\bibinfo {author} {\bibfnamefont {S.}~\bibnamefont {M{\"
  u}hlbauer}}, \bibinfo {author} {\bibfnamefont {B.}~\bibnamefont {Binz}},
  \bibinfo {author} {\bibfnamefont {F.}~\bibnamefont {Jonietz}}, \bibinfo
  {author} {\bibfnamefont {C.}~\bibnamefont {Pfleiderer}}, \bibinfo {author}
  {\bibfnamefont {A.}~\bibnamefont {Rosch}}, \bibinfo {author} {\bibfnamefont
  {A.}~\bibnamefont {Neubauer}}, \bibinfo {author} {\bibfnamefont
  {R.}~\bibnamefont {Georgii}}, \ and\ \bibinfo {author} {\bibfnamefont
  {P.}~\bibnamefont {B{\" o}ni}},\ }\bibfield  {title} {\enquote {\bibinfo
  {title} {Skyrmion lattice in a chiral magnet},}\ }\href {\doibase
  10.1126/science.1166767} {\bibfield  {journal} {\bibinfo  {journal}
  {Science}\ }\textbf {\bibinfo {volume} {323}},\ \bibinfo {pages} {915--919}
  (\bibinfo {year} {2009})}\BibitemShut {NoStop}%
\bibitem [{\citenamefont {Yu}\ \emph {et~al.}(2010)\citenamefont {Yu},
  \citenamefont {Onose}, \citenamefont {Kanazawa}, \citenamefont {Park},
  \citenamefont {Han}, \citenamefont {Matsui}, \citenamefont {Nagaosa},\ and\
  \citenamefont {Tokura}}]{Yu10}%
  \BibitemOpen
  \bibfield  {author} {\bibinfo {author} {\bibfnamefont {X.~Z.}\ \bibnamefont
  {Yu}}, \bibinfo {author} {\bibfnamefont {Y.}~\bibnamefont {Onose}}, \bibinfo
  {author} {\bibfnamefont {N.}~\bibnamefont {Kanazawa}}, \bibinfo {author}
  {\bibfnamefont {J.~H.}\ \bibnamefont {Park}}, \bibinfo {author}
  {\bibfnamefont {J.~H.}\ \bibnamefont {Han}}, \bibinfo {author} {\bibfnamefont
  {Y.}~\bibnamefont {Matsui}}, \bibinfo {author} {\bibfnamefont
  {N.}~\bibnamefont {Nagaosa}}, \ and\ \bibinfo {author} {\bibfnamefont
  {Y.}~\bibnamefont {Tokura}},\ }\bibfield  {title} {\enquote {\bibinfo {title}
  {Real-space observation of a two-dimensional skyrmion crystal},}\ }\href
  {\doibase 10.1038/nature09124} {\bibfield  {journal} {\bibinfo  {journal}
  {Nature (London)}\ }\textbf {\bibinfo {volume} {465}},\ \bibinfo {pages}
  {901--904} (\bibinfo {year} {2010})}\BibitemShut {NoStop}%
\bibitem [{\citenamefont {Nagaosa}\ and\ \citenamefont
  {Tokura}(2013)}]{Nagaosa13}%
  \BibitemOpen
  \bibfield  {author} {\bibinfo {author} {\bibfnamefont {N.}~\bibnamefont
  {Nagaosa}}\ and\ \bibinfo {author} {\bibfnamefont {Y.}~\bibnamefont
  {Tokura}},\ }\bibfield  {title} {\enquote {\bibinfo {title} {Topological
  properties and dynamics of magnetic skyrmions},}\ }\href {\doibase
  10.1038/NNANO.2013.243} {\bibfield  {journal} {\bibinfo  {journal} {Nature
  Nanotechnol.}\ }\textbf {\bibinfo {volume} {8}},\ \bibinfo {pages} {899--911}
  (\bibinfo {year} {2013})}\BibitemShut {NoStop}%
\bibitem [{\citenamefont {Everschor-Sitte}\ \emph {et~al.}(2018)\citenamefont
  {Everschor-Sitte}, \citenamefont {Masell}, \citenamefont {Reeve},\ and\
  \citenamefont {Kla{\" u}i}}]{EverschorSitte18}%
  \BibitemOpen
  \bibfield  {author} {\bibinfo {author} {\bibfnamefont {K.}~\bibnamefont
  {Everschor-Sitte}}, \bibinfo {author} {\bibfnamefont {J.}~\bibnamefont
  {Masell}}, \bibinfo {author} {\bibfnamefont {R.~M.}\ \bibnamefont {Reeve}}, \
  and\ \bibinfo {author} {\bibfnamefont {M.}~\bibnamefont {Kla{\" u}i}},\
  }\bibfield  {title} {\enquote {\bibinfo {title} {Perspective: Magnetic
  skyrmions - {O}verview of recent progress in an active research field},}\
  }\href {\doibase 10.1063/1.5048972} {\bibfield  {journal} {\bibinfo
  {journal} {J. Appl. Phys.}\ }\textbf {\bibinfo {volume} {124}},\ \bibinfo
  {pages} {240901} (\bibinfo {year} {2018})}\BibitemShut {NoStop}%
\bibitem [{\citenamefont {Reichhardt}\ \emph {et~al.}(2021)\citenamefont
  {Reichhardt}, \citenamefont {Reichhardt},\ and\ \citenamefont
  {Milosevic}}]{Reichhardt21}%
  \BibitemOpen
  \bibfield  {author} {\bibinfo {author} {\bibfnamefont {C.}~\bibnamefont
  {Reichhardt}}, \bibinfo {author} {\bibfnamefont {C.~J.~O.}\ \bibnamefont
  {Reichhardt}}, \ and\ \bibinfo {author} {\bibfnamefont {M.~V.}\ \bibnamefont
  {Milosevic}},\ }\bibfield  {title} {\enquote {\bibinfo {title} {Statics and
  dynamics of skyrmions interacting with pinning: a review},}\ }\href@noop {}
  {\bibfield  {journal} {\bibinfo  {journal} {arXiv e-prints}\ ,\ \bibinfo
  {pages} {arXiv:2102.10464}} (\bibinfo {year} {2021})}\BibitemShut {NoStop}%
\bibitem [{\citenamefont {Woo}\ \emph {et~al.}(2016)\citenamefont {Woo},
  \citenamefont {Litzius}, \citenamefont {Kr{\" u}ger}, \citenamefont {Im},
  \citenamefont {Caretta}, \citenamefont {Richter}, \citenamefont {Mann},
  \citenamefont {Krone}, \citenamefont {Reeve}, \citenamefont {Weigand},
  \citenamefont {Agrawal}, \citenamefont {Lemesh}, \citenamefont {Mawass},
  \citenamefont {Fischer}, \citenamefont {Kl{\" a}ui},\ and\ \citenamefont
  {Beach}}]{Woo16}%
  \BibitemOpen
  \bibfield  {author} {\bibinfo {author} {\bibfnamefont {S.}~\bibnamefont
  {Woo}}, \bibinfo {author} {\bibfnamefont {K.}~\bibnamefont {Litzius}},
  \bibinfo {author} {\bibfnamefont {B.}~\bibnamefont {Kr{\" u}ger}}, \bibinfo
  {author} {\bibfnamefont {M.-Y.}\ \bibnamefont {Im}}, \bibinfo {author}
  {\bibfnamefont {L.}~\bibnamefont {Caretta}}, \bibinfo {author} {\bibfnamefont
  {K.}~\bibnamefont {Richter}}, \bibinfo {author} {\bibfnamefont
  {M.}~\bibnamefont {Mann}}, \bibinfo {author} {\bibfnamefont {A.}~\bibnamefont
  {Krone}}, \bibinfo {author} {\bibfnamefont {R.~M.}\ \bibnamefont {Reeve}},
  \bibinfo {author} {\bibfnamefont {M.}~\bibnamefont {Weigand}}, \bibinfo
  {author} {\bibfnamefont {P.}~\bibnamefont {Agrawal}}, \bibinfo {author}
  {\bibfnamefont {I.}~\bibnamefont {Lemesh}}, \bibinfo {author} {\bibfnamefont
  {M.-A.}\ \bibnamefont {Mawass}}, \bibinfo {author} {\bibfnamefont
  {P.}~\bibnamefont {Fischer}}, \bibinfo {author} {\bibfnamefont
  {M.}~\bibnamefont {Kl{\" a}ui}}, \ and\ \bibinfo {author} {\bibfnamefont
  {G.~S.~D.}\ \bibnamefont {Beach}},\ }\bibfield  {title} {\enquote {\bibinfo
  {title} {Observation of room-temperature magnetic skyrmions and their
  current-driven dynamics in ultrathin metallic ferromagnets},}\ }\href
  {\doibase 10.1038/NMAT4593} {\bibfield  {journal} {\bibinfo  {journal}
  {Nature Mater.}\ }\textbf {\bibinfo {volume} {15}},\ \bibinfo {pages} {501}
  (\bibinfo {year} {2016})}\BibitemShut {NoStop}%
\bibitem [{\citenamefont {Soumyanarayanan}\ \emph {et~al.}(2017)\citenamefont
  {Soumyanarayanan}, \citenamefont {Raju}, \citenamefont {Oyarce},
  \citenamefont {Tan}, \citenamefont {Im}, \citenamefont {Petrovic},
  \citenamefont {Ho}, \citenamefont {Khoo}, \citenamefont {Tran}, \citenamefont
  {Gan}, \citenamefont {Ernult},\ and\ \citenamefont
  {Panagopoulos}}]{Soumyanarayanan17}%
  \BibitemOpen
  \bibfield  {author} {\bibinfo {author} {\bibfnamefont {A.}~\bibnamefont
  {Soumyanarayanan}}, \bibinfo {author} {\bibfnamefont {M.}~\bibnamefont
  {Raju}}, \bibinfo {author} {\bibfnamefont {A.~L.~G.}\ \bibnamefont {Oyarce}},
  \bibinfo {author} {\bibfnamefont {A.~K.~C.}\ \bibnamefont {Tan}}, \bibinfo
  {author} {\bibfnamefont {M.-Y.}\ \bibnamefont {Im}}, \bibinfo {author}
  {\bibfnamefont {A.~P.}\ \bibnamefont {Petrovic}}, \bibinfo {author}
  {\bibfnamefont {P.}~\bibnamefont {Ho}}, \bibinfo {author} {\bibfnamefont
  {K.~H.}\ \bibnamefont {Khoo}}, \bibinfo {author} {\bibfnamefont
  {M.}~\bibnamefont {Tran}}, \bibinfo {author} {\bibfnamefont {C.~K.}\
  \bibnamefont {Gan}}, \bibinfo {author} {\bibfnamefont {F.}~\bibnamefont
  {Ernult}}, \ and\ \bibinfo {author} {\bibfnamefont {C.}~\bibnamefont
  {Panagopoulos}},\ }\bibfield  {title} {\enquote {\bibinfo {title} {Tunable
  room-temperature magnetic skyrmions in {Ir/Fe/Co/Pt} multilayers},}\ }\href
  {\doibase 10.1038/NMAT4934} {\bibfield  {journal} {\bibinfo  {journal}
  {Nature Mater.}\ }\textbf {\bibinfo {volume} {16}},\ \bibinfo {pages} {898}
  (\bibinfo {year} {2017})}\BibitemShut {NoStop}%
\bibitem [{\citenamefont {Montoya}\ \emph {et~al.}(2018)\citenamefont
  {Montoya}, \citenamefont {Tolley}, \citenamefont {Gilbert}, \citenamefont
  {Je}, \citenamefont {Im},\ and\ \citenamefont {Fullerton}}]{Montoya18}%
  \BibitemOpen
  \bibfield  {author} {\bibinfo {author} {\bibfnamefont {S.~A.}\ \bibnamefont
  {Montoya}}, \bibinfo {author} {\bibfnamefont {R.}~\bibnamefont {Tolley}},
  \bibinfo {author} {\bibfnamefont {I.}~\bibnamefont {Gilbert}}, \bibinfo
  {author} {\bibfnamefont {S.-G.}\ \bibnamefont {Je}}, \bibinfo {author}
  {\bibfnamefont {M.-Y.}\ \bibnamefont {Im}}, \ and\ \bibinfo {author}
  {\bibfnamefont {E.~E.}\ \bibnamefont {Fullerton}},\ }\bibfield  {title}
  {\enquote {\bibinfo {title} {Spin-orbit torque induced dipole skyrmion motion
  at room temperature},}\ }\href {\doibase 10.1103/PhysRevB.98.104432}
  {\bibfield  {journal} {\bibinfo  {journal} {Phys. Rev. B}\ }\textbf {\bibinfo
  {volume} {98}},\ \bibinfo {pages} {104432} (\bibinfo {year}
  {2018})}\BibitemShut {NoStop}%
\bibitem [{\citenamefont {Jonietz}\ \emph {et~al.}(2010)\citenamefont
  {Jonietz}, \citenamefont {M{\" u}hlbauer}, \citenamefont {Pfleiderer},
  \citenamefont {Neubauer}, \citenamefont {M{\" u}nzer}, \citenamefont {Bauer},
  \citenamefont {Adams}, \citenamefont {Georgii}, \citenamefont {B{\" o}ni},
  \citenamefont {Duine}, \citenamefont {Everschor}, \citenamefont {Garst},\
  and\ \citenamefont {Rosch}}]{Jonietz10}%
  \BibitemOpen
  \bibfield  {author} {\bibinfo {author} {\bibfnamefont {F.}~\bibnamefont
  {Jonietz}}, \bibinfo {author} {\bibfnamefont {S.}~\bibnamefont {M{\"
  u}hlbauer}}, \bibinfo {author} {\bibfnamefont {C.}~\bibnamefont
  {Pfleiderer}}, \bibinfo {author} {\bibfnamefont {A.}~\bibnamefont
  {Neubauer}}, \bibinfo {author} {\bibfnamefont {W.}~\bibnamefont {M{\"
  u}nzer}}, \bibinfo {author} {\bibfnamefont {A.}~\bibnamefont {Bauer}},
  \bibinfo {author} {\bibfnamefont {T.}~\bibnamefont {Adams}}, \bibinfo
  {author} {\bibfnamefont {R.}~\bibnamefont {Georgii}}, \bibinfo {author}
  {\bibfnamefont {P.}~\bibnamefont {B{\" o}ni}}, \bibinfo {author}
  {\bibfnamefont {R.~A.}\ \bibnamefont {Duine}}, \bibinfo {author}
  {\bibfnamefont {K.}~\bibnamefont {Everschor}}, \bibinfo {author}
  {\bibfnamefont {M.}~\bibnamefont {Garst}}, \ and\ \bibinfo {author}
  {\bibfnamefont {A.}~\bibnamefont {Rosch}},\ }\bibfield  {title} {\enquote
  {\bibinfo {title} {Spin transfer torques in {MnSi} at ultralow current
  densities},}\ }\href {\doibase 10.1126/science.1195709} {\bibfield  {journal}
  {\bibinfo  {journal} {Science}\ }\textbf {\bibinfo {volume} {330}},\ \bibinfo
  {pages} {1648--1651} (\bibinfo {year} {2010})}\BibitemShut {NoStop}%
\bibitem [{\citenamefont {Yu}\ \emph {et~al.}(2012)\citenamefont {Yu},
  \citenamefont {Kanazawa}, \citenamefont {Zhang}, \citenamefont {Nagai},
  \citenamefont {Hara}, \citenamefont {Kimoto}, \citenamefont {Matsui},
  \citenamefont {Onose},\ and\ \citenamefont {Tokura}}]{Yu12}%
  \BibitemOpen
  \bibfield  {author} {\bibinfo {author} {\bibfnamefont {X.~Z.}\ \bibnamefont
  {Yu}}, \bibinfo {author} {\bibfnamefont {N.}~\bibnamefont {Kanazawa}},
  \bibinfo {author} {\bibfnamefont {W.~Z.}\ \bibnamefont {Zhang}}, \bibinfo
  {author} {\bibfnamefont {T.}~\bibnamefont {Nagai}}, \bibinfo {author}
  {\bibfnamefont {T.}~\bibnamefont {Hara}}, \bibinfo {author} {\bibfnamefont
  {K.}~\bibnamefont {Kimoto}}, \bibinfo {author} {\bibfnamefont
  {Y.}~\bibnamefont {Matsui}}, \bibinfo {author} {\bibfnamefont
  {Y.}~\bibnamefont {Onose}}, \ and\ \bibinfo {author} {\bibfnamefont
  {Y.}~\bibnamefont {Tokura}},\ }\bibfield  {title} {\enquote {\bibinfo {title}
  {Skyrmion flow near room temperature in an ultralow current density},}\
  }\href {\doibase 10.1038/ncomms1990} {\bibfield  {journal} {\bibinfo
  {journal} {Nature Commun.}\ }\textbf {\bibinfo {volume} {3}},\ \bibinfo
  {pages} {988} (\bibinfo {year} {2012})}\BibitemShut {NoStop}%
\bibitem [{\citenamefont {Iwasaki}\ \emph {et~al.}(2013)\citenamefont
  {Iwasaki}, \citenamefont {Mochizuki},\ and\ \citenamefont
  {Nagaosa}}]{Iwasaki13}%
  \BibitemOpen
  \bibfield  {author} {\bibinfo {author} {\bibfnamefont {J.}~\bibnamefont
  {Iwasaki}}, \bibinfo {author} {\bibfnamefont {M.}~\bibnamefont {Mochizuki}},
  \ and\ \bibinfo {author} {\bibfnamefont {N.}~\bibnamefont {Nagaosa}},\
  }\bibfield  {title} {\enquote {\bibinfo {title} {Universal current-velocity
  relation of skyrmion motion in chiral magnets},}\ }\href {\doibase
  10.1038/ncomms2442} {\bibfield  {journal} {\bibinfo  {journal} {Nature
  Commun.}\ }\textbf {\bibinfo {volume} {4}},\ \bibinfo {pages} {1463}
  (\bibinfo {year} {2013})}\BibitemShut {NoStop}%
\bibitem [{\citenamefont {Zhang}\ \emph {et~al.}(2018)\citenamefont {Zhang},
  \citenamefont {Wang}, \citenamefont {Burn}, \citenamefont {Peng},
  \citenamefont {Berger}, \citenamefont {Bauer}, \citenamefont {Pfleiderer},
  \citenamefont {van~der Laan},\ and\ \citenamefont {Hesjedal}}]{Zhang18}%
  \BibitemOpen
  \bibfield  {author} {\bibinfo {author} {\bibfnamefont {S.~L.}\ \bibnamefont
  {Zhang}}, \bibinfo {author} {\bibfnamefont {W.~W.}\ \bibnamefont {Wang}},
  \bibinfo {author} {\bibfnamefont {D.~M.}\ \bibnamefont {Burn}}, \bibinfo
  {author} {\bibfnamefont {H.}~\bibnamefont {Peng}}, \bibinfo {author}
  {\bibfnamefont {H.}~\bibnamefont {Berger}}, \bibinfo {author} {\bibfnamefont
  {A.}~\bibnamefont {Bauer}}, \bibinfo {author} {\bibfnamefont
  {C.}~\bibnamefont {Pfleiderer}}, \bibinfo {author} {\bibfnamefont
  {G.}~\bibnamefont {van~der Laan}}, \ and\ \bibinfo {author} {\bibfnamefont
  {T.}~\bibnamefont {Hesjedal}},\ }\bibfield  {title} {\enquote {\bibinfo
  {title} {Manipulation of skyrmion motion by magnetic field gradients},}\
  }\href {\doibase 10.1038/s41467-018-04563-4} {\bibfield  {journal} {\bibinfo
  {journal} {Nature Commun.}\ }\textbf {\bibinfo {volume} {9}},\ \bibinfo
  {pages} {2115} (\bibinfo {year} {2018})}\BibitemShut {NoStop}%
\bibitem [{\citenamefont {Mochizuki}\ \emph {et~al.}(2014)\citenamefont
  {Mochizuki}, \citenamefont {Yu}, \citenamefont {Seki}, \citenamefont
  {Kanazawa}, \citenamefont {Koshibae}, \citenamefont {Zang}, \citenamefont
  {Mostovoy}, \citenamefont {Tokura},\ and\ \citenamefont
  {Nagaosa}}]{Mochizuki14}%
  \BibitemOpen
  \bibfield  {author} {\bibinfo {author} {\bibfnamefont {M.}~\bibnamefont
  {Mochizuki}}, \bibinfo {author} {\bibfnamefont {X.~Z.}\ \bibnamefont {Yu}},
  \bibinfo {author} {\bibfnamefont {S.}~\bibnamefont {Seki}}, \bibinfo {author}
  {\bibfnamefont {N.}~\bibnamefont {Kanazawa}}, \bibinfo {author}
  {\bibfnamefont {W.}~\bibnamefont {Koshibae}}, \bibinfo {author}
  {\bibfnamefont {J.}~\bibnamefont {Zang}}, \bibinfo {author} {\bibfnamefont
  {M.}~\bibnamefont {Mostovoy}}, \bibinfo {author} {\bibfnamefont
  {Y.}~\bibnamefont {Tokura}}, \ and\ \bibinfo {author} {\bibfnamefont
  {N.}~\bibnamefont {Nagaosa}},\ }\bibfield  {title} {\enquote {\bibinfo
  {title} {Thermally driven ratchet motion of a skyrmion microcrystal and
  topological magnon {H}all effect},}\ }\href {\doibase 10.1038/NMAT3862}
  {\bibfield  {journal} {\bibinfo  {journal} {Nature Mater.}\ }\textbf
  {\bibinfo {volume} {13}},\ \bibinfo {pages} {241--246} (\bibinfo {year}
  {2014})}\BibitemShut {NoStop}%
\bibitem [{\citenamefont {Fert}\ \emph {et~al.}(2013)\citenamefont {Fert},
  \citenamefont {Cros},\ and\ \citenamefont {Sampaio}}]{Fert13}%
  \BibitemOpen
  \bibfield  {author} {\bibinfo {author} {\bibfnamefont {A.}~\bibnamefont
  {Fert}}, \bibinfo {author} {\bibfnamefont {V.}~\bibnamefont {Cros}}, \ and\
  \bibinfo {author} {\bibfnamefont {J.}~\bibnamefont {Sampaio}},\ }\bibfield
  {title} {\enquote {\bibinfo {title} {Skyrmions on the track},}\ }\href
  {\doibase 10.1038/nnano.2013.29} {\bibfield  {journal} {\bibinfo  {journal}
  {Nature Nanotechnol.}\ }\textbf {\bibinfo {volume} {8}},\ \bibinfo {pages}
  {152--156} (\bibinfo {year} {2013})}\BibitemShut {NoStop}%
\bibitem [{\citenamefont {Tomasello}\ \emph {et~al.}(2014)\citenamefont
  {Tomasello}, \citenamefont {Martinez}, \citenamefont {Zivieri}, \citenamefont
  {Torres}, \citenamefont {Carpentieri},\ and\ \citenamefont
  {Finocchio}}]{Tomasello14}%
  \BibitemOpen
  \bibfield  {author} {\bibinfo {author} {\bibfnamefont {R.}~\bibnamefont
  {Tomasello}}, \bibinfo {author} {\bibfnamefont {E.}~\bibnamefont {Martinez}},
  \bibinfo {author} {\bibfnamefont {R.}~\bibnamefont {Zivieri}}, \bibinfo
  {author} {\bibfnamefont {L.}~\bibnamefont {Torres}}, \bibinfo {author}
  {\bibfnamefont {M.}~\bibnamefont {Carpentieri}}, \ and\ \bibinfo {author}
  {\bibfnamefont {G.}~\bibnamefont {Finocchio}},\ }\bibfield  {title} {\enquote
  {\bibinfo {title} {A strategy for the design of skyrmion racetrack
  memories},}\ }\href {\doibase 10.1038/srep06784} {\bibfield  {journal}
  {\bibinfo  {journal} {Sci. Rep.}\ }\textbf {\bibinfo {volume} {4}},\ \bibinfo
  {pages} {6784} (\bibinfo {year} {2014})}\BibitemShut {NoStop}%
\bibitem [{\citenamefont {Fert}\ \emph {et~al.}(2017)\citenamefont {Fert},
  \citenamefont {Reyren},\ and\ \citenamefont {Cros}}]{Fert17}%
  \BibitemOpen
  \bibfield  {author} {\bibinfo {author} {\bibfnamefont {A.}~\bibnamefont
  {Fert}}, \bibinfo {author} {\bibfnamefont {N.}~\bibnamefont {Reyren}}, \ and\
  \bibinfo {author} {\bibfnamefont {V.}~\bibnamefont {Cros}},\ }\bibfield
  {title} {\enquote {\bibinfo {title} {Magnetic skyrmions: advances in physics
  and potential applications},}\ }\href {\doibase 10.1038/natrevmats.2017.31}
  {\bibfield  {journal} {\bibinfo  {journal} {Nature Rev. Mater.}\ }\textbf
  {\bibinfo {volume} {2}},\ \bibinfo {pages} {17031} (\bibinfo {year}
  {2017})}\BibitemShut {NoStop}%
\bibitem [{\citenamefont {Z{\' a}zvorka}\ \emph {et~al.}(2019)\citenamefont
  {Z{\' a}zvorka}, \citenamefont {Jakobs}, \citenamefont {Heinze},
  \citenamefont {Keil}, \citenamefont {Kromin}, \citenamefont {Jaiswal},
  \citenamefont {Litzius}, \citenamefont {Jakob}, \citenamefont {Virnau},
  \citenamefont {Pinna}, \citenamefont {Everschor-Sitte}, \citenamefont {R{\'
  o}zsa}, \citenamefont {Donges}, \citenamefont {Nowak},\ and\ \citenamefont
  {Kl{\"a}ui}}]{Zazvorka19}%
  \BibitemOpen
  \bibfield  {author} {\bibinfo {author} {\bibfnamefont {J.}~\bibnamefont {Z{\'
  a}zvorka}}, \bibinfo {author} {\bibfnamefont {F.}~\bibnamefont {Jakobs}},
  \bibinfo {author} {\bibfnamefont {D.}~\bibnamefont {Heinze}}, \bibinfo
  {author} {\bibfnamefont {N.}~\bibnamefont {Keil}}, \bibinfo {author}
  {\bibfnamefont {S.}~\bibnamefont {Kromin}}, \bibinfo {author} {\bibfnamefont
  {S.}~\bibnamefont {Jaiswal}}, \bibinfo {author} {\bibfnamefont
  {K.}~\bibnamefont {Litzius}}, \bibinfo {author} {\bibfnamefont
  {G.}~\bibnamefont {Jakob}}, \bibinfo {author} {\bibfnamefont
  {P.}~\bibnamefont {Virnau}}, \bibinfo {author} {\bibfnamefont
  {D.}~\bibnamefont {Pinna}}, \bibinfo {author} {\bibfnamefont
  {K.}~\bibnamefont {Everschor-Sitte}}, \bibinfo {author} {\bibfnamefont
  {L.}~\bibnamefont {R{\' o}zsa}}, \bibinfo {author} {\bibfnamefont
  {A.}~\bibnamefont {Donges}}, \bibinfo {author} {\bibfnamefont
  {U.}~\bibnamefont {Nowak}}, \ and\ \bibinfo {author} {\bibfnamefont
  {M.}~\bibnamefont {Kl{\"a}ui}},\ }\bibfield  {title} {\enquote {\bibinfo
  {title} {Thermal skyrmion diffusion used in a reshuffler device},}\ }\href
  {\doibase 10.1038/s41565-019-0436-8} {\bibfield  {journal} {\bibinfo
  {journal} {Nature Nanotechnol.}\ }\textbf {\bibinfo {volume} {14}},\ \bibinfo
  {pages} {658--661} (\bibinfo {year} {2019})}\BibitemShut {NoStop}%
\bibitem [{\citenamefont {Zhang}\ \emph {et~al.}(2020)\citenamefont {Zhang},
  \citenamefont {Zhu}, \citenamefont {Kang}, \citenamefont {Zhang},\ and\
  \citenamefont {Zhao}}]{Zhang20a}%
  \BibitemOpen
  \bibfield  {author} {\bibinfo {author} {\bibfnamefont {H.}~\bibnamefont
  {Zhang}}, \bibinfo {author} {\bibfnamefont {D.}~\bibnamefont {Zhu}}, \bibinfo
  {author} {\bibfnamefont {W.}~\bibnamefont {Kang}}, \bibinfo {author}
  {\bibfnamefont {Y.}~\bibnamefont {Zhang}}, \ and\ \bibinfo {author}
  {\bibfnamefont {W.}~\bibnamefont {Zhao}},\ }\bibfield  {title} {\enquote
  {\bibinfo {title} {Stochastic computing implemented by skyrmionic logic
  devices},}\ }\href {\doibase 10.1103/PhysRevApplied.13.054049} {\bibfield
  {journal} {\bibinfo  {journal} {Phys. Rev. Applied}\ }\textbf {\bibinfo
  {volume} {13}},\ \bibinfo {pages} {054049} (\bibinfo {year}
  {2020})}\BibitemShut {NoStop}%
\bibitem [{\citenamefont {Pinna}\ \emph {et~al.}(2020)\citenamefont {Pinna},
  \citenamefont {Bourianoff},\ and\ \citenamefont {Everschor-Sitte}}]{Pinna20}%
  \BibitemOpen
  \bibfield  {author} {\bibinfo {author} {\bibfnamefont {D.}~\bibnamefont
  {Pinna}}, \bibinfo {author} {\bibfnamefont {G.}~\bibnamefont {Bourianoff}}, \
  and\ \bibinfo {author} {\bibfnamefont {K.}~\bibnamefont {Everschor-Sitte}},\
  }\bibfield  {title} {\enquote {\bibinfo {title} {Reservoir computing with
  random skyrmion textures},}\ }\href {\doibase
  10.1103/PhysRevApplied.14.054020} {\bibfield  {journal} {\bibinfo  {journal}
  {Phys. Rev. Applied}\ }\textbf {\bibinfo {volume} {14}},\ \bibinfo {pages}
  {054020} (\bibinfo {year} {2020})}\BibitemShut {NoStop}%
\bibitem [{\citenamefont {Fisher}(1998)}]{Fisher98}%
  \BibitemOpen
  \bibfield  {author} {\bibinfo {author} {\bibfnamefont {D.~S.}\ \bibnamefont
  {Fisher}},\ }\bibfield  {title} {\enquote {\bibinfo {title} {Collective
  transport in random media: from superconductors to earthquakes},}\ }\href
  {\doibase 10.1016/S0370-1573(98)00008-8} {\bibfield  {journal} {\bibinfo
  {journal} {Phys. Rep.}\ }\textbf {\bibinfo {volume} {301}},\ \bibinfo {pages}
  {113--150} (\bibinfo {year} {1998})}\BibitemShut {NoStop}%
\bibitem [{\citenamefont {Reichhardt}\ and\ \citenamefont
  {Reichhardt}(2017)}]{Reichhardt17}%
  \BibitemOpen
  \bibfield  {author} {\bibinfo {author} {\bibfnamefont {C.}~\bibnamefont
  {Reichhardt}}\ and\ \bibinfo {author} {\bibfnamefont {C.~J.~Olson}\
  \bibnamefont {Reichhardt}},\ }\bibfield  {title} {\enquote {\bibinfo {title}
  {Depinning and nonequilibrium dynamic phases of particle assemblies driven
  over random and ordered substrates: a review},}\ }\href {\doibase
  10.1088/1361-6633/80/2/026501} {\bibfield  {journal} {\bibinfo  {journal}
  {Rep. Prog. Phys.}\ }\textbf {\bibinfo {volume} {80}},\ \bibinfo {pages}
  {026501} (\bibinfo {year} {2017})}\BibitemShut {NoStop}%
\bibitem [{\citenamefont {Bhattacharya}\ and\ \citenamefont
  {Higgins}(1993)}]{Bhattacharya93}%
  \BibitemOpen
  \bibfield  {author} {\bibinfo {author} {\bibfnamefont {S.}~\bibnamefont
  {Bhattacharya}}\ and\ \bibinfo {author} {\bibfnamefont {M.~J.}\ \bibnamefont
  {Higgins}},\ }\bibfield  {title} {\enquote {\bibinfo {title} {Dynamics of a
  disordered flux line lattice},}\ }\href {\doibase
  10.1103/PhysRevLett.70.2617} {\bibfield  {journal} {\bibinfo  {journal}
  {Phys. Rev. Lett.}\ }\textbf {\bibinfo {volume} {70}},\ \bibinfo {pages}
  {2617--2620} (\bibinfo {year} {1993})}\BibitemShut {NoStop}%
\bibitem [{\citenamefont {Blatter}\ \emph {et~al.}(1994)\citenamefont
  {Blatter}, \citenamefont {Feigel'man}, \citenamefont {Geshkenbein},
  \citenamefont {Larkin},\ and\ \citenamefont {Vinokur}}]{Blatter94}%
  \BibitemOpen
  \bibfield  {author} {\bibinfo {author} {\bibfnamefont {G.}~\bibnamefont
  {Blatter}}, \bibinfo {author} {\bibfnamefont {M.~V.}\ \bibnamefont
  {Feigel'man}}, \bibinfo {author} {\bibfnamefont {V.~B.}\ \bibnamefont
  {Geshkenbein}}, \bibinfo {author} {\bibfnamefont {A.~I.}\ \bibnamefont
  {Larkin}}, \ and\ \bibinfo {author} {\bibfnamefont {V.~M.}\ \bibnamefont
  {Vinokur}},\ }\bibfield  {title} {\enquote {\bibinfo {title} {Vortices in
  high-temperature superconductors},}\ }\href {\doibase
  10.1103/RevModPhys.66.1125} {\bibfield  {journal} {\bibinfo  {journal} {Rev.
  Mod. Phys.}\ }\textbf {\bibinfo {volume} {66}},\ \bibinfo {pages}
  {1125--1388} (\bibinfo {year} {1994})}\BibitemShut {NoStop}%
\bibitem [{\citenamefont {Tierno}(2012)}]{Tierno12a}%
  \BibitemOpen
  \bibfield  {author} {\bibinfo {author} {\bibfnamefont {P.}~\bibnamefont
  {Tierno}},\ }\bibfield  {title} {\enquote {\bibinfo {title} {Depinning and
  collective dynamics of magnetically driven colloidal monolayers},}\ }\href
  {\doibase 10.1103/PhysRevLett.109.198304} {\bibfield  {journal} {\bibinfo
  {journal} {Phys. Rev. Lett.}\ }\textbf {\bibinfo {volume} {109}},\ \bibinfo
  {pages} {198304} (\bibinfo {year} {2012})}\BibitemShut {NoStop}%
\bibitem [{\citenamefont {Reichhardt}\ \emph {et~al.}(2001)\citenamefont
  {Reichhardt}, \citenamefont {Olson}, \citenamefont {Gr\o{}nbech-Jensen},\
  and\ \citenamefont {Nori}}]{Reichhardt01}%
  \BibitemOpen
  \bibfield  {author} {\bibinfo {author} {\bibfnamefont {C.}~\bibnamefont
  {Reichhardt}}, \bibinfo {author} {\bibfnamefont {C.~J.}\ \bibnamefont
  {Olson}}, \bibinfo {author} {\bibfnamefont {N.}~\bibnamefont
  {Gr\o{}nbech-Jensen}}, \ and\ \bibinfo {author} {\bibfnamefont
  {F.}~\bibnamefont {Nori}},\ }\bibfield  {title} {\enquote {\bibinfo {title}
  {Moving {Wigner} glasses and smectics: Dynamics of disordered {Wigner}
  crystals},}\ }\href {\doibase 10.1103/PhysRevLett.86.4354} {\bibfield
  {journal} {\bibinfo  {journal} {Phys. Rev. Lett.}\ }\textbf {\bibinfo
  {volume} {86}},\ \bibinfo {pages} {4354--4357} (\bibinfo {year}
  {2001})}\BibitemShut {NoStop}%
\bibitem [{\citenamefont {Vanossi}\ \emph {et~al.}(2013)\citenamefont
  {Vanossi}, \citenamefont {Manini}, \citenamefont {Urbakh}, \citenamefont
  {Zapperi},\ and\ \citenamefont {Tosatti}}]{Vanossi13}%
  \BibitemOpen
  \bibfield  {author} {\bibinfo {author} {\bibfnamefont {A.}~\bibnamefont
  {Vanossi}}, \bibinfo {author} {\bibfnamefont {N.}~\bibnamefont {Manini}},
  \bibinfo {author} {\bibfnamefont {M.}~\bibnamefont {Urbakh}}, \bibinfo
  {author} {\bibfnamefont {S.}~\bibnamefont {Zapperi}}, \ and\ \bibinfo
  {author} {\bibfnamefont {E.}~\bibnamefont {Tosatti}},\ }\bibfield  {title}
  {\enquote {\bibinfo {title} {Colloquium: Modeling friction: From nanoscale to
  mesoscale},}\ }\href {\doibase 10.1103/RevModPhys.85.529} {\bibfield
  {journal} {\bibinfo  {journal} {Rev. Mod. Phys.}\ }\textbf {\bibinfo {volume}
  {85}},\ \bibinfo {pages} {529--552} (\bibinfo {year} {2013})}\BibitemShut
  {NoStop}%
\bibitem [{\citenamefont {O'Hern}\ \emph {et~al.}(2003)\citenamefont {O'Hern},
  \citenamefont {Silbert}, \citenamefont {Liu},\ and\ \citenamefont
  {Nagel}}]{OHern03}%
  \BibitemOpen
  \bibfield  {author} {\bibinfo {author} {\bibfnamefont {C.~S.}\ \bibnamefont
  {O'Hern}}, \bibinfo {author} {\bibfnamefont {L.~E.}\ \bibnamefont {Silbert}},
  \bibinfo {author} {\bibfnamefont {A.~J.}\ \bibnamefont {Liu}}, \ and\
  \bibinfo {author} {\bibfnamefont {S.~R.}\ \bibnamefont {Nagel}},\ }\bibfield
  {title} {\enquote {\bibinfo {title} {Jamming at zero temperature and zero
  applied stress: The epitome of disorder},}\ }\href {\doibase
  10.1103/PhysRevE.68.011306} {\bibfield  {journal} {\bibinfo  {journal} {Phys.
  Rev. E}\ }\textbf {\bibinfo {volume} {68}},\ \bibinfo {pages} {011306}
  (\bibinfo {year} {2003})}\BibitemShut {NoStop}%
\bibitem [{\citenamefont {Everschor-Sitte}\ and\ \citenamefont
  {Sitte}(2014)}]{EverschorSitte14}%
  \BibitemOpen
  \bibfield  {author} {\bibinfo {author} {\bibfnamefont {K.}~\bibnamefont
  {Everschor-Sitte}}\ and\ \bibinfo {author} {\bibfnamefont {M.}~\bibnamefont
  {Sitte}},\ }\bibfield  {title} {\enquote {\bibinfo {title} {Real-space
  {B}erry phases: Skyrmion soccer (invited)},}\ }\href {\doibase
  10.1063/1.4870695} {\bibfield  {journal} {\bibinfo  {journal} {J. Appl.
  Phys.}\ }\textbf {\bibinfo {volume} {115}},\ \bibinfo {pages} {172602}
  (\bibinfo {year} {2014})}\BibitemShut {NoStop}%
\bibitem [{\citenamefont {Reichhardt}\ \emph
  {et~al.}(2015{\natexlab{a}})\citenamefont {Reichhardt}, \citenamefont {Ray},\
  and\ \citenamefont {Reichhardt}}]{Reichhardt15a}%
  \BibitemOpen
  \bibfield  {author} {\bibinfo {author} {\bibfnamefont {C.}~\bibnamefont
  {Reichhardt}}, \bibinfo {author} {\bibfnamefont {D.}~\bibnamefont {Ray}}, \
  and\ \bibinfo {author} {\bibfnamefont {C.~J.~Olson}\ \bibnamefont
  {Reichhardt}},\ }\bibfield  {title} {\enquote {\bibinfo {title} {Quantized
  transport for a skyrmion moving on a two-dimensional periodic substrate},}\
  }\href {\doibase 10.1103/PhysRevB.91.104426} {\bibfield  {journal} {\bibinfo
  {journal} {Phys. Rev. B}\ }\textbf {\bibinfo {volume} {91}},\ \bibinfo
  {pages} {104426} (\bibinfo {year} {2015}{\natexlab{a}})}\BibitemShut
  {NoStop}%
\bibitem [{\citenamefont {Reichhardt}\ \emph
  {et~al.}(2015{\natexlab{b}})\citenamefont {Reichhardt}, \citenamefont {Ray},\
  and\ \citenamefont {Reichhardt}}]{Reichhardt15}%
  \BibitemOpen
  \bibfield  {author} {\bibinfo {author} {\bibfnamefont {C.}~\bibnamefont
  {Reichhardt}}, \bibinfo {author} {\bibfnamefont {D.}~\bibnamefont {Ray}}, \
  and\ \bibinfo {author} {\bibfnamefont {C.~J.~Olson}\ \bibnamefont
  {Reichhardt}},\ }\bibfield  {title} {\enquote {\bibinfo {title} {Collective
  transport properties of driven skyrmions with random disorder},}\ }\href
  {\doibase 10.1103/PhysRevLett.114.217202} {\bibfield  {journal} {\bibinfo
  {journal} {Phys. Rev. Lett.}\ }\textbf {\bibinfo {volume} {114}},\ \bibinfo
  {pages} {217202} (\bibinfo {year} {2015}{\natexlab{b}})}\BibitemShut
  {NoStop}%
\bibitem [{\citenamefont {Jiang}\ \emph {et~al.}(2017)\citenamefont {Jiang},
  \citenamefont {Zhang}, \citenamefont {Yu}, \citenamefont {Zhang},
  \citenamefont {Wang}, \citenamefont {Jungfleisch}, \citenamefont {Pearson},
  \citenamefont {Cheng}, \citenamefont {Heinonen}, \citenamefont {Wang},
  \citenamefont {Zhou}, \citenamefont {Hoffmann},\ and\ \citenamefont
  {te~Velthuis}}]{Jiang17}%
  \BibitemOpen
  \bibfield  {author} {\bibinfo {author} {\bibfnamefont {W.}~\bibnamefont
  {Jiang}}, \bibinfo {author} {\bibfnamefont {X.}~\bibnamefont {Zhang}},
  \bibinfo {author} {\bibfnamefont {G.}~\bibnamefont {Yu}}, \bibinfo {author}
  {\bibfnamefont {W.}~\bibnamefont {Zhang}}, \bibinfo {author} {\bibfnamefont
  {X.}~\bibnamefont {Wang}}, \bibinfo {author} {\bibfnamefont {M.~B.}\
  \bibnamefont {Jungfleisch}}, \bibinfo {author} {\bibfnamefont {J.~E.}\
  \bibnamefont {Pearson}}, \bibinfo {author} {\bibfnamefont {X.}~\bibnamefont
  {Cheng}}, \bibinfo {author} {\bibfnamefont {O.}~\bibnamefont {Heinonen}},
  \bibinfo {author} {\bibfnamefont {K.~L.}\ \bibnamefont {Wang}}, \bibinfo
  {author} {\bibfnamefont {Y.}~\bibnamefont {Zhou}}, \bibinfo {author}
  {\bibfnamefont {A.}~\bibnamefont {Hoffmann}}, \ and\ \bibinfo {author}
  {\bibfnamefont {S.~G.~E.}\ \bibnamefont {te~Velthuis}},\ }\bibfield  {title}
  {\enquote {\bibinfo {title} {Direct observation of the skyrmion {H}all
  effect},}\ }\href {\doibase 10.1038/NPHYS3883} {\bibfield  {journal}
  {\bibinfo  {journal} {Nature Phys.}\ }\textbf {\bibinfo {volume} {13}},\
  \bibinfo {pages} {162--169} (\bibinfo {year} {2017})}\BibitemShut {NoStop}%
\bibitem [{\citenamefont {Litzius}\ \emph {et~al.}(2017)\citenamefont
  {Litzius}, \citenamefont {Lemesh}, \citenamefont {Kr{\" u}ger}, \citenamefont
  {Bassirian}, \citenamefont {Caretta}, \citenamefont {Richter}, \citenamefont
  {B{\" u}ttner}, \citenamefont {Sato}, \citenamefont {Tretiakov},
  \citenamefont {F{\" o}rster}, \citenamefont {Reeve}, \citenamefont {Weigand},
  \citenamefont {Bykova}, \citenamefont {Stoll}, \citenamefont {Sch{\" u}tz},
  \citenamefont {Beach},\ and\ \citenamefont {Kl{\" a}ui}}]{Litzius17}%
  \BibitemOpen
  \bibfield  {author} {\bibinfo {author} {\bibfnamefont {K.}~\bibnamefont
  {Litzius}}, \bibinfo {author} {\bibfnamefont {I.}~\bibnamefont {Lemesh}},
  \bibinfo {author} {\bibfnamefont {B.}~\bibnamefont {Kr{\" u}ger}}, \bibinfo
  {author} {\bibfnamefont {P.}~\bibnamefont {Bassirian}}, \bibinfo {author}
  {\bibfnamefont {L.}~\bibnamefont {Caretta}}, \bibinfo {author} {\bibfnamefont
  {K.}~\bibnamefont {Richter}}, \bibinfo {author} {\bibfnamefont
  {F.}~\bibnamefont {B{\" u}ttner}}, \bibinfo {author} {\bibfnamefont
  {K.}~\bibnamefont {Sato}}, \bibinfo {author} {\bibfnamefont {O.~A.}\
  \bibnamefont {Tretiakov}}, \bibinfo {author} {\bibfnamefont {J.}~\bibnamefont
  {F{\" o}rster}}, \bibinfo {author} {\bibfnamefont {R.~M.}\ \bibnamefont
  {Reeve}}, \bibinfo {author} {\bibfnamefont {M.}~\bibnamefont {Weigand}},
  \bibinfo {author} {\bibfnamefont {L.}~\bibnamefont {Bykova}}, \bibinfo
  {author} {\bibfnamefont {H.}~\bibnamefont {Stoll}}, \bibinfo {author}
  {\bibfnamefont {G.}~\bibnamefont {Sch{\" u}tz}}, \bibinfo {author}
  {\bibfnamefont {G.~S.~D.}\ \bibnamefont {Beach}}, \ and\ \bibinfo {author}
  {\bibfnamefont {M.}~\bibnamefont {Kl{\" a}ui}},\ }\bibfield  {title}
  {\enquote {\bibinfo {title} {Skyrmion {H}all effect revealed by direct
  time-resolved {X}-ray microscopy},}\ }\href {\doibase 10.1038/NPHYS4000}
  {\bibfield  {journal} {\bibinfo  {journal} {Nature Phys.}\ }\textbf {\bibinfo
  {volume} {13}},\ \bibinfo {pages} {170--175} (\bibinfo {year}
  {2017})}\BibitemShut {NoStop}%
\bibitem [{\citenamefont {Iwasaki}\ \emph {et~al.}(2014)\citenamefont
  {Iwasaki}, \citenamefont {Koshibae},\ and\ \citenamefont
  {Nagaosa}}]{Iwasaki14}%
  \BibitemOpen
  \bibfield  {author} {\bibinfo {author} {\bibfnamefont {J.}~\bibnamefont
  {Iwasaki}}, \bibinfo {author} {\bibfnamefont {W.}~\bibnamefont {Koshibae}}, \
  and\ \bibinfo {author} {\bibfnamefont {N.}~\bibnamefont {Nagaosa}},\
  }\bibfield  {title} {\enquote {\bibinfo {title} {Colossal spin transfer
  torque effect on skyrmion along the edge},}\ }\href {\doibase
  10.1021/nl501379k} {\bibfield  {journal} {\bibinfo  {journal} {Nano Lett.}\
  }\textbf {\bibinfo {volume} {14}},\ \bibinfo {pages} {4432} (\bibinfo {year}
  {2014})}\BibitemShut {NoStop}%
\bibitem [{\citenamefont {Zhang}\ \emph {et~al.}(2015)\citenamefont {Zhang},
  \citenamefont {Zhao}, \citenamefont {Fangohr}, \citenamefont {Liu},
  \citenamefont {Xia}, \citenamefont {Xia},\ and\ \citenamefont
  {Morvan}}]{Zhang15a}%
  \BibitemOpen
  \bibfield  {author} {\bibinfo {author} {\bibfnamefont {X.}~\bibnamefont
  {Zhang}}, \bibinfo {author} {\bibfnamefont {G.~P.}\ \bibnamefont {Zhao}},
  \bibinfo {author} {\bibfnamefont {H.}~\bibnamefont {Fangohr}}, \bibinfo
  {author} {\bibfnamefont {J.~P.}\ \bibnamefont {Liu}}, \bibinfo {author}
  {\bibfnamefont {W.~X.}\ \bibnamefont {Xia}}, \bibinfo {author} {\bibfnamefont
  {J.}~\bibnamefont {Xia}}, \ and\ \bibinfo {author} {\bibfnamefont {F.~J.}\
  \bibnamefont {Morvan}},\ }\bibfield  {title} {\enquote {\bibinfo {title}
  {Skyrmion-skyrmion and skyrmion-edge repulsions in skyrmion-based racetrack
  memory},}\ }\href {\doibase 10.1038/srep07643} {\bibfield  {journal}
  {\bibinfo  {journal} {Sci. Rep.}\ }\textbf {\bibinfo {volume} {5}},\ \bibinfo
  {pages} {7643} (\bibinfo {year} {2015})}\BibitemShut {NoStop}%
\bibitem [{\citenamefont {Reichhardt}\ and\ \citenamefont
  {Reichhardt}(2016{\natexlab{a}})}]{Reichhardt16a}%
  \BibitemOpen
  \bibfield  {author} {\bibinfo {author} {\bibfnamefont {C.}~\bibnamefont
  {Reichhardt}}\ and\ \bibinfo {author} {\bibfnamefont {C.~J.~Olson}\
  \bibnamefont {Reichhardt}},\ }\bibfield  {title} {\enquote {\bibinfo {title}
  {Magnus-induced dynamics of driven skyrmions on a quasi-one-dimensional
  periodic substrate},}\ }\href {\doibase 10.1103/PhysRevB.94.094413}
  {\bibfield  {journal} {\bibinfo  {journal} {Phys. Rev. B}\ }\textbf {\bibinfo
  {volume} {94}},\ \bibinfo {pages} {094413} (\bibinfo {year}
  {2016}{\natexlab{a}})}\BibitemShut {NoStop}%
\bibitem [{\citenamefont {Castell-Queralt}\ \emph {et~al.}(2019)\citenamefont
  {Castell-Queralt}, \citenamefont {Gonzalez-Gomez}, \citenamefont {Del-Valle},
  \citenamefont {Sanchez},\ and\ \citenamefont {Navau}}]{CastellQueralt19}%
  \BibitemOpen
  \bibfield  {author} {\bibinfo {author} {\bibfnamefont {J.}~\bibnamefont
  {Castell-Queralt}}, \bibinfo {author} {\bibfnamefont {L.}~\bibnamefont
  {Gonzalez-Gomez}}, \bibinfo {author} {\bibfnamefont {N.}~\bibnamefont
  {Del-Valle}}, \bibinfo {author} {\bibfnamefont {A.}~\bibnamefont {Sanchez}},
  \ and\ \bibinfo {author} {\bibfnamefont {C.}~\bibnamefont {Navau}},\
  }\bibfield  {title} {\enquote {\bibinfo {title} {Accelerating, guiding, and
  compressing skyrmions by defect rails},}\ }\href {\doibase
  10.1039/c9nr02171j} {\bibfield  {journal} {\bibinfo  {journal} {Nanoscale}\
  }\textbf {\bibinfo {volume} {11}},\ \bibinfo {pages} {12589--12594} (\bibinfo
  {year} {2019})}\BibitemShut {NoStop}%
\bibitem [{\citenamefont {Xing}\ \emph {et~al.}(2020)\citenamefont {Xing},
  \citenamefont {\AA{}kerman},\ and\ \citenamefont {Zhou}}]{Xing20}%
  \BibitemOpen
  \bibfield  {author} {\bibinfo {author} {\bibfnamefont {X.}~\bibnamefont
  {Xing}}, \bibinfo {author} {\bibfnamefont {J.}~\bibnamefont {\AA{}kerman}}, \
  and\ \bibinfo {author} {\bibfnamefont {Y.}~\bibnamefont {Zhou}},\ }\bibfield
  {title} {\enquote {\bibinfo {title} {Enhanced skyrmion motion via strip
  domain wall},}\ }\href {\doibase 10.1103/PhysRevB.101.214432} {\bibfield
  {journal} {\bibinfo  {journal} {Phys. Rev. B}\ }\textbf {\bibinfo {volume}
  {101}},\ \bibinfo {pages} {214432} (\bibinfo {year} {2020})}\BibitemShut
  {NoStop}%
\bibitem [{\citenamefont {Reichhardt}\ \emph
  {et~al.}(2015{\natexlab{c}})\citenamefont {Reichhardt}, \citenamefont {Ray},\
  and\ \citenamefont {Reichhardt}}]{Reichhardt15aa}%
  \BibitemOpen
  \bibfield  {author} {\bibinfo {author} {\bibfnamefont {C.}~\bibnamefont
  {Reichhardt}}, \bibinfo {author} {\bibfnamefont {D.}~\bibnamefont {Ray}}, \
  and\ \bibinfo {author} {\bibfnamefont {C.~J.~Olson}\ \bibnamefont
  {Reichhardt}},\ }\bibfield  {title} {\enquote {\bibinfo {title}
  {Magnus-induced ratchet effects for skyrmions interacting with asymmetric
  substrates},}\ }\href {\doibase 10.1088/1367-2630/17/7/073034} {\bibfield
  {journal} {\bibinfo  {journal} {New J. Phys.}\ }\textbf {\bibinfo {volume}
  {17}},\ \bibinfo {pages} {073034} (\bibinfo {year}
  {2015}{\natexlab{c}})}\BibitemShut {NoStop}%
\bibitem [{\citenamefont {Ma}\ \emph {et~al.}(2017)\citenamefont {Ma},
  \citenamefont {Reichhardt},\ and\ \citenamefont {Reichhardt}}]{Ma17}%
  \BibitemOpen
  \bibfield  {author} {\bibinfo {author} {\bibfnamefont {X.}~\bibnamefont
  {Ma}}, \bibinfo {author} {\bibfnamefont {C.~J.~Olson}\ \bibnamefont
  {Reichhardt}}, \ and\ \bibinfo {author} {\bibfnamefont {C.}~\bibnamefont
  {Reichhardt}},\ }\bibfield  {title} {\enquote {\bibinfo {title} {Reversible
  vector ratchets for skyrmion systems},}\ }\href {\doibase
  10.1103/PhysRevB.95.104401} {\bibfield  {journal} {\bibinfo  {journal} {Phys.
  Rev. B}\ }\textbf {\bibinfo {volume} {95}},\ \bibinfo {pages} {104401}
  (\bibinfo {year} {2017})}\BibitemShut {NoStop}%
\bibitem [{\citenamefont {Chen}\ \emph {et~al.}(2020)\citenamefont {Chen},
  \citenamefont {Liu},\ and\ \citenamefont {Zheng}}]{Chen20}%
  \BibitemOpen
  \bibfield  {author} {\bibinfo {author} {\bibfnamefont {W.}~\bibnamefont
  {Chen}}, \bibinfo {author} {\bibfnamefont {L.}~\bibnamefont {Liu}}, \ and\
  \bibinfo {author} {\bibfnamefont {Y.}~\bibnamefont {Zheng}},\ }\bibfield
  {title} {\enquote {\bibinfo {title} {Ultrafast ratchet dynamics of skyrmions
  by defect engineering in materials with poor conductivity under gigahertz
  magnetic fields},}\ }\href {\doibase 10.1103/PhysRevApplied.14.064014}
  {\bibfield  {journal} {\bibinfo  {journal} {Phys. Rev. Applied}\ }\textbf
  {\bibinfo {volume} {14}},\ \bibinfo {pages} {064014} (\bibinfo {year}
  {2020})}\BibitemShut {NoStop}%
\bibitem [{\citenamefont {G{\" o}bel}\ and\ \citenamefont
  {Mertig}(2021)}]{Gobel21}%
  \BibitemOpen
  \bibfield  {author} {\bibinfo {author} {\bibfnamefont {B.}~\bibnamefont {G{\"
  o}bel}}\ and\ \bibinfo {author} {\bibfnamefont {I.}~\bibnamefont {Mertig}},\
  }\bibfield  {title} {\enquote {\bibinfo {title} {Skyrmion ratchet
  propagation: utilizing the skyrmion {H}all effect in {AC} racetrack storage
  devices},}\ }\href {\doibase 10.1038/s41598-021-81992-0} {\bibfield
  {journal} {\bibinfo  {journal} {Sci. Rep.}\ }\textbf {\bibinfo {volume}
  {11}},\ \bibinfo {pages} {3020} (\bibinfo {year} {2021})}\BibitemShut
  {NoStop}%
\bibitem [{\citenamefont {B{\" u}ttner}\ \emph {et~al.}(2015)\citenamefont
  {B{\" u}ttner}, \citenamefont {Moutafis}, \citenamefont {Schneider},
  \citenamefont {Kr{\" u}ger}, \citenamefont {G{\" u}nther}, \citenamefont
  {Geilhufe}, \citenamefont {Schmising}, \citenamefont {Mohanty}, \citenamefont
  {Pfau}, \citenamefont {Schaffert}, \citenamefont {Bisig}, \citenamefont
  {Foerster}, \citenamefont {Schulz}, \citenamefont {Vaz}, \citenamefont
  {Franken}, \citenamefont {Swagten}, \citenamefont {Kl{\" a}ui},\ and\
  \citenamefont {Eisebitt}}]{Buttner15}%
  \BibitemOpen
  \bibfield  {author} {\bibinfo {author} {\bibfnamefont {F.}~\bibnamefont {B{\"
  u}ttner}}, \bibinfo {author} {\bibfnamefont {C.}~\bibnamefont {Moutafis}},
  \bibinfo {author} {\bibfnamefont {M.}~\bibnamefont {Schneider}}, \bibinfo
  {author} {\bibfnamefont {B.}~\bibnamefont {Kr{\" u}ger}}, \bibinfo {author}
  {\bibfnamefont {C.~M.}\ \bibnamefont {G{\" u}nther}}, \bibinfo {author}
  {\bibfnamefont {J.}~\bibnamefont {Geilhufe}}, \bibinfo {author}
  {\bibfnamefont {C.~von~Kor}\ \bibnamefont {Schmising}}, \bibinfo {author}
  {\bibfnamefont {J.}~\bibnamefont {Mohanty}}, \bibinfo {author} {\bibfnamefont
  {B.}~\bibnamefont {Pfau}}, \bibinfo {author} {\bibfnamefont {S.}~\bibnamefont
  {Schaffert}}, \bibinfo {author} {\bibfnamefont {A.}~\bibnamefont {Bisig}},
  \bibinfo {author} {\bibfnamefont {M.}~\bibnamefont {Foerster}}, \bibinfo
  {author} {\bibfnamefont {T.}~\bibnamefont {Schulz}}, \bibinfo {author}
  {\bibfnamefont {C.~A.~F.}\ \bibnamefont {Vaz}}, \bibinfo {author}
  {\bibfnamefont {J.~H.}\ \bibnamefont {Franken}}, \bibinfo {author}
  {\bibfnamefont {H.~J.~M.}\ \bibnamefont {Swagten}}, \bibinfo {author}
  {\bibfnamefont {M.}~\bibnamefont {Kl{\" a}ui}}, \ and\ \bibinfo {author}
  {\bibfnamefont {S.}~\bibnamefont {Eisebitt}},\ }\bibfield  {title} {\enquote
  {\bibinfo {title} {Dynamics and inertia of skyrmionic spin structures},}\
  }\href {\doibase 10.1038/NPHYS3234} {\bibfield  {journal} {\bibinfo
  {journal} {Nature Phys.}\ }\textbf {\bibinfo {volume} {11}},\ \bibinfo
  {pages} {225--228} (\bibinfo {year} {2015})}\BibitemShut {NoStop}%
\bibitem [{\citenamefont {Fernandes}\ \emph {et~al.}(2020)\citenamefont
  {Fernandes}, \citenamefont {Chico},\ and\ \citenamefont
  {Lounis}}]{Fernandes20}%
  \BibitemOpen
  \bibfield  {author} {\bibinfo {author} {\bibfnamefont {I.~L.}\ \bibnamefont
  {Fernandes}}, \bibinfo {author} {\bibfnamefont {J.}~\bibnamefont {Chico}}, \
  and\ \bibinfo {author} {\bibfnamefont {S.}~\bibnamefont {Lounis}},\
  }\bibfield  {title} {\enquote {\bibinfo {title} {Impurity-dependent
  gyrotropic motion, deflection and pinning of current-driven ultrasmall
  skyrmions in pdfe/ir(111) surface},}\ }\href {\doibase
  10.1088/1361-648X/ab9cf0} {\bibfield  {journal} {\bibinfo  {journal} {J.
  Phys.: Condens. Matter}\ }\textbf {\bibinfo {volume} {32}},\ \bibinfo {pages}
  {425802} (\bibinfo {year} {2020})}\BibitemShut {NoStop}%
\bibitem [{\citenamefont {Brown}\ \emph {et~al.}(2018)\citenamefont {Brown},
  \citenamefont {T\"auber},\ and\ \citenamefont {Pleimling}}]{Brown18}%
  \BibitemOpen
  \bibfield  {author} {\bibinfo {author} {\bibfnamefont {B.~L.}\ \bibnamefont
  {Brown}}, \bibinfo {author} {\bibfnamefont {U.~C.}\ \bibnamefont {T\"auber}},
  \ and\ \bibinfo {author} {\bibfnamefont {M.}~\bibnamefont {Pleimling}},\
  }\bibfield  {title} {\enquote {\bibinfo {title} {Effect of the {M}agnus force
  on skyrmion relaxation dynamics},}\ }\href {\doibase
  10.1103/PhysRevB.97.020405} {\bibfield  {journal} {\bibinfo  {journal} {Phys.
  Rev. B}\ }\textbf {\bibinfo {volume} {97}},\ \bibinfo {pages} {020405}
  (\bibinfo {year} {2018})}\BibitemShut {NoStop}%
\bibitem [{\citenamefont {Hanneken}\ \emph {et~al.}(2016)\citenamefont
  {Hanneken}, \citenamefont {Kubetzka}, \citenamefont {von Bergmann},\ and\
  \citenamefont {Wiesendanger}}]{Hanneken16}%
  \BibitemOpen
  \bibfield  {author} {\bibinfo {author} {\bibfnamefont {C.}~\bibnamefont
  {Hanneken}}, \bibinfo {author} {\bibfnamefont {A.}~\bibnamefont {Kubetzka}},
  \bibinfo {author} {\bibfnamefont {K.}~\bibnamefont {von Bergmann}}, \ and\
  \bibinfo {author} {\bibfnamefont {R.}~\bibnamefont {Wiesendanger}},\
  }\bibfield  {title} {\enquote {\bibinfo {title} {Pinning and movement of
  individual nanoscale magnetic skyrmions via defects},}\ }\href {\doibase
  10.1088/1367-2630/18/5/055009} {\bibfield  {journal} {\bibinfo  {journal}
  {New J. Phys.}\ }\textbf {\bibinfo {volume} {18}},\ \bibinfo {pages} {055009}
  (\bibinfo {year} {2016})}\BibitemShut {NoStop}%
\bibitem [{\citenamefont {Wang}\ \emph {et~al.}(2017)\citenamefont {Wang},
  \citenamefont {Xiao}, \citenamefont {Chen}, \citenamefont {Zhou},\ and\
  \citenamefont {Liu}}]{Wang17}%
  \BibitemOpen
  \bibfield  {author} {\bibinfo {author} {\bibfnamefont {C.}~\bibnamefont
  {Wang}}, \bibinfo {author} {\bibfnamefont {D.}~\bibnamefont {Xiao}}, \bibinfo
  {author} {\bibfnamefont {X.}~\bibnamefont {Chen}}, \bibinfo {author}
  {\bibfnamefont {Y.}~\bibnamefont {Zhou}}, \ and\ \bibinfo {author}
  {\bibfnamefont {Y.}~\bibnamefont {Liu}},\ }\bibfield  {title} {\enquote
  {\bibinfo {title} {Manipulating and trapping skyrmions by magnetic field
  gradients},}\ }\href {\doibase 10.1088/1367-2630/aa7812} {\bibfield
  {journal} {\bibinfo  {journal} {New J. Phys.}\ }\textbf {\bibinfo {volume}
  {19}},\ \bibinfo {pages} {083008} (\bibinfo {year} {2017})}\BibitemShut
  {NoStop}%
\bibitem [{\citenamefont {Casiraghi}\ \emph {et~al.}(2019)\citenamefont
  {Casiraghi}, \citenamefont {Corte-Le{\' o}n}, \citenamefont {Vafaee},
  \citenamefont {Garcia-Sanchez}, \citenamefont {Durin}, \citenamefont
  {Pasquale}, \citenamefont {Jakob}, \citenamefont {Kl{\" a}ui},\ and\
  \citenamefont {Kazakova}}]{Casiraghi19}%
  \BibitemOpen
  \bibfield  {author} {\bibinfo {author} {\bibfnamefont {A.}~\bibnamefont
  {Casiraghi}}, \bibinfo {author} {\bibfnamefont {H.}~\bibnamefont {Corte-Le{\'
  o}n}}, \bibinfo {author} {\bibfnamefont {M.}~\bibnamefont {Vafaee}}, \bibinfo
  {author} {\bibfnamefont {F.}~\bibnamefont {Garcia-Sanchez}}, \bibinfo
  {author} {\bibfnamefont {G.}~\bibnamefont {Durin}}, \bibinfo {author}
  {\bibfnamefont {M.}~\bibnamefont {Pasquale}}, \bibinfo {author}
  {\bibfnamefont {G.}~\bibnamefont {Jakob}}, \bibinfo {author} {\bibfnamefont
  {M.}~\bibnamefont {Kl{\" a}ui}}, \ and\ \bibinfo {author} {\bibfnamefont
  {O.}~\bibnamefont {Kazakova}},\ }\bibfield  {title} {\enquote {\bibinfo
  {title} {Individual skyrmion manipulation by local magnetic field
  gradients},}\ }\href {\doibase 10.1038/s42005-019-0242-5} {\bibfield
  {journal} {\bibinfo  {journal} {Commun. Phys.}\ }\textbf {\bibinfo {volume}
  {2}},\ \bibinfo {pages} {145} (\bibinfo {year} {2019})}\BibitemShut {NoStop}%
\bibitem [{\citenamefont {Wang}\ \emph {et~al.}(2020)\citenamefont {Wang},
  \citenamefont {Chotorlishvili}, \citenamefont {Dugaev}, \citenamefont
  {Ernst}, \citenamefont {Maznichenko}, \citenamefont {Arnold}, \citenamefont
  {Jia}, \citenamefont {Berakdar}, \citenamefont {Mertig},\ and\ \citenamefont
  {Barna{\` s}}}]{Wang20b}%
  \BibitemOpen
  \bibfield  {author} {\bibinfo {author} {\bibfnamefont {X.-G.}\ \bibnamefont
  {Wang}}, \bibinfo {author} {\bibfnamefont {L.}~\bibnamefont
  {Chotorlishvili}}, \bibinfo {author} {\bibfnamefont {V.~K.}\ \bibnamefont
  {Dugaev}}, \bibinfo {author} {\bibfnamefont {A.}~\bibnamefont {Ernst}},
  \bibinfo {author} {\bibfnamefont {I.~V.}\ \bibnamefont {Maznichenko}},
  \bibinfo {author} {\bibfnamefont {N.}~\bibnamefont {Arnold}}, \bibinfo
  {author} {\bibfnamefont {C.}~\bibnamefont {Jia}}, \bibinfo {author}
  {\bibfnamefont {J.}~\bibnamefont {Berakdar}}, \bibinfo {author}
  {\bibfnamefont {I.}~\bibnamefont {Mertig}}, \ and\ \bibinfo {author}
  {\bibfnamefont {J.}~\bibnamefont {Barna{\` s}}},\ }\bibfield  {title}
  {\enquote {\bibinfo {title} {The optical tweezer of skyrmions},}\ }\href
  {\doibase 10.1038/s41524-020-00402-7} {\bibfield  {journal} {\bibinfo
  {journal} {npj Comput. Mater.}\ }\textbf {\bibinfo {volume} {6}},\ \bibinfo
  {pages} {140} (\bibinfo {year} {2020})}\BibitemShut {NoStop}%
\bibitem [{\citenamefont {Hastings}\ \emph {et~al.}(2003)\citenamefont
  {Hastings}, \citenamefont {Olson~Reichhardt},\ and\ \citenamefont
  {Reichhardt}}]{Hastings03}%
  \BibitemOpen
  \bibfield  {author} {\bibinfo {author} {\bibfnamefont {M.~B.}\ \bibnamefont
  {Hastings}}, \bibinfo {author} {\bibfnamefont {C.~J.}\ \bibnamefont
  {Olson~Reichhardt}}, \ and\ \bibinfo {author} {\bibfnamefont
  {C.}~\bibnamefont {Reichhardt}},\ }\bibfield  {title} {\enquote {\bibinfo
  {title} {Depinning by fracture in a glassy background},}\ }\href {\doibase
  10.1103/PhysRevLett.90.098302} {\bibfield  {journal} {\bibinfo  {journal}
  {Phys. Rev. Lett.}\ }\textbf {\bibinfo {volume} {90}},\ \bibinfo {pages}
  {098302} (\bibinfo {year} {2003})}\BibitemShut {NoStop}%
\bibitem [{\citenamefont {Habdas}\ \emph {et~al.}(2004)\citenamefont {Habdas},
  \citenamefont {Schaar}, \citenamefont {Levitt},\ and\ \citenamefont
  {Weeks}}]{Habdas04}%
  \BibitemOpen
  \bibfield  {author} {\bibinfo {author} {\bibfnamefont {P.}~\bibnamefont
  {Habdas}}, \bibinfo {author} {\bibfnamefont {D.}~\bibnamefont {Schaar}},
  \bibinfo {author} {\bibfnamefont {A.~C.}\ \bibnamefont {Levitt}}, \ and\
  \bibinfo {author} {\bibfnamefont {E.~R.}\ \bibnamefont {Weeks}},\ }\bibfield
  {title} {\enquote {\bibinfo {title} {Forced motion of a probe particle near
  the colloidal glass transition},}\ }\href {\doibase
  10.1209/epl/i2004-10075-y} {\bibfield  {journal} {\bibinfo  {journal}
  {Europhys. Lett.}\ }\textbf {\bibinfo {volume} {67}},\ \bibinfo {pages}
  {477--483} (\bibinfo {year} {2004})}\BibitemShut {NoStop}%
\bibitem [{\citenamefont {Squires}\ and\ \citenamefont
  {Brady}(2005)}]{Squires05}%
  \BibitemOpen
  \bibfield  {author} {\bibinfo {author} {\bibfnamefont {T.~M.}\ \bibnamefont
  {Squires}}\ and\ \bibinfo {author} {\bibfnamefont {J.~F.}\ \bibnamefont
  {Brady}},\ }\bibfield  {title} {\enquote {\bibinfo {title} {A simple paradigm
  for active and nonlinear microrheology},}\ }\href {\doibase
  10.1063/1.1960607} {\bibfield  {journal} {\bibinfo  {journal} {Phys. Fluids}\
  }\textbf {\bibinfo {volume} {17}},\ \bibinfo {pages} {073101} (\bibinfo
  {year} {2005})}\BibitemShut {NoStop}%
\bibitem [{\citenamefont {Gazuz}\ \emph {et~al.}(2009)\citenamefont {Gazuz},
  \citenamefont {Puertas}, \citenamefont {Voigtmann},\ and\ \citenamefont
  {Fuchs}}]{Gazuz09}%
  \BibitemOpen
  \bibfield  {author} {\bibinfo {author} {\bibfnamefont {I.}~\bibnamefont
  {Gazuz}}, \bibinfo {author} {\bibfnamefont {A.~M.}\ \bibnamefont {Puertas}},
  \bibinfo {author} {\bibfnamefont {Th.}\ \bibnamefont {Voigtmann}}, \ and\
  \bibinfo {author} {\bibfnamefont {M.}~\bibnamefont {Fuchs}},\ }\bibfield
  {title} {\enquote {\bibinfo {title} {Active and nonlinear microrheology in
  dense colloidal suspensions},}\ }\href {\doibase
  10.1103/PhysRevLett.102.248302} {\bibfield  {journal} {\bibinfo  {journal}
  {Phys. Rev. Lett.}\ }\textbf {\bibinfo {volume} {102}},\ \bibinfo {pages}
  {248302} (\bibinfo {year} {2009})}\BibitemShut {NoStop}%
\bibitem [{\citenamefont {Winter}\ \emph {et~al.}(2012)\citenamefont {Winter},
  \citenamefont {Horbach}, \citenamefont {Virnau},\ and\ \citenamefont
  {Binder}}]{Winter12}%
  \BibitemOpen
  \bibfield  {author} {\bibinfo {author} {\bibfnamefont {D.}~\bibnamefont
  {Winter}}, \bibinfo {author} {\bibfnamefont {J.}~\bibnamefont {Horbach}},
  \bibinfo {author} {\bibfnamefont {P.}~\bibnamefont {Virnau}}, \ and\ \bibinfo
  {author} {\bibfnamefont {K.}~\bibnamefont {Binder}},\ }\bibfield  {title}
  {\enquote {\bibinfo {title} {Active nonlinear microrheology in a
  glass-forming {Y}ukawa fluid},}\ }\href {\doibase
  10.1103/PhysRevLett.108.028303} {\bibfield  {journal} {\bibinfo  {journal}
  {Phys. Rev. Lett.}\ }\textbf {\bibinfo {volume} {108}},\ \bibinfo {pages}
  {028303} (\bibinfo {year} {2012})}\BibitemShut {NoStop}%
\bibitem [{\citenamefont {Drocco}\ \emph {et~al.}(2005)\citenamefont {Drocco},
  \citenamefont {Hastings}, \citenamefont {Reichhardt},\ and\ \citenamefont
  {Reichhardt}}]{Drocco05}%
  \BibitemOpen
  \bibfield  {author} {\bibinfo {author} {\bibfnamefont {J.~A.}\ \bibnamefont
  {Drocco}}, \bibinfo {author} {\bibfnamefont {M.~B.}\ \bibnamefont
  {Hastings}}, \bibinfo {author} {\bibfnamefont {C.~J.~Olson}\ \bibnamefont
  {Reichhardt}}, \ and\ \bibinfo {author} {\bibfnamefont {C.}~\bibnamefont
  {Reichhardt}},\ }\bibfield  {title} {\enquote {\bibinfo {title} {Multiscaling
  at point {$J$}: Jamming is a critical phenomenon},}\ }\href {\doibase
  10.1103/PhysRevLett.95.088001} {\bibfield  {journal} {\bibinfo  {journal}
  {Phys. Rev. Lett.}\ }\textbf {\bibinfo {volume} {95}},\ \bibinfo {pages}
  {088001} (\bibinfo {year} {2005})}\BibitemShut {NoStop}%
\bibitem [{\citenamefont {Candelier}\ and\ \citenamefont
  {Dauchot}(2010)}]{Candelier10}%
  \BibitemOpen
  \bibfield  {author} {\bibinfo {author} {\bibfnamefont {R.}~\bibnamefont
  {Candelier}}\ and\ \bibinfo {author} {\bibfnamefont {O.}~\bibnamefont
  {Dauchot}},\ }\bibfield  {title} {\enquote {\bibinfo {title} {Journey of an
  intruder through the fluidization and jamming transitions of a dense granular
  media},}\ }\href {\doibase 10.1103/PhysRevE.81.011304} {\bibfield  {journal}
  {\bibinfo  {journal} {Phys. Rev. E}\ }\textbf {\bibinfo {volume} {81}},\
  \bibinfo {pages} {011304} (\bibinfo {year} {2010})}\BibitemShut {NoStop}%
\bibitem [{\citenamefont {Olson~Reichhardt}\ and\ \citenamefont
  {Reichhardt}(2010)}]{Reichhardt10}%
  \BibitemOpen
  \bibfield  {author} {\bibinfo {author} {\bibfnamefont {C.~J.}\ \bibnamefont
  {Olson~Reichhardt}}\ and\ \bibinfo {author} {\bibfnamefont {C.}~\bibnamefont
  {Reichhardt}},\ }\bibfield  {title} {\enquote {\bibinfo {title}
  {Fluctuations, jamming, and yielding for a driven probe particle in
  disordered disk assemblies},}\ }\href {\doibase 10.1103/PhysRevE.82.051306}
  {\bibfield  {journal} {\bibinfo  {journal} {Phys. Rev. E}\ }\textbf {\bibinfo
  {volume} {82}},\ \bibinfo {pages} {051306} (\bibinfo {year}
  {2010})}\BibitemShut {NoStop}%
\bibitem [{\citenamefont {Kolb}\ \emph {et~al.}(2013)\citenamefont {Kolb},
  \citenamefont {Cixous}, \citenamefont {Gaudouen},\ and\ \citenamefont
  {Darnige}}]{Kolb13}%
  \BibitemOpen
  \bibfield  {author} {\bibinfo {author} {\bibfnamefont {E.}~\bibnamefont
  {Kolb}}, \bibinfo {author} {\bibfnamefont {P.}~\bibnamefont {Cixous}},
  \bibinfo {author} {\bibfnamefont {N.}~\bibnamefont {Gaudouen}}, \ and\
  \bibinfo {author} {\bibfnamefont {T.}~\bibnamefont {Darnige}},\ }\bibfield
  {title} {\enquote {\bibinfo {title} {Rigid intruder inside a two-dimensional
  dense granular flow: Drag force and cavity formation},}\ }\href {\doibase
  10.1103/PhysRevE.87.032207} {\bibfield  {journal} {\bibinfo  {journal} {Phys.
  Rev. E}\ }\textbf {\bibinfo {volume} {87}},\ \bibinfo {pages} {032207}
  (\bibinfo {year} {2013})}\BibitemShut {NoStop}%
\bibitem [{\citenamefont {Reichhardt}\ and\ \citenamefont
  {Reichhardt}(2019{\natexlab{a}})}]{Reichhardt19a}%
  \BibitemOpen
  \bibfield  {author} {\bibinfo {author} {\bibfnamefont {C.}~\bibnamefont
  {Reichhardt}}\ and\ \bibinfo {author} {\bibfnamefont {C.~J.~O.}\ \bibnamefont
  {Reichhardt}},\ }\bibfield  {title} {\enquote {\bibinfo {title} {Active
  microrheology, {H}all effect, and jamming in chiral fluids},}\ }\href
  {\doibase 10.1103/PhysRevE.100.012604} {\bibfield  {journal} {\bibinfo
  {journal} {Phys. Rev. E}\ }\textbf {\bibinfo {volume} {100}},\ \bibinfo
  {pages} {012604} (\bibinfo {year} {2019}{\natexlab{a}})}\BibitemShut
  {NoStop}%
\bibitem [{\citenamefont {Reichhardt}\ and\ \citenamefont
  {Reichhardt}(2015)}]{Reichhardt15aaa}%
  \BibitemOpen
  \bibfield  {author} {\bibinfo {author} {\bibfnamefont {C.}~\bibnamefont
  {Reichhardt}}\ and\ \bibinfo {author} {\bibfnamefont {C.~J.~Olson}\
  \bibnamefont {Reichhardt}},\ }\bibfield  {title} {\enquote {\bibinfo {title}
  {Active microrheology in active matter systems: Mobility, intermittency, and
  avalanches},}\ }\href {\doibase 10.1103/PhysRevE.91.032313} {\bibfield
  {journal} {\bibinfo  {journal} {Phys. Rev. E}\ }\textbf {\bibinfo {volume}
  {91}},\ \bibinfo {pages} {032313} (\bibinfo {year} {2015})}\BibitemShut
  {NoStop}%
\bibitem [{\citenamefont {Straver}\ \emph {et~al.}(2008)\citenamefont
  {Straver}, \citenamefont {Hoffman}, \citenamefont {Auslaender}, \citenamefont
  {Rugar},\ and\ \citenamefont {Moler}}]{Straver08}%
  \BibitemOpen
  \bibfield  {author} {\bibinfo {author} {\bibfnamefont {E.~W.~J.}\
  \bibnamefont {Straver}}, \bibinfo {author} {\bibfnamefont {J.~E.}\
  \bibnamefont {Hoffman}}, \bibinfo {author} {\bibfnamefont {O.~M.}\
  \bibnamefont {Auslaender}}, \bibinfo {author} {\bibfnamefont
  {D.}~\bibnamefont {Rugar}}, \ and\ \bibinfo {author} {\bibfnamefont {K.~A.}\
  \bibnamefont {Moler}},\ }\bibfield  {title} {\enquote {\bibinfo {title}
  {Controlled manipulation of individual vortices in a superconductor},}\
  }\href {\doibase 10.1063/1.3000963} {\bibfield  {journal} {\bibinfo
  {journal} {Appl. Phys. Lett.}\ }\textbf {\bibinfo {volume} {93}},\ \bibinfo
  {pages} {172514} (\bibinfo {year} {2008})}\BibitemShut {NoStop}%
\bibitem [{\citenamefont {Reichhardt}(2009)}]{Reichhardt09a}%
  \BibitemOpen
  \bibfield  {author} {\bibinfo {author} {\bibfnamefont {C.}~\bibnamefont
  {Reichhardt}},\ }\bibfield  {title} {\enquote {\bibinfo {title} {Vortices
  wiggled and dragged},}\ }\href {\doibase 10.1038/nphys1169} {\bibfield
  {journal} {\bibinfo  {journal} {Nature Phys.}\ }\textbf {\bibinfo {volume}
  {5}},\ \bibinfo {pages} {15--16} (\bibinfo {year} {2009})}\BibitemShut
  {NoStop}%
\bibitem [{\citenamefont {Auslaender}\ \emph {et~al.}(2009)\citenamefont
  {Auslaender}, \citenamefont {Luan}, \citenamefont {Straver}, \citenamefont
  {Hoffman}, \citenamefont {Koshnick}, \citenamefont {Zeldov}, \citenamefont
  {Bonn}, \citenamefont {Liang}, \citenamefont {Hardy},\ and\ \citenamefont
  {Moler}}]{Auslaender09}%
  \BibitemOpen
  \bibfield  {author} {\bibinfo {author} {\bibfnamefont {O.~M.}\ \bibnamefont
  {Auslaender}}, \bibinfo {author} {\bibfnamefont {L.}~\bibnamefont {Luan}},
  \bibinfo {author} {\bibfnamefont {E.~W.~J.}\ \bibnamefont {Straver}},
  \bibinfo {author} {\bibfnamefont {J.~E.}\ \bibnamefont {Hoffman}}, \bibinfo
  {author} {\bibfnamefont {N.~C.}\ \bibnamefont {Koshnick}}, \bibinfo {author}
  {\bibfnamefont {E.}~\bibnamefont {Zeldov}}, \bibinfo {author} {\bibfnamefont
  {D.~A.}\ \bibnamefont {Bonn}}, \bibinfo {author} {\bibfnamefont
  {R.}~\bibnamefont {Liang}}, \bibinfo {author} {\bibfnamefont {W.~N.}\
  \bibnamefont {Hardy}}, \ and\ \bibinfo {author} {\bibfnamefont {K.~A.}\
  \bibnamefont {Moler}},\ }\bibfield  {title} {\enquote {\bibinfo {title}
  {Mechanics of individual isolated vortices in a cuprate superconductor},}\
  }\href {\doibase 10.1038/NPHYS1127} {\bibfield  {journal} {\bibinfo
  {journal} {Nature Phys.}\ }\textbf {\bibinfo {volume} {5}},\ \bibinfo {pages}
  {35--39} (\bibinfo {year} {2009})}\BibitemShut {NoStop}%
\bibitem [{\citenamefont {Veshchunov}\ \emph {et~al.}(2016)\citenamefont
  {Veshchunov}, \citenamefont {Magrini}, \citenamefont {Mironov}, \citenamefont
  {Godin}, \citenamefont {Trebbia}, \citenamefont {Buzdin}, \citenamefont
  {Tamarat},\ and\ \citenamefont {Lounis}}]{Veshchunov16}%
  \BibitemOpen
  \bibfield  {author} {\bibinfo {author} {\bibfnamefont {I.~S.}\ \bibnamefont
  {Veshchunov}}, \bibinfo {author} {\bibfnamefont {W.}~\bibnamefont {Magrini}},
  \bibinfo {author} {\bibfnamefont {S.~V.}\ \bibnamefont {Mironov}}, \bibinfo
  {author} {\bibfnamefont {A.~G.}\ \bibnamefont {Godin}}, \bibinfo {author}
  {\bibfnamefont {J.~B.}\ \bibnamefont {Trebbia}}, \bibinfo {author}
  {\bibfnamefont {A.~I.}\ \bibnamefont {Buzdin}}, \bibinfo {author}
  {\bibfnamefont {Ph.}\ \bibnamefont {Tamarat}}, \ and\ \bibinfo {author}
  {\bibfnamefont {B.}~\bibnamefont {Lounis}},\ }\bibfield  {title} {\enquote
  {\bibinfo {title} {Optical manipulation of single flux quanta},}\ }\href
  {\doibase 10.1038/ncomms12801} {\bibfield  {journal} {\bibinfo  {journal}
  {Nature Commun.}\ }\textbf {\bibinfo {volume} {7}},\ \bibinfo {pages} {12801}
  (\bibinfo {year} {2016})}\BibitemShut {NoStop}%
\bibitem [{\citenamefont {Kremen}\ \emph {et~al.}(2016)\citenamefont {Kremen},
  \citenamefont {Wissberg}, \citenamefont {Haham}, \citenamefont {Persky},
  \citenamefont {Frenkel},\ and\ \citenamefont {Kalisky}}]{Kremen16}%
  \BibitemOpen
  \bibfield  {author} {\bibinfo {author} {\bibfnamefont {A.}~\bibnamefont
  {Kremen}}, \bibinfo {author} {\bibfnamefont {S.}~\bibnamefont {Wissberg}},
  \bibinfo {author} {\bibfnamefont {N.}~\bibnamefont {Haham}}, \bibinfo
  {author} {\bibfnamefont {E.}~\bibnamefont {Persky}}, \bibinfo {author}
  {\bibfnamefont {Y.}~\bibnamefont {Frenkel}}, \ and\ \bibinfo {author}
  {\bibfnamefont {B.}~\bibnamefont {Kalisky}},\ }\bibfield  {title} {\enquote
  {\bibinfo {title} {Mechanical control of individual superconducting
  vortices},}\ }\href {\doibase 10.1021/acs.nanolett.5b04444} {\bibfield
  {journal} {\bibinfo  {journal} {Nano Lett.}\ }\textbf {\bibinfo {volume}
  {16}},\ \bibinfo {pages} {1626--1630} (\bibinfo {year} {2016})}\BibitemShut
  {NoStop}%
\bibitem [{\citenamefont {Crassous}\ \emph {et~al.}(2011)\citenamefont
  {Crassous}, \citenamefont {Bernard}, \citenamefont {Fusil}, \citenamefont
  {Bouzehouane}, \citenamefont {Le~Bourdais}, \citenamefont {Enouz-Vedrenne},
  \citenamefont {Briatico}, \citenamefont {Bibes}, \citenamefont
  {Barth\'el\'emy},\ and\ \citenamefont {Villegas}}]{Crassous11}%
  \BibitemOpen
  \bibfield  {author} {\bibinfo {author} {\bibfnamefont {A.}~\bibnamefont
  {Crassous}}, \bibinfo {author} {\bibfnamefont {R.}~\bibnamefont {Bernard}},
  \bibinfo {author} {\bibfnamefont {S.}~\bibnamefont {Fusil}}, \bibinfo
  {author} {\bibfnamefont {K.}~\bibnamefont {Bouzehouane}}, \bibinfo {author}
  {\bibfnamefont {D.}~\bibnamefont {Le~Bourdais}}, \bibinfo {author}
  {\bibfnamefont {S.}~\bibnamefont {Enouz-Vedrenne}}, \bibinfo {author}
  {\bibfnamefont {J.}~\bibnamefont {Briatico}}, \bibinfo {author}
  {\bibfnamefont {M.}~\bibnamefont {Bibes}}, \bibinfo {author} {\bibfnamefont
  {A.}~\bibnamefont {Barth\'el\'emy}}, \ and\ \bibinfo {author} {\bibfnamefont
  {J.~E.}\ \bibnamefont {Villegas}},\ }\bibfield  {title} {\enquote {\bibinfo
  {title} {Nanoscale electrostatic manipulation of magnetic flux quanta in
  ferroelectric/superconductor {BiFeO$_3$/YBa$_2$Cu$_3$O$_{7-\delta}$}
  heterostructures},}\ }\href {\doibase 10.1103/PhysRevLett.107.247002}
  {\bibfield  {journal} {\bibinfo  {journal} {Phys. Rev. Lett.}\ }\textbf
  {\bibinfo {volume} {107}},\ \bibinfo {pages} {247002} (\bibinfo {year}
  {2011})}\BibitemShut {NoStop}%
\bibitem [{\citenamefont {Olson~Reichhardt}\ and\ \citenamefont
  {Reichhardt}(2008)}]{Reichhardt08}%
  \BibitemOpen
  \bibfield  {author} {\bibinfo {author} {\bibfnamefont {C.~J.}\ \bibnamefont
  {Olson~Reichhardt}}\ and\ \bibinfo {author} {\bibfnamefont {C.}~\bibnamefont
  {Reichhardt}},\ }\bibfield  {title} {\enquote {\bibinfo {title} {Viscous
  decoupling transitions for individually dragged particles in systems with
  quenched disorder},}\ }\href {\doibase 10.1103/PhysRevE.78.011402} {\bibfield
   {journal} {\bibinfo  {journal} {Phys. Rev. E}\ }\textbf {\bibinfo {volume}
  {78}},\ \bibinfo {pages} {011402} (\bibinfo {year} {2008})}\BibitemShut
  {NoStop}%
\bibitem [{\citenamefont {Wulfert}\ \emph {et~al.}(2017)\citenamefont
  {Wulfert}, \citenamefont {Seifert},\ and\ \citenamefont {Speck}}]{Wulfert17}%
  \BibitemOpen
  \bibfield  {author} {\bibinfo {author} {\bibfnamefont {R.}~\bibnamefont
  {Wulfert}}, \bibinfo {author} {\bibfnamefont {U.}~\bibnamefont {Seifert}}, \
  and\ \bibinfo {author} {\bibfnamefont {T.}~\bibnamefont {Speck}},\ }\bibfield
   {title} {\enquote {\bibinfo {title} {Nonequilibrium depletion interactions
  in active microrheology},}\ }\href {\doibase 10.1039/C7SM01737E} {\bibfield
  {journal} {\bibinfo  {journal} {Soft Matter}\ }\textbf {\bibinfo {volume}
  {13}},\ \bibinfo {pages} {9093} (\bibinfo {year} {2017})}\BibitemShut
  {NoStop}%
\bibitem [{\citenamefont {Zia}(2018)}]{Zia18}%
  \BibitemOpen
  \bibfield  {author} {\bibinfo {author} {\bibfnamefont {R.~N.}\ \bibnamefont
  {Zia}},\ }\bibfield  {title} {\enquote {\bibinfo {title} {Active and passive
  microrheology: Theory and simulation},}\ }\href {\doibase
  10.1146/annurev-fluid-122316-044514} {\bibfield  {journal} {\bibinfo
  {journal} {Ann. Rev. Fluid Mech.}\ }\textbf {\bibinfo {volume} {50}},\
  \bibinfo {pages} {371} (\bibinfo {year} {2018})}\BibitemShut {NoStop}%
\bibitem [{\citenamefont {Wang}\ \emph {et~al.}(2019)\citenamefont {Wang},
  \citenamefont {Mohori{\v c}}, \citenamefont {Zhang}, \citenamefont
  {Dobnikar},\ and\ \citenamefont {Horbach}}]{Wang19}%
  \BibitemOpen
  \bibfield  {author} {\bibinfo {author} {\bibfnamefont {H.}~\bibnamefont
  {Wang}}, \bibinfo {author} {\bibfnamefont {T.}~\bibnamefont {Mohori{\v c}}},
  \bibinfo {author} {\bibfnamefont {X.}~\bibnamefont {Zhang}}, \bibinfo
  {author} {\bibfnamefont {J.}~\bibnamefont {Dobnikar}}, \ and\ \bibinfo
  {author} {\bibfnamefont {J.}~\bibnamefont {Horbach}},\ }\bibfield  {title}
  {\enquote {\bibinfo {title} {Active microrheology in two-dimensional magnetic
  networks},}\ }\href {\doibase 10.1039/C9SM00085B} {\bibfield  {journal}
  {\bibinfo  {journal} {Soft Matter}\ }\textbf {\bibinfo {volume} {15}},\
  \bibinfo {pages} {4437} (\bibinfo {year} {2019})}\BibitemShut {NoStop}%
\bibitem [{\citenamefont {Yu}\ \emph {et~al.}(2020)\citenamefont {Yu},
  \citenamefont {Rahbari}, \citenamefont {Kawasaki}, \citenamefont {Park},\
  and\ \citenamefont {Lee}}]{Yu20}%
  \BibitemOpen
  \bibfield  {author} {\bibinfo {author} {\bibfnamefont {J.~W.}\ \bibnamefont
  {Yu}}, \bibinfo {author} {\bibfnamefont {S.~H.~E.}\ \bibnamefont {Rahbari}},
  \bibinfo {author} {\bibfnamefont {T.}~\bibnamefont {Kawasaki}}, \bibinfo
  {author} {\bibfnamefont {H.}~\bibnamefont {Park}}, \ and\ \bibinfo {author}
  {\bibfnamefont {W.~B.}\ \bibnamefont {Lee}},\ }\bibfield  {title} {\enquote
  {\bibinfo {title} {Active microrheology of a bulk metallic glass},}\ }\href
  {\doibase 10.1126/sciadv.aba8766} {\bibfield  {journal} {\bibinfo  {journal}
  {Sci. Adv.}\ }\textbf {\bibinfo {volume} {6}} (\bibinfo {year} {2020}),\
  10.1126/sciadv.aba8766}\BibitemShut {NoStop}%
\bibitem [{\citenamefont {Reichhardt}\ and\ \citenamefont
  {Reichhardt}(2004)}]{Reichhardt04a}%
  \BibitemOpen
  \bibfield  {author} {\bibinfo {author} {\bibfnamefont {C.}~\bibnamefont
  {Reichhardt}}\ and\ \bibinfo {author} {\bibfnamefont {C.~J.~Olson}\
  \bibnamefont {Reichhardt}},\ }\bibfield  {title} {\enquote {\bibinfo {title}
  {Local melting and drag for a particle driven through a colloidal crystal},}\
  }\href {\doibase 10.1103/PhysRevLett.92.108301} {\bibfield  {journal}
  {\bibinfo  {journal} {Phys. Rev. Lett.}\ }\textbf {\bibinfo {volume} {92}},\
  \bibinfo {pages} {108301} (\bibinfo {year} {2004})}\BibitemShut {NoStop}%
\bibitem [{\citenamefont {Dullens}\ and\ \citenamefont
  {Bechinger}(2011)}]{Dullens11}%
  \BibitemOpen
  \bibfield  {author} {\bibinfo {author} {\bibfnamefont {R.~P.~A.}\
  \bibnamefont {Dullens}}\ and\ \bibinfo {author} {\bibfnamefont
  {C.}~\bibnamefont {Bechinger}},\ }\bibfield  {title} {\enquote {\bibinfo
  {title} {Shear thinning and local melting of colloidal crystals},}\ }\href
  {\doibase 10.1103/PhysRevLett.107.138301} {\bibfield  {journal} {\bibinfo
  {journal} {Phys. Rev. Lett.}\ }\textbf {\bibinfo {volume} {107}},\ \bibinfo
  {pages} {138301} (\bibinfo {year} {2011})}\BibitemShut {NoStop}%
\bibitem [{\citenamefont {\ifmmode~\mbox{\c{S}}\else \c{S}\fi{}enbil}\ \emph
  {et~al.}(2019)\citenamefont {\ifmmode~\mbox{\c{S}}\else \c{S}\fi{}enbil},
  \citenamefont {Gruber}, \citenamefont {Zhang}, \citenamefont {Fuchs},\ and\
  \citenamefont {Scheffold}}]{Senbil19}%
  \BibitemOpen
  \bibfield  {author} {\bibinfo {author} {\bibfnamefont {N.}~\bibnamefont
  {\ifmmode~\mbox{\c{S}}\else \c{S}\fi{}enbil}}, \bibinfo {author}
  {\bibfnamefont {M.}~\bibnamefont {Gruber}}, \bibinfo {author} {\bibfnamefont
  {C.}~\bibnamefont {Zhang}}, \bibinfo {author} {\bibfnamefont
  {M.}~\bibnamefont {Fuchs}}, \ and\ \bibinfo {author} {\bibfnamefont
  {F.}~\bibnamefont {Scheffold}},\ }\bibfield  {title} {\enquote {\bibinfo
  {title} {Observation of strongly heterogeneous dynamics at the depinning
  transition in a colloidal glass},}\ }\href {\doibase
  10.1103/PhysRevLett.122.108002} {\bibfield  {journal} {\bibinfo  {journal}
  {Phys. Rev. Lett.}\ }\textbf {\bibinfo {volume} {122}},\ \bibinfo {pages}
  {108002} (\bibinfo {year} {2019})}\BibitemShut {NoStop}%
\bibitem [{\citenamefont {Gruber}\ \emph {et~al.}(2020)\citenamefont {Gruber},
  \citenamefont {Puertas},\ and\ \citenamefont {Fuchs}}]{Gruber20}%
  \BibitemOpen
  \bibfield  {author} {\bibinfo {author} {\bibfnamefont {M.}~\bibnamefont
  {Gruber}}, \bibinfo {author} {\bibfnamefont {A.~M.}\ \bibnamefont {Puertas}},
  \ and\ \bibinfo {author} {\bibfnamefont {M.}~\bibnamefont {Fuchs}},\
  }\bibfield  {title} {\enquote {\bibinfo {title} {Critical force in active
  microrheology},}\ }\href {\doibase 10.1103/PhysRevE.101.012612} {\bibfield
  {journal} {\bibinfo  {journal} {Phys. Rev. E}\ }\textbf {\bibinfo {volume}
  {101}},\ \bibinfo {pages} {012612} (\bibinfo {year} {2020})}\BibitemShut
  {NoStop}%
\bibitem [{\citenamefont {Benichou}\ \emph {et~al.}(2013)\citenamefont
  {Benichou}, \citenamefont {Illien}, \citenamefont {Mejia-Monasterio},\ and\
  \citenamefont {Oshanin}}]{Benichou13a}%
  \BibitemOpen
  \bibfield  {author} {\bibinfo {author} {\bibfnamefont {O.}~\bibnamefont
  {Benichou}}, \bibinfo {author} {\bibfnamefont {P.}~\bibnamefont {Illien}},
  \bibinfo {author} {\bibfnamefont {C.}~\bibnamefont {Mejia-Monasterio}}, \
  and\ \bibinfo {author} {\bibfnamefont {G.}~\bibnamefont {Oshanin}},\
  }\bibfield  {title} {\enquote {\bibinfo {title} {A biased intruder in a dense
  quiescent medium: looking beyond the force-velocity relation},}\ }\href
  {\doibase 10.1088/1742-5468/2013/05/P05008} {\bibfield  {journal} {\bibinfo
  {journal} {J. Stat. Mech.}\ }\textbf {\bibinfo {volume} {2013}},\ \bibinfo
  {pages} {P05008} (\bibinfo {year} {2013})}\BibitemShut {NoStop}%
\bibitem [{\citenamefont {Illien}\ \emph {et~al.}(2018)\citenamefont {Illien},
  \citenamefont {B\'enichou}, \citenamefont {Oshanin}, \citenamefont
  {Sarracino},\ and\ \citenamefont {Voituriez}}]{Illien18}%
  \BibitemOpen
  \bibfield  {author} {\bibinfo {author} {\bibfnamefont {P.}~\bibnamefont
  {Illien}}, \bibinfo {author} {\bibfnamefont {O.}~\bibnamefont {B\'enichou}},
  \bibinfo {author} {\bibfnamefont {G.}~\bibnamefont {Oshanin}}, \bibinfo
  {author} {\bibfnamefont {A.}~\bibnamefont {Sarracino}}, \ and\ \bibinfo
  {author} {\bibfnamefont {R.}~\bibnamefont {Voituriez}},\ }\bibfield  {title}
  {\enquote {\bibinfo {title} {Nonequilibrium fluctuations and enhanced
  diffusion of a driven particle in a dense environment},}\ }\href {\doibase
  10.1103/PhysRevLett.120.200606} {\bibfield  {journal} {\bibinfo  {journal}
  {Phys. Rev. Lett.}\ }\textbf {\bibinfo {volume} {120}},\ \bibinfo {pages}
  {200606} (\bibinfo {year} {2018})}\BibitemShut {NoStop}%
\bibitem [{\citenamefont {Kokot}\ \emph {et~al.}(2017)\citenamefont {Kokot},
  \citenamefont {Das}, \citenamefont {Winkler}, \citenamefont {Gompper},
  \citenamefont {Aranson},\ and\ \citenamefont {Snezhko}}]{Kokot17}%
  \BibitemOpen
  \bibfield  {author} {\bibinfo {author} {\bibfnamefont {G.}~\bibnamefont
  {Kokot}}, \bibinfo {author} {\bibfnamefont {S.}~\bibnamefont {Das}}, \bibinfo
  {author} {\bibfnamefont {R.~G.}\ \bibnamefont {Winkler}}, \bibinfo {author}
  {\bibfnamefont {G.}~\bibnamefont {Gompper}}, \bibinfo {author} {\bibfnamefont
  {I.~S.}\ \bibnamefont {Aranson}}, \ and\ \bibinfo {author} {\bibfnamefont
  {A.}~\bibnamefont {Snezhko}},\ }\bibfield  {title} {\enquote {\bibinfo
  {title} {Active turbulence in a gas of self-assembled spinners},}\ }\href
  {\doibase 10.1073/pnas.1710188114} {\bibfield  {journal} {\bibinfo  {journal}
  {Proc. Natl. Acad. Sci. (USA)}\ }\textbf {\bibinfo {volume} {114}},\ \bibinfo
  {pages} {12870--12875} (\bibinfo {year} {2017})}\BibitemShut {NoStop}%
\bibitem [{\citenamefont {Banerjee}\ \emph {et~al.}(2017)\citenamefont
  {Banerjee}, \citenamefont {Souslov}, \citenamefont {Abanov},\ and\
  \citenamefont {Vitelli}}]{Banerjee17}%
  \BibitemOpen
  \bibfield  {author} {\bibinfo {author} {\bibfnamefont {D.}~\bibnamefont
  {Banerjee}}, \bibinfo {author} {\bibfnamefont {A.}~\bibnamefont {Souslov}},
  \bibinfo {author} {\bibfnamefont {A.~G.}\ \bibnamefont {Abanov}}, \ and\
  \bibinfo {author} {\bibfnamefont {V.}~\bibnamefont {Vitelli}},\ }\bibfield
  {title} {\enquote {\bibinfo {title} {Odd viscosity in chiral active
  fluids},}\ }\href {\doibase 10.1038/s41467-017-01378-7} {\bibfield  {journal}
  {\bibinfo  {journal} {Nature Commun.}\ }\textbf {\bibinfo {volume} {8}},\
  \bibinfo {pages} {1573} (\bibinfo {year} {2017})}\BibitemShut {NoStop}%
\bibitem [{\citenamefont {Soni}\ \emph {et~al.}(2019)\citenamefont {Soni},
  \citenamefont {Bililign}, \citenamefont {Magkiriadou}, \citenamefont
  {Sacanna}, \citenamefont {Bartolo}, \citenamefont {Shelley},\ and\
  \citenamefont {Irvine}}]{Soni19}%
  \BibitemOpen
  \bibfield  {author} {\bibinfo {author} {\bibfnamefont {V.}~\bibnamefont
  {Soni}}, \bibinfo {author} {\bibfnamefont {E.~S.}\ \bibnamefont {Bililign}},
  \bibinfo {author} {\bibfnamefont {S.}~\bibnamefont {Magkiriadou}}, \bibinfo
  {author} {\bibfnamefont {S.}~\bibnamefont {Sacanna}}, \bibinfo {author}
  {\bibfnamefont {D.}~\bibnamefont {Bartolo}}, \bibinfo {author} {\bibfnamefont
  {M.~J.}\ \bibnamefont {Shelley}}, \ and\ \bibinfo {author} {\bibfnamefont
  {W.~T.~M.}\ \bibnamefont {Irvine}},\ }\bibfield  {title} {\enquote {\bibinfo
  {title} {The odd free surface flows of a colloidal chiral fluid},}\ }\href
  {\doibase 10.1038/s41567-019-0603-8} {\bibfield  {journal} {\bibinfo
  {journal} {Nature Phys.}\ }\textbf {\bibinfo {volume} {15}},\ \bibinfo
  {pages} {1188} (\bibinfo {year} {2019})}\BibitemShut {NoStop}%
\bibitem [{\citenamefont {Scholz}\ \emph {et~al.}(2021)\citenamefont {Scholz},
  \citenamefont {Ldov}, \citenamefont {P{\" o}schel}, \citenamefont {Engel},\
  and\ \citenamefont {L{\" o}wen}}]{Scholz21}%
  \BibitemOpen
  \bibfield  {author} {\bibinfo {author} {\bibfnamefont {C.}~\bibnamefont
  {Scholz}}, \bibinfo {author} {\bibfnamefont {A.}~\bibnamefont {Ldov}},
  \bibinfo {author} {\bibfnamefont {T.}~\bibnamefont {P{\" o}schel}}, \bibinfo
  {author} {\bibfnamefont {M.}~\bibnamefont {Engel}}, \ and\ \bibinfo {author}
  {\bibfnamefont {H.}~\bibnamefont {L{\" o}wen}},\ }\bibfield  {title}
  {\enquote {\bibinfo {title} {Surfactants and rotelles in active chiral
  fluids},}\ }\href {\doibase 10.1126/sciadv.abf8998} {\bibfield  {journal}
  {\bibinfo  {journal} {Sci. Adv.}\ }\textbf {\bibinfo {volume} {7}},\ \bibinfo
  {pages} {eabf8998} (\bibinfo {year} {2021})}\BibitemShut {NoStop}%
\bibitem [{\citenamefont {Reichhardt}\ and\ \citenamefont
  {Reichhardt}(2020)}]{Reichhardt20b}%
  \BibitemOpen
  \bibfield  {author} {\bibinfo {author} {\bibfnamefont {C.}~\bibnamefont
  {Reichhardt}}\ and\ \bibinfo {author} {\bibfnamefont {C.~J.~O.}\ \bibnamefont
  {Reichhardt}},\ }\bibfield  {title} {\enquote {\bibinfo {title} {Dynamics of
  {M}agnus-dominated particle clusters, collisions, pinning, and ratchets},}\
  }\href {\doibase 10.1103/PhysRevE.101.062602} {\bibfield  {journal} {\bibinfo
   {journal} {Phys. Rev. E}\ }\textbf {\bibinfo {volume} {101}},\ \bibinfo
  {pages} {062602} (\bibinfo {year} {2020})}\BibitemShut {NoStop}%
\bibitem [{\citenamefont {Schecter}\ \emph {et~al.}(1999)\citenamefont
  {Schecter}, \citenamefont {Dubin}, \citenamefont {Fine},\ and\ \citenamefont
  {Driscoll}}]{Schecter99}%
  \BibitemOpen
  \bibfield  {author} {\bibinfo {author} {\bibfnamefont {D.~A.}\ \bibnamefont
  {Schecter}}, \bibinfo {author} {\bibfnamefont {D.~H.~E.}\ \bibnamefont
  {Dubin}}, \bibinfo {author} {\bibfnamefont {K.~S.}\ \bibnamefont {Fine}}, \
  and\ \bibinfo {author} {\bibfnamefont {C.~F.}\ \bibnamefont {Driscoll}},\
  }\bibfield  {title} {\enquote {\bibinfo {title} {Vortex crystals from 2{D}
  {E}uler flow: Experiment and simulation},}\ }\href {\doibase
  10.1063/1.869961} {\bibfield  {journal} {\bibinfo  {journal} {Phys. Fluids}\
  }\textbf {\bibinfo {volume} {11}},\ \bibinfo {pages} {905} (\bibinfo {year}
  {1999})}\BibitemShut {NoStop}%
\bibitem [{\citenamefont {Schirmacher}\ \emph {et~al.}(2015)\citenamefont
  {Schirmacher}, \citenamefont {Fuchs}, \citenamefont {H\"ofling},\ and\
  \citenamefont {Franosch}}]{Schirmacher15}%
  \BibitemOpen
  \bibfield  {author} {\bibinfo {author} {\bibfnamefont {W.}~\bibnamefont
  {Schirmacher}}, \bibinfo {author} {\bibfnamefont {B.}~\bibnamefont {Fuchs}},
  \bibinfo {author} {\bibfnamefont {F.}~\bibnamefont {H\"ofling}}, \ and\
  \bibinfo {author} {\bibfnamefont {T.}~\bibnamefont {Franosch}},\ }\bibfield
  {title} {\enquote {\bibinfo {title} {Anomalous magnetotransport in disordered
  structures: Classical edge-state percolation},}\ }\href {\doibase
  10.1103/PhysRevLett.115.240602} {\bibfield  {journal} {\bibinfo  {journal}
  {Phys. Rev. Lett.}\ }\textbf {\bibinfo {volume} {115}},\ \bibinfo {pages}
  {240602} (\bibinfo {year} {2015})}\BibitemShut {NoStop}%
\bibitem [{\citenamefont {Tierno}\ \emph {et~al.}(2007)\citenamefont {Tierno},
  \citenamefont {Johansen},\ and\ \citenamefont {Fischer}}]{Tierno07}%
  \BibitemOpen
  \bibfield  {author} {\bibinfo {author} {\bibfnamefont {P.}~\bibnamefont
  {Tierno}}, \bibinfo {author} {\bibfnamefont {T.~H.}\ \bibnamefont
  {Johansen}}, \ and\ \bibinfo {author} {\bibfnamefont {T.~M.}\ \bibnamefont
  {Fischer}},\ }\bibfield  {title} {\enquote {\bibinfo {title} {Localized and
  delocalized motion of colloidal particles on a magnetic bubble lattice},}\
  }\href {\doibase 10.1103/PhysRevLett.99.038303} {\bibfield  {journal}
  {\bibinfo  {journal} {Phys. Rev. Lett.}\ }\textbf {\bibinfo {volume} {99}},\
  \bibinfo {pages} {038303} (\bibinfo {year} {2007})}\BibitemShut {NoStop}%
\bibitem [{\citenamefont {Grzybowski}\ \emph {et~al.}(2001)\citenamefont
  {Grzybowski}, \citenamefont {Jiang}, \citenamefont {Stone},\ and\
  \citenamefont {Whitesides}}]{Grzybowski01}%
  \BibitemOpen
  \bibfield  {author} {\bibinfo {author} {\bibfnamefont {B.~A.}\ \bibnamefont
  {Grzybowski}}, \bibinfo {author} {\bibfnamefont {X.}~\bibnamefont {Jiang}},
  \bibinfo {author} {\bibfnamefont {H.~A.}\ \bibnamefont {Stone}}, \ and\
  \bibinfo {author} {\bibfnamefont {G~M.}\ \bibnamefont {Whitesides}},\
  }\bibfield  {title} {\enquote {\bibinfo {title} {Dynamic, self-assembled
  aggregates of magnetized, millimeter-sized objects rotating at the liquid-air
  interface: Macroscopic, two-dimensional classical artificial atoms and
  molecules},}\ }\href {\doibase 10.1103/PhysRevE.64.011603} {\bibfield
  {journal} {\bibinfo  {journal} {Phys. Rev. E}\ }\textbf {\bibinfo {volume}
  {64}},\ \bibinfo {pages} {011603} (\bibinfo {year} {2001})}\BibitemShut
  {NoStop}%
\bibitem [{\citenamefont {Denisov}\ \emph {et~al.}(2018)\citenamefont
  {Denisov}, \citenamefont {Lyutyy}, \citenamefont {Reva},\ and\ \citenamefont
  {Yermolenko}}]{Denisov18}%
  \BibitemOpen
  \bibfield  {author} {\bibinfo {author} {\bibfnamefont {S.~I.}\ \bibnamefont
  {Denisov}}, \bibinfo {author} {\bibfnamefont {T.~V.}\ \bibnamefont {Lyutyy}},
  \bibinfo {author} {\bibfnamefont {V.~V.}\ \bibnamefont {Reva}}, \ and\
  \bibinfo {author} {\bibfnamefont {A.~S.}\ \bibnamefont {Yermolenko}},\
  }\bibfield  {title} {\enquote {\bibinfo {title} {Temperature effects on drift
  of suspended single-domain particles induced by the {M}agnus force},}\ }\href
  {\doibase 10.1103/PhysRevE.97.032608} {\bibfield  {journal} {\bibinfo
  {journal} {Phys. Rev. E}\ }\textbf {\bibinfo {volume} {97}},\ \bibinfo
  {pages} {032608} (\bibinfo {year} {2018})}\BibitemShut {NoStop}%
\bibitem [{\citenamefont {Ryzhov}\ and\ \citenamefont
  {Koshel}(2013)}]{Ryzhov13}%
  \BibitemOpen
  \bibfield  {author} {\bibinfo {author} {\bibfnamefont {E.~A.}\ \bibnamefont
  {Ryzhov}}\ and\ \bibinfo {author} {\bibfnamefont {K.~V.}\ \bibnamefont
  {Koshel}},\ }\bibfield  {title} {\enquote {\bibinfo {title} {Dynamics of a
  vortex pair interacting with a fixed point vortex},}\ }\href {\doibase
  10.1209/0295-5075/102/44004} {\bibfield  {journal} {\bibinfo  {journal}
  {EPL}\ }\textbf {\bibinfo {volume} {102}},\ \bibinfo {pages} {44004}
  (\bibinfo {year} {2013})}\BibitemShut {NoStop}%
\bibitem [{\citenamefont {Aref}(2007)}]{Aref07}%
  \BibitemOpen
  \bibfield  {author} {\bibinfo {author} {\bibfnamefont {H.}~\bibnamefont
  {Aref}},\ }\bibfield  {title} {\enquote {\bibinfo {title} {Point vortex
  dynamics: A classical mathematics playground},}\ }\href {\doibase
  10.1063/1.2425103} {\bibfield  {journal} {\bibinfo  {journal} {J. Math.
  Phys.}\ }\textbf {\bibinfo {volume} {48}},\ \bibinfo {pages} {065401}
  (\bibinfo {year} {2007})}\BibitemShut {NoStop}%
\bibitem [{\citenamefont {Doshi}\ and\ \citenamefont {Gromov}(2021)}]{Doshi21}%
  \BibitemOpen
  \bibfield  {author} {\bibinfo {author} {\bibfnamefont {D.}~\bibnamefont
  {Doshi}}\ and\ \bibinfo {author} {\bibfnamefont {A.}~\bibnamefont {Gromov}},\
  }\bibfield  {title} {\enquote {\bibinfo {title} {Vortices as fractons},}\
  }\href {\doibase 10.1038/s42005-021-00540-4} {\bibfield  {journal} {\bibinfo
  {journal} {Commun. Phys.}\ }\textbf {\bibinfo {volume} {4}},\ \bibinfo
  {pages} {44} (\bibinfo {year} {2021})}\BibitemShut {NoStop}%
\bibitem [{\citenamefont {Reichhardt}\ and\ \citenamefont
  {Reichhardt}(2016{\natexlab{b}})}]{Reichhardt16}%
  \BibitemOpen
  \bibfield  {author} {\bibinfo {author} {\bibfnamefont {C.}~\bibnamefont
  {Reichhardt}}\ and\ \bibinfo {author} {\bibfnamefont {C.~J.~Olson}\
  \bibnamefont {Reichhardt}},\ }\bibfield  {title} {\enquote {\bibinfo {title}
  {Noise fluctuations and drive dependence of the skyrmion {H}all effect in
  disordered systems},}\ }\href {\doibase 10.1088/1367-2630/18/9/095005}
  {\bibfield  {journal} {\bibinfo  {journal} {New J. Phys.}\ }\textbf {\bibinfo
  {volume} {18}},\ \bibinfo {pages} {095005} (\bibinfo {year}
  {2016}{\natexlab{b}})}\BibitemShut {NoStop}%
\bibitem [{\citenamefont {Legrand}\ \emph {et~al.}(2017)\citenamefont
  {Legrand}, \citenamefont {Maccariello}, \citenamefont {Reyren}, \citenamefont
  {Garcia}, \citenamefont {Moutafis}, \citenamefont {Moreau-Luchaire},
  \citenamefont {Coffin}, \citenamefont {Bouzehouane}, \citenamefont {Cros},\
  and\ \citenamefont {Fert}}]{Legrand17}%
  \BibitemOpen
  \bibfield  {author} {\bibinfo {author} {\bibfnamefont {W.}~\bibnamefont
  {Legrand}}, \bibinfo {author} {\bibfnamefont {D.}~\bibnamefont
  {Maccariello}}, \bibinfo {author} {\bibfnamefont {N.}~\bibnamefont {Reyren}},
  \bibinfo {author} {\bibfnamefont {K.}~\bibnamefont {Garcia}}, \bibinfo
  {author} {\bibfnamefont {C.}~\bibnamefont {Moutafis}}, \bibinfo {author}
  {\bibfnamefont {C.}~\bibnamefont {Moreau-Luchaire}}, \bibinfo {author}
  {\bibfnamefont {S.}~\bibnamefont {Coffin}}, \bibinfo {author} {\bibfnamefont
  {K.}~\bibnamefont {Bouzehouane}}, \bibinfo {author} {\bibfnamefont
  {V.}~\bibnamefont {Cros}}, \ and\ \bibinfo {author} {\bibfnamefont
  {A.}~\bibnamefont {Fert}},\ }\bibfield  {title} {\enquote {\bibinfo {title}
  {Room-temperature current-induced generation and motion of sub-100 nm
  skyrmions},}\ }\href {\doibase 10.1021/acs.nanolett.7b00649} {\bibfield
  {journal} {\bibinfo  {journal} {Nano Lett.}\ }\textbf {\bibinfo {volume}
  {17}},\ \bibinfo {pages} {2703--2712} (\bibinfo {year} {2017})}\BibitemShut
  {NoStop}%
\bibitem [{\citenamefont {D\'{\i}az}\ \emph {et~al.}(2017)\citenamefont
  {D\'{\i}az}, \citenamefont {Reichhardt}, \citenamefont {Arovas},
  \citenamefont {Saxena},\ and\ \citenamefont {Reichhardt}}]{Diaz17}%
  \BibitemOpen
  \bibfield  {author} {\bibinfo {author} {\bibfnamefont {S.~A.}\ \bibnamefont
  {D\'{\i}az}}, \bibinfo {author} {\bibfnamefont {C.~J.~O.}\ \bibnamefont
  {Reichhardt}}, \bibinfo {author} {\bibfnamefont {D.~P.}\ \bibnamefont
  {Arovas}}, \bibinfo {author} {\bibfnamefont {A.}~\bibnamefont {Saxena}}, \
  and\ \bibinfo {author} {\bibfnamefont {C.}~\bibnamefont {Reichhardt}},\
  }\bibfield  {title} {\enquote {\bibinfo {title} {Fluctuations and noise
  signatures of driven magnetic skyrmions},}\ }\href {\doibase
  10.1103/PhysRevB.96.085106} {\bibfield  {journal} {\bibinfo  {journal} {Phys.
  Rev. B}\ }\textbf {\bibinfo {volume} {96}},\ \bibinfo {pages} {085106}
  (\bibinfo {year} {2017})}\BibitemShut {NoStop}%
\bibitem [{\citenamefont {Juge}\ \emph {et~al.}(2019)\citenamefont {Juge},
  \citenamefont {Je}, \citenamefont {Chaves}, \citenamefont {Buda-Prejbeanu},
  \citenamefont {Pe\~na Garcia}, \citenamefont {Nath}, \citenamefont {Miron},
  \citenamefont {Rana}, \citenamefont {Aballe}, \citenamefont {Foerster},
  \citenamefont {Genuzio}, \citenamefont {Mente\ifmmode~\mbox{\c{s}}\else
  \c{s}\fi{}}, \citenamefont {Locatelli}, \citenamefont {Maccherozzi},
  \citenamefont {Dhesi}, \citenamefont {Belmeguenai}, \citenamefont
  {Roussign\'e}, \citenamefont {Auffret}, \citenamefont {Pizzini},
  \citenamefont {Gaudin}, \citenamefont {Vogel},\ and\ \citenamefont
  {Boulle}}]{Juge19}%
  \BibitemOpen
  \bibfield  {author} {\bibinfo {author} {\bibfnamefont {R.}~\bibnamefont
  {Juge}}, \bibinfo {author} {\bibfnamefont {S.-G.}\ \bibnamefont {Je}},
  \bibinfo {author} {\bibfnamefont {D.~de~Souza}\ \bibnamefont {Chaves}},
  \bibinfo {author} {\bibfnamefont {L.~D.}\ \bibnamefont {Buda-Prejbeanu}},
  \bibinfo {author} {\bibfnamefont {J.}~\bibnamefont {Pe\~na Garcia}}, \bibinfo
  {author} {\bibfnamefont {J.}~\bibnamefont {Nath}}, \bibinfo {author}
  {\bibfnamefont {I.~M.}\ \bibnamefont {Miron}}, \bibinfo {author}
  {\bibfnamefont {K.~G.}\ \bibnamefont {Rana}}, \bibinfo {author}
  {\bibfnamefont {L.}~\bibnamefont {Aballe}}, \bibinfo {author} {\bibfnamefont
  {M.}~\bibnamefont {Foerster}}, \bibinfo {author} {\bibfnamefont
  {F.}~\bibnamefont {Genuzio}}, \bibinfo {author} {\bibfnamefont {T.~O.}\
  \bibnamefont {Mente\ifmmode~\mbox{\c{s}}\else \c{s}\fi{}}}, \bibinfo {author}
  {\bibfnamefont {A.}~\bibnamefont {Locatelli}}, \bibinfo {author}
  {\bibfnamefont {F.}~\bibnamefont {Maccherozzi}}, \bibinfo {author}
  {\bibfnamefont {S.~S.}\ \bibnamefont {Dhesi}}, \bibinfo {author}
  {\bibfnamefont {M.}~\bibnamefont {Belmeguenai}}, \bibinfo {author}
  {\bibfnamefont {Y.}~\bibnamefont {Roussign\'e}}, \bibinfo {author}
  {\bibfnamefont {S.}~\bibnamefont {Auffret}}, \bibinfo {author} {\bibfnamefont
  {S.}~\bibnamefont {Pizzini}}, \bibinfo {author} {\bibfnamefont
  {G.}~\bibnamefont {Gaudin}}, \bibinfo {author} {\bibfnamefont
  {J.}~\bibnamefont {Vogel}}, \ and\ \bibinfo {author} {\bibfnamefont
  {O.}~\bibnamefont {Boulle}},\ }\bibfield  {title} {\enquote {\bibinfo {title}
  {Current-driven skyrmion dynamics and drive-dependent skyrmion {H}all effect
  in an ultrathin film},}\ }\href {\doibase 10.1103/PhysRevApplied.12.044007}
  {\bibfield  {journal} {\bibinfo  {journal} {Phys. Rev. Applied}\ }\textbf
  {\bibinfo {volume} {12}},\ \bibinfo {pages} {044007} (\bibinfo {year}
  {2019})}\BibitemShut {NoStop}%
\bibitem [{\citenamefont {Zeissler}\ \emph {et~al.}(2020)\citenamefont
  {Zeissler}, \citenamefont {Finizio}, \citenamefont {Barton}, \citenamefont
  {Huxtable}, \citenamefont {Massey}, \citenamefont {Raabe}, \citenamefont
  {Sadovnikov}, \citenamefont {Nikitov}, \citenamefont {Brearton},
  \citenamefont {Hesjedal}, \citenamefont {van~der Laan}, \citenamefont
  {Rosamond}, \citenamefont {Linfield}, \citenamefont {Burnell},\ and\
  \citenamefont {Marrows}}]{Zeissler20}%
  \BibitemOpen
  \bibfield  {author} {\bibinfo {author} {\bibfnamefont {K.}~\bibnamefont
  {Zeissler}}, \bibinfo {author} {\bibfnamefont {S.}~\bibnamefont {Finizio}},
  \bibinfo {author} {\bibfnamefont {C.}~\bibnamefont {Barton}}, \bibinfo
  {author} {\bibfnamefont {A.~J.}\ \bibnamefont {Huxtable}}, \bibinfo {author}
  {\bibfnamefont {J.}~\bibnamefont {Massey}}, \bibinfo {author} {\bibfnamefont
  {J.}~\bibnamefont {Raabe}}, \bibinfo {author} {\bibfnamefont {A.~V.}\
  \bibnamefont {Sadovnikov}}, \bibinfo {author} {\bibfnamefont {S.~A.}\
  \bibnamefont {Nikitov}}, \bibinfo {author} {\bibfnamefont {R.}~\bibnamefont
  {Brearton}}, \bibinfo {author} {\bibfnamefont {T.}~\bibnamefont {Hesjedal}},
  \bibinfo {author} {\bibfnamefont {G.}~\bibnamefont {van~der Laan}}, \bibinfo
  {author} {\bibfnamefont {M.~C.}\ \bibnamefont {Rosamond}}, \bibinfo {author}
  {\bibfnamefont {E.~H.}\ \bibnamefont {Linfield}}, \bibinfo {author}
  {\bibfnamefont {G.}~\bibnamefont {Burnell}}, \ and\ \bibinfo {author}
  {\bibfnamefont {C.~H.}\ \bibnamefont {Marrows}},\ }\bibfield  {title}
  {\enquote {\bibinfo {title} {Diameter-independent skyrmion {H}all angle
  observed in chiral magnetic multilayers},}\ }\href {\doibase
  10.1038/s41467-019-14232-9} {\bibfield  {journal} {\bibinfo  {journal}
  {Nature Commun.}\ }\textbf {\bibinfo {volume} {11}},\ \bibinfo {pages} {428}
  (\bibinfo {year} {2020})}\BibitemShut {NoStop}%
\bibitem [{\citenamefont {Litzius}\ \emph {et~al.}(2020)\citenamefont
  {Litzius}, \citenamefont {Leliaert}, \citenamefont {Bassirian}, \citenamefont
  {Rodrigues}, \citenamefont {Kromin}, \citenamefont {Lemesh}, \citenamefont
  {Zazvorka}, \citenamefont {Lee}, \citenamefont {Mulkers}, \citenamefont
  {Kerber}, \citenamefont {Heinze}, \citenamefont {Keil}, \citenamefont
  {Reeve}, \citenamefont {Weigand}, \citenamefont {Van~Waeyenberge},
  \citenamefont {Sch{\" u}tz}, \citenamefont {Everschor-Sitte}, \citenamefont
  {Beach},\ and\ \citenamefont {Kla{\" u}i}}]{Litzius20}%
  \BibitemOpen
  \bibfield  {author} {\bibinfo {author} {\bibfnamefont {K.}~\bibnamefont
  {Litzius}}, \bibinfo {author} {\bibfnamefont {J.}~\bibnamefont {Leliaert}},
  \bibinfo {author} {\bibfnamefont {P.}~\bibnamefont {Bassirian}}, \bibinfo
  {author} {\bibfnamefont {D.}~\bibnamefont {Rodrigues}}, \bibinfo {author}
  {\bibfnamefont {S.}~\bibnamefont {Kromin}}, \bibinfo {author} {\bibfnamefont
  {I.}~\bibnamefont {Lemesh}}, \bibinfo {author} {\bibfnamefont
  {J.}~\bibnamefont {Zazvorka}}, \bibinfo {author} {\bibfnamefont {K.-J.}\
  \bibnamefont {Lee}}, \bibinfo {author} {\bibfnamefont {J.}~\bibnamefont
  {Mulkers}}, \bibinfo {author} {\bibfnamefont {N.}~\bibnamefont {Kerber}},
  \bibinfo {author} {\bibfnamefont {D.}~\bibnamefont {Heinze}}, \bibinfo
  {author} {\bibfnamefont {N.}~\bibnamefont {Keil}}, \bibinfo {author}
  {\bibfnamefont {R.~M.}\ \bibnamefont {Reeve}}, \bibinfo {author}
  {\bibfnamefont {M.}~\bibnamefont {Weigand}}, \bibinfo {author} {\bibfnamefont
  {B.}~\bibnamefont {Van~Waeyenberge}}, \bibinfo {author} {\bibfnamefont
  {G.}~\bibnamefont {Sch{\" u}tz}}, \bibinfo {author} {\bibfnamefont
  {K.}~\bibnamefont {Everschor-Sitte}}, \bibinfo {author} {\bibfnamefont
  {G.~S.~D.}\ \bibnamefont {Beach}}, \ and\ \bibinfo {author} {\bibfnamefont
  {M.}~\bibnamefont {Kla{\" u}i}},\ }\bibfield  {title} {\enquote {\bibinfo
  {title} {The role of temperature and drive current in skyrmion dynamics},}\
  }\href {\doibase 10.1038/s41928-019-0359-2} {\bibfield  {journal} {\bibinfo
  {journal} {Nature Electron.}\ }\textbf {\bibinfo {volume} {3}},\ \bibinfo
  {pages} {30--36} (\bibinfo {year} {2020})}\BibitemShut {NoStop}%
\bibitem [{\citenamefont {Lin}\ \emph {et~al.}(2013)\citenamefont {Lin},
  \citenamefont {Reichhardt}, \citenamefont {Batista},\ and\ \citenamefont
  {Saxena}}]{Lin13}%
  \BibitemOpen
  \bibfield  {author} {\bibinfo {author} {\bibfnamefont {S.-Z.}\ \bibnamefont
  {Lin}}, \bibinfo {author} {\bibfnamefont {C.}~\bibnamefont {Reichhardt}},
  \bibinfo {author} {\bibfnamefont {C.~D.}\ \bibnamefont {Batista}}, \ and\
  \bibinfo {author} {\bibfnamefont {A.}~\bibnamefont {Saxena}},\ }\bibfield
  {title} {\enquote {\bibinfo {title} {Particle model for skyrmions in metallic
  chiral magnets: Dynamics, pinning, and creep},}\ }\href {\doibase
  10.1103/PhysRevB.87.214419} {\bibfield  {journal} {\bibinfo  {journal} {Phys.
  Rev. B}\ }\textbf {\bibinfo {volume} {87}},\ \bibinfo {pages} {214419}
  (\bibinfo {year} {2013})}\BibitemShut {NoStop}%
\bibitem [{\citenamefont {Reichhardt}\ and\ \citenamefont
  {Reichhardt}(2009)}]{Reichhardt09}%
  \BibitemOpen
  \bibfield  {author} {\bibinfo {author} {\bibfnamefont {C.}~\bibnamefont
  {Reichhardt}}\ and\ \bibinfo {author} {\bibfnamefont {C.~J.~Olson}\
  \bibnamefont {Reichhardt}},\ }\bibfield  {title} {\enquote {\bibinfo {title}
  {Random organization and plastic depinning},}\ }\href {\doibase
  10.1103/PhysRevLett.103.168301} {\bibfield  {journal} {\bibinfo  {journal}
  {Phys. Rev. Lett.}\ }\textbf {\bibinfo {volume} {103}},\ \bibinfo {pages}
  {168301} (\bibinfo {year} {2009})}\BibitemShut {NoStop}%
\bibitem [{\citenamefont {Brown}\ \emph {et~al.}(2019)\citenamefont {Brown},
  \citenamefont {T\"auber},\ and\ \citenamefont {Pleimling}}]{Brown19}%
  \BibitemOpen
  \bibfield  {author} {\bibinfo {author} {\bibfnamefont {B.~L.}\ \bibnamefont
  {Brown}}, \bibinfo {author} {\bibfnamefont {U.~C.}\ \bibnamefont {T\"auber}},
  \ and\ \bibinfo {author} {\bibfnamefont {M.}~\bibnamefont {Pleimling}},\
  }\bibfield  {title} {\enquote {\bibinfo {title} {Skyrmion relaxation dynamics
  in the presence of quenched disorder},}\ }\href {\doibase
  10.1103/PhysRevB.100.024410} {\bibfield  {journal} {\bibinfo  {journal}
  {Phys. Rev. B}\ }\textbf {\bibinfo {volume} {100}},\ \bibinfo {pages}
  {024410} (\bibinfo {year} {2019})}\BibitemShut {NoStop}%
\bibitem [{\citenamefont {Gong}\ \emph {et~al.}(2020)\citenamefont {Gong},
  \citenamefont {Yuan},\ and\ \citenamefont {Wang}}]{Gong20}%
  \BibitemOpen
  \bibfield  {author} {\bibinfo {author} {\bibfnamefont {X.}~\bibnamefont
  {Gong}}, \bibinfo {author} {\bibfnamefont {H.~Y.}\ \bibnamefont {Yuan}}, \
  and\ \bibinfo {author} {\bibfnamefont {X.~R.}\ \bibnamefont {Wang}},\
  }\bibfield  {title} {\enquote {\bibinfo {title} {Current-driven skyrmion
  motion in granular films},}\ }\href {\doibase 10.1103/PhysRevB.101.064421}
  {\bibfield  {journal} {\bibinfo  {journal} {Phys. Rev. B}\ }\textbf {\bibinfo
  {volume} {101}},\ \bibinfo {pages} {064421} (\bibinfo {year}
  {2020})}\BibitemShut {NoStop}%
\bibitem [{\citenamefont {Reichhardt}\ and\ \citenamefont
  {Reichhardt}(2019{\natexlab{b}})}]{Reichhardt19}%
  \BibitemOpen
  \bibfield  {author} {\bibinfo {author} {\bibfnamefont {C.}~\bibnamefont
  {Reichhardt}}\ and\ \bibinfo {author} {\bibfnamefont {C.~J.~O.}\ \bibnamefont
  {Reichhardt}},\ }\bibfield  {title} {\enquote {\bibinfo {title} {Thermal
  creep and the skyrmion {H}all angle in driven skyrmion crystals},}\ }\href
  {\doibase 10.1088/1361-648X/aaefd7} {\bibfield  {journal} {\bibinfo
  {journal} {J. Phys.: Condens. Matter}\ }\textbf {\bibinfo {volume} {31}},\
  \bibinfo {pages} {07LT01} (\bibinfo {year} {2019}{\natexlab{b}})}\BibitemShut
  {NoStop}%
\bibitem [{\citenamefont {Reichhardt}\ and\ \citenamefont
  {Reichhardt}(2021)}]{Reichhardt21a}%
  \BibitemOpen
  \bibfield  {author} {\bibinfo {author} {\bibfnamefont {C.}~\bibnamefont
  {Reichhardt}}\ and\ \bibinfo {author} {\bibfnamefont {C.~J.~O.}\ \bibnamefont
  {Reichhardt}},\ }\bibfield  {title} {\enquote {\bibinfo {title} {Drive
  dependence of the {H}all angle for a sliding {W}igner crystal in a magnetic
  field},}\ }\href {\doibase 10.1103/PhysRevB.103.125107} {\bibfield  {journal}
  {\bibinfo  {journal} {Phys. Rev. B}\ }\textbf {\bibinfo {volume} {103}},\
  \bibinfo {pages} {125107} (\bibinfo {year} {2021})}\BibitemShut {NoStop}%
\bibitem [{\citenamefont {Sato}\ \emph {et~al.}(2019)\citenamefont {Sato},
  \citenamefont {Koshibae}, \citenamefont {Kikkawa}, \citenamefont {Yokouchi},
  \citenamefont {Oike}, \citenamefont {Taguchi}, \citenamefont {Nagaosa},
  \citenamefont {Tokura},\ and\ \citenamefont {Kagawa}}]{Sato19}%
  \BibitemOpen
  \bibfield  {author} {\bibinfo {author} {\bibfnamefont {T.}~\bibnamefont
  {Sato}}, \bibinfo {author} {\bibfnamefont {W.}~\bibnamefont {Koshibae}},
  \bibinfo {author} {\bibfnamefont {A.}~\bibnamefont {Kikkawa}}, \bibinfo
  {author} {\bibfnamefont {T.}~\bibnamefont {Yokouchi}}, \bibinfo {author}
  {\bibfnamefont {H.}~\bibnamefont {Oike}}, \bibinfo {author} {\bibfnamefont
  {Y.}~\bibnamefont {Taguchi}}, \bibinfo {author} {\bibfnamefont
  {N.}~\bibnamefont {Nagaosa}}, \bibinfo {author} {\bibfnamefont
  {Y.}~\bibnamefont {Tokura}}, \ and\ \bibinfo {author} {\bibfnamefont
  {F.}~\bibnamefont {Kagawa}},\ }\bibfield  {title} {\enquote {\bibinfo {title}
  {Slow steady flow of a skyrmion lattice in a confined geometry probed by
  narrow-band resistance noise},}\ }\href {\doibase
  10.1103/PhysRevB.100.094410} {\bibfield  {journal} {\bibinfo  {journal}
  {Phys. Rev. B}\ }\textbf {\bibinfo {volume} {100}},\ \bibinfo {pages}
  {094410} (\bibinfo {year} {2019})}\BibitemShut {NoStop}%
\bibitem [{\citenamefont {Sato}\ \emph {et~al.}(2020)\citenamefont {Sato},
  \citenamefont {Kikkawa}, \citenamefont {Taguchi}, \citenamefont {Tokura},\
  and\ \citenamefont {Kagawa}}]{Sato20}%
  \BibitemOpen
  \bibfield  {author} {\bibinfo {author} {\bibfnamefont {T.}~\bibnamefont
  {Sato}}, \bibinfo {author} {\bibfnamefont {A.}~\bibnamefont {Kikkawa}},
  \bibinfo {author} {\bibfnamefont {Y.}~\bibnamefont {Taguchi}}, \bibinfo
  {author} {\bibfnamefont {Y.}~\bibnamefont {Tokura}}, \ and\ \bibinfo {author}
  {\bibfnamefont {F.}~\bibnamefont {Kagawa}},\ }\bibfield  {title} {\enquote
  {\bibinfo {title} {Mode locking phenomena of the current-induced
  skyrmion-lattice motion in microfabricated {MnSi}},}\ }\href {\doibase
  10.1103/PhysRevB.102.180411} {\bibfield  {journal} {\bibinfo  {journal}
  {Phys. Rev. B}\ }\textbf {\bibinfo {volume} {102}},\ \bibinfo {pages}
  {180411} (\bibinfo {year} {2020})}\BibitemShut {NoStop}%
\end{thebibliography}%

\end{document}